\documentclass[11pt,reprint,showpacs,showkeys,nofootinbib,ssfamily]{article}
\usepackage{axodraw2}
\usepackage{amsmath,fontenc,calrsfs}
\usepackage{amsfonts,amssymb,mathtools,slashed,empheq}
\usepackage{cite} 
\usepackage{hyperref} 
\hypersetup{backref, colorlinks=true  } 
\usepackage{multirow}
\usepackage{float}
\usepackage{appendix}
\usepackage{color}
\usepackage{url}
\usepackage{subfigure}
\usepackage{footnote}
\usepackage{authblk} 
\def\beq{\begin{equation}}
\def\eeq{\end{equation}}
\def\bea{\begin{eqnarray}}
\def\eea{\end{eqnarray}}
\def\nn{\nonumber}

\def\chic1{\chi_{c1}}

\newcommand{\ep}{\epsilon}

\newcommand{\gs}{\sigma}

\newcommand{\sis}{\sigma}
\newcommand{\ts}{\tilde{\sigma}}

\newcommand{\ve}{\varepsilon}
\newcommand{\vep}{\varepsilon}

\newcommand{\la}{\langle}
\newcommand{\ra}{\rangle}

\def \Im{\text{Im}\,}

\def\Xint#1{\mathchoice
   {\XXint\displaystyle\textstyle{#1}}%
   {\XXint\textstyle\scriptstyle{#1}}%
   {\XXint\scriptstyle\scriptscriptstyle{#1}}%
   {\XXint\scriptscriptstyle\scriptscriptstyle{#1}}%
   \!\int}
\def\XXint#1#2#3{{\setbox0=\hbox{$#1{#2#3}{\int}$}
    \vcenter{\hbox{$#2#3$}}\kern-.5\wd0}}

\def\dashint{\Xint-}

\def\Xint#1{\mathchoice
   {\XXint\displaystyle\textstyle{#1}}%
   {\XXint\textstyle\scriptstyle{#1}}%
   {\XXint\scriptstyle\scriptscriptstyle{#1}}%
   {\XXint\scriptscriptstyle\scriptscriptstyle{#1}}%
   \!\int}

\newcommand{\hva}{\hat{\mathbf{a}}}
\newcommand{\va}{\mathbf{a}}

\newcommand{\hvk}{\hat{\mathbf{k}}}
\newcommand{\vk}{\mathbf{k}}
\newcommand{\hvp}{\hat{\mathbf{p}}}
\newcommand{\vp}{\mathbf{p}}
\newcommand{\hvq}{\hat{\mathbf{q}}}
\newcommand{\vq}{\mathbf{q}}

\newcommand{\vx}{\mathbf{x}}
\newcommand{\vy}{\mathbf{y}}

\newcommand{\hvz}{\hat{\mathbf{z}}}
\newcommand{\vz}{\mathbf{z}}

\newcommand{\cZ}{\cal{Z}}

\newcommand{\cL}{{\cal{L}}}

\newcommand{\cE}{\cal{E}}
\newcommand{\cef}{\cal{E_F}}
\newcommand{\ceh}{\cal{E_H}}

\newcommand{\IS}{{\mathbb{I}_S}}

\def\tcr#1{\textcolor{black}{#1}}
\def\trr#1{\textcolor{black}{#1}}

\usepackage{hyperref}
\hypersetup{%
  colorlinks = true,
  linkcolor  = black
}

\title{Ladder resummation of spin 1/2 fermion many-body systems with arbitrary partial-wave content} 
\author[]{J.~M.~Alarc\'on\thanks{jmanuel.alarcon@uah.es}}
\affil[]{{\it Universidad de Alcal\'a, Grupo de F\'{\i}sica Nuclear y de Part\'{\i}culas,} 
\\{\it  Departamento de F\'{\i}sica y Matem\'aticas,  28805 Alcal\'a de Henares (Madrid), Spain}}
\author[]{J.~A. Oller\thanks{oller@um.es}}
\affil[]{\it Departamento de F\'{\i}sica, Universidad de Murcia, 30071 Murcia, Spain}

\begin{document}
\maketitle
\begin{abstract}

  We resum the ladder diagrams for the calculation of the energy density $\cE$  of a spin 1/2 fermion many-body system in terms of arbitrary vacuum two-body  scattering amplitudes.
The partial-wave decomposition of the in-medium two-body scattering amplitudes is developed, and the expression for calculating $\cE$  in a partial-wave amplitude expansion is also given.
  The case of contact interactions is completely solved for any content in partial waves  and it is shown to provide renormalized results, expressed directly in terms of scattering data parameters, within cutoff regularization in a wide class of schemes.  The  
  $S$- and $P$-wave interactions are considered up to including the first three-terms in the effective-range expansion,
  paying special attention to the parametric region around the unitary limit.
\end{abstract}


\newpage
\tableofcontents

\section{Introduction}
\label{sec.210319.1}

The study of strongly-interacting many-body Fermi systems is a major challenge in physics. It has implications in particle, nuclear, molecular and atomic physics as well as in condensed matter. From the highest densities reached in natural systems at the interior of neutron stars to the dilute systems of ultracold fermion atoms in optical traps. The use of cold atomic Fermi gases, whose interactions can be manipulated by means of a magnetic field near a  Feshbach resonance, has made possible to study these systems over many strength scales \cite{ketterle.200831.1}, including both the strong and weak interacting regimes (attractive and repulsive ones)  by fine-tuning the $S$-wave scattering length $a_0$. 
In this way, the smooth crossover between the Bardeen-Cooper-Schrieffer (BCS) superfluidity ($a_0<0$) \cite{Zwierlein2005VorticesAS,Schunck2007SuperfluidEO} to the molecular Bose-Einstein Condensate (BEC) \cite{Anderson1995ObservationOB,davis.200901.1} ($a_0>0$) could be settled and studied by changing the sign of $a_0$ while keeping it large in modulus.
In between, the scattering length diverges and we have the so-called unitary limit ($a_0=\infty$) \cite{zwerger.200901.1,Giorgini2008TheoryOU,Randeria2013BCSBECCA}.  
Denoting by $k_F$ the Fermi momentum of the system and by $R$ the range of the interactions, the unitary limit takes place when $|k_F a_0|\to \infty$ and $R k_F\to 0$, so that one can neglect higher order terms in powers of $k_F$ from the effective-range expansion (ERE) \cite{Bethe.200901.1} in $S$-wave,\footnote{This definition of the unitary limit can be considered analogously to any other partial-wave amplitude taken as reference, cf. Sec.~\ref{sec.201229.1}.} as well as the higher partial waves. Its properties have been studied in great detail experimentally during the last years \cite{Navon2010TheEO,Ku2012RevealingTS,Zuern2013PreciseCO}.

In the unitary limit there is scale invariance \cite{Kolck2017UnitarityAD}, which implies that the only energy scale that is available is the Fermi energy $E_F$ of a free Fermi gas, $E_F=k_F^2/2m$, and the physics is said to be universal without depending on any interaction parameter \cite{ketterle.200831.1,tan27}. As a result the energy per fermion $\bar{\cE}=E/N$ is proportional to $E_F$,
\begin{align}
\bar{\cE}&=\xi \frac{3k_F^2}{10 m}~,
\end{align}
where $\xi$ is the so-called Bertsch parameter. According to experimental determinations from ultracold fermion atoms $\xi=0.370(5)(8)$ \cite{Zuern2013PreciseCO}. Similarly, the critical temperature  ($T_C$) for the superfluid transition and the binding energy of a pair of fermions are also proportional to $E_F$.

The theoretical calculation of the number $\xi$ is a non-perturbative problem. Numerical calculations within quantum Monte Carlo Methods  \cite{Forbes:2010gt,Forbes:2012ku} provide $\xi\approx 0.38$, which is compatible with the experimental determinations in Refs.~\cite{Ku2012RevealingTS,Zuern2013PreciseCO}. In particular Ref.~\cite{Carlson:2011kv} gives the value $0.372\pm 0.005$ employing auxiliary-field quantum Monte Carlo method. Interesting results are also provided by a perturbative expansion in the number of spatial dimensions $d$ that can be developed  around the even values $d=4$ or 2 \cite{Nishida:2006br,Nishida:2006eu}.  
Here the small parameter is $\epsilon=4-d$ or $d-2$, respectively, being  finally fixed to 1 in order to reach $d=3$ spatial dimensions. 
Calculations are performed up to next-to-leading order in Ref.~\cite{Nishida:2006br} with the result $\xi\approx 0.475$. By considering simultaneously the expansion around $d=4$ and $d=2$ and connecting them to $d=3$ by the use of interpolators \cite{Nishida:2006eu,Nishida:2008mh} a smaller value $\xi=0.377\pm0.014$ is concluded in the next-to-next-to-leading order calculation of Ref.~\cite{Nishida:2008mh}.  
Density functional theory has also been  applied for the study of the unitary limit \cite{Papenbrock:2005bd,Lacroix:2016dfs,Boulet:2019wfd,Grasso:2018pen}.

The many-body calculations within  perturbation theory \cite{fetter} 
are well-known   
since long, giving rise to the low-density expansion
for a hard-sphere scattering in the classical
papers by Huang, Yang and Lee \cite{Huang:1957im,Lee:1957zza} in powers of the hard-sphere radius. 
These calculations on a systematic low-density expansion for the ground-state energy were extended later on to higher orders in Ref.~\cite{efimov:1965,efimov:1968,Baker:1971vm,bishop:1973}, and rederived within the context of effective field theory (EFT) in Ref.~\cite{Hammer:2000xg}. 
However, for larger scattering lengths
the perturbative expansion in powers of $a_0 k_F$ fails and, in particular, these expressions cannot be used in the unitary limit $|a_0 k_F|\to\infty$. This is closely the case for neutron matter due to the large and negative neutron-neutron ($nn$)  scattering length   $a_{nn}=-18.95\pm 0.40$~fm \cite{Chen:2008zzj}, such that $|a_{nn}|\gg m_\pi^{-1}$, with the pion mass denoted by $m_\pi$ and whose inverse controls typically the longest range of strong interactions. At the same  time $k_F$ is several times $m_\pi$ in the region of interest for such systems \cite{Lacour:2009ej,Dobado:2011gd}. 

In such circumstances one possibility to reach a finite result for $a_0 k_F\to\pm \infty$ (i.e. meaningful for large scattering lengths in general) is to resum the two-body interactions in the medium. 
According to the Brueckner theory \cite{bethebrueckner,Brueckner:1954zz,Brueckner:1955nst,Brueckner:1955zze}  one sums over  particle-particle intermediate states, in which the two particles always have momenta above the Fermi momenta (in short we say that they lie above the Fermi sea), while they are allowed to rescatter any number of times. This theory was generalized by Thouless in Ref.~\cite{Thouless:1960anp} to allow for two-fermion intermediate states below the Fermi sea (or hole-hole states). He also introduced the notation of ladder diagrams to refer to the associated Feynman graphs, so that both particle-particle and hole-hole intermediate states interact between two consecutive \tcr{rungs of the ladder series.}

The ladder resummation at zero temperature  is undertaken e.g. in Refs.~\cite{Steele:2000qt,Lacour:2009ej,Kaiser:2011cg,Kaiser:2012sr} with in-medium propagators accounting for Pauli blocking without including self-energy effects.   
 Remarkably, an algebraic renormalized form for this resummation is obtained by Kaiser in Ref.~\cite{Kaiser:2011cg} for the case of a pure contact $S$-wave interaction  between two spin 1/2 fermions.
It was shown in Ref.~\cite{Thouless:1960anp} that the ladder resummation converges above the critical temperature for describing the normal matter.
In the case of zero temperature ($T=0$) this implies that in the unitary limit this kind of resummation cannot describe the true superfluid ground state which happens below a critical temperature \cite{Ku2012RevealingTS}.\footnote{\tcr{Needless to say,  many-body theory for infinite systems  at $T=0$ has long been  extensively applied to describe \mbox{(non-)perturbative} interacting electrons and nucleons, describing ground-state properties,  fermion correlations and pairing, transport properties,   interactions with external particles and probes, phase transitions as a function of increasing density from gaseous to liquid and (Wigner) solids governed by the Coulomb interactions, supersolids (dipole-dipole interactions), etc. For classical textbooks the reader can consult e.g. \cite{fetter,pines,migdal}.}} Indeed, the BCS theory shows that interactions of fermion pairs with null total spin and momentum contribute a finite amount to the energy per particle, which of course does not occur in the ladder resummation (cf. Eq.~\eqref{190615.1}).  
This is indeed the case in Ref.~\cite{Kaiser:2011cg} where the Bertsch parameter found $\xi\simeq 0.507$
is close to the   experimentally measured value in Ref.~\cite{Ku2012RevealingTS} by extrapolating to $T=0$ the results above $T_C$.
  The connection between the ladder resummation in many-body calculations and the density-functional theory has been discussed in  \cite{Boulet:2019wfd,Grasso:2018pen}. 
Since $|a_{nn}|$ is very large compared to the range of strong interactions, around 1~fm, the sophisticated many-body calculations in dilute neutron matter, including quantum Monte Carlo techniques \cite{Carlson:2003wm,Gezerlis:2009iw}, 
are of interest in relation with the unitary limit yielding a value $\xi_{nn}\simeq 0.5$, with the subscript $nn$ referring to neutron matter. 
 The effective-range  in $S$-wave is expected to provide sizeable contributions to the energy per particle in realistic neutron matter \cite{Schwenk:2005ka,Carlson:2011kv}, as we also check here in Sec.~\ref{sec.201230.1}.    There is an ongoing effort in EFT to study the properties of nuclear and atomic systems whose two-body subsystems are near the unitary limit, with the binding energy of the three-body system that is conjectured to establish essentially the relevant scale for the low-energy observables, like the energy per particle  \cite{Kolck2017UnitarityAD,Konig:2016utl}.

The extension of the ladder resummation of Ref.~\cite{Kaiser:2011cg} so as to account for the contributions  of an $S$-wave  effective-range ($r_0$)  
was addressed by the same author in Ref.~\cite{Kaiser:2012sr}, where due to off-shell effects, the arctangent-function
formula obtained was conjectured and checked up to some finite order with diagrammatic methods. The extension to treat a $P$-wave scattering volume $a_1$ was discussed in the same reference \cite{Kaiser:2012sr}  as well. 

In this work we undertake the generalization of the ladder resummation of Refs.~\cite{Kaiser:2011cg,Kaiser:2012sr} in the calculation of $\bar{\cE}$ at zero temperature so as 
to include arbitrary many higher orders in the ERE of a given partial-wave amplitude (PWA), as well as any number of PWAs. 
In order to achieve this result we have found crucial to use the derivation 
of many-body field theory achieved in Ref.~\cite{Oller:2001sn},
since it offers a reordering of the diagrams involved in the calculation
of $\bar{\cE}$ that allows the solution of the non-trivial combinatoric problem \cite{Kaiser:2011cg,Kaiser:2012sr} 
associated with a general expression for the resummation of ladders diagrams including
both particle-particle and hole-hole states.

 The case of contact interactions is further studied and renormalized results for $\bar{\cE}$ are obtained, being expressed directly in terms of vacuum scattering parameters. 
 We proceed by considering a generic cutoff regularization scheme characterized by the evaluation of loop integrals in powers of a momentum cutoff $\Lambda$, 
as in Ref.~\cite{vanKolck:1998bw}. Within this language dimensional regularization (DR) is a particular case in which all cutoff powers are absent.  
Kaiser \cite{Kaiser:2011cg,Kaiser:2012sr} performs his calculations in DR and obtains that the value of the Bertsch parameter in the unitary limit depends of the order in which the limit $a_0 \to \infty$ and $r_0\to 0$ are taken. If $a_0\to\infty$ is taken first and then $r_0\to 0$ the Bertsch parameter increases to 0.876 \cite{Kaiser:2012sr}, that is very different to the value  $\xi\approx 0.51$ obtained previously in \cite{Kaiser:2011cg}, with $r_0=0$ first and then letting $a_0\to \infty$.
For $P$-waves  Ref.~\cite{Kaiser:2012sr} obtains that in the limit $a_1\to\infty$ the energy per particle $\bar{\cE}\simeq -E_F$, which indicates an overwhelming attraction. In Ref.~\cite{Schafer:2005kg} it was shown that to resum separately  the particle-particle intermediate states, on the one hand, and  the hole-hole ones, on the other hand, implies a dependence for the 
effects of including $r_0$ on the value taken for the renormalization scale of the power-divergence subtraction scheme used \cite{Kaplan:1998we}. Then, one wonders whether a renormalization issue by using  DR  is affecting the results of Ref.~\cite{Kaiser:2012sr} concerning the unitary limit when $r_0\neq 0$ in $S$ wave, and  the $P$ waves with $a_1\neq 0$.
This question is addressed in detail in Secs.~\ref{sec.201230.1} and \ref{sec.210104.1} where renormalized solutions  employing a cutoff regulator are obtained, and we show that: i) The results are independent of the cutoff scheme used, ii) they are perturbative regarding the $S$-wave effective range $r_0$, and iii) one cannot obtain renormalized results in $P$-wave with only the scattering volume and the effective range is also needed. \tcr{Furthermore, we also include up to three terms in the ERE, one parameter more in $S$ wave and two more in $P$ wave than Ref.~\cite{Kaiser:2012sr}}. For the case of a vanishing $P$-wave effective range, $r_1=0$, we find that $\bar{\cE}\simeq - E_F$ in the limit $a_1\to \infty$, similarly as in Ref.~\cite{Kaiser:2012sr}.

The contents of the article are presented as follows. In the next Sec.~\ref{sec.200904.1} we introduce the in-medium quantum field theory and apply it to the evaluation of the Fock and Hartree contributions within the ladder approximation for an arbitrary fermion-fermion scattering amplitude in vacuum. An important point discussed in Sec.~\ref{sec.200907.2} is that the resulting energy density is real, that is not a trivial resulting property because both particle-particle and hole-hole contributions are resummed. The formalism of partial-wave amplitudes to calculate the  energy density and the in-medium scattering amplitude is derived along Secs.~\ref{sec.190630.1} and \ref{sec.190808.1}, respectively. 
Several applications of this formalism for contact interactions are developed in Sec.~\ref{sec.201229.1} to calculate the energy density by considering $S$- and $P$-wave potentials with couplings renormalized with cutoff regularization in order to match the effective-range expansion up to some order. In particular, we  pay special attention to the impact of including the effective range.  
  Some more technical material is relegated to the Appendices \ref{app.201228.1} and \ref{app.200925.1}, while in Appendix \ref{app.210102.1} we apply  our generic formalism  in DR for the $S$ and $P$ waves. As a by product, we algebraically reproduce   Kaiser's results there given in Ref.~\cite{Kaiser:2012sr}.\footnote{The Appendix \ref{app.210102.1} is specially suited for a pedagogical illustration of the methods  exposed.} \trr{We discuss the emergence of poles in the border of the Fermi seas of two in-medium fermions interacting in $S$ wave, a characteristic feature for pairing, in Appendix~\ref{app.211029.1}. Finally, the values for the scattering length and effective range in $P$ wave giving rise to unacceptable resonant poles with positive imaginary part in the complex-$p$ plane are characterized in Appendix \ref{app.211027.1}.}

\section{Resummation of ladder diagrams for the energy density $\cE$}
\label{sec.200904.1}

\def\theequation{\arabic{section}.\arabic{equation}}
\setcounter{equation}{0}

We develop in this work a calculation of the energy density ${\cal E}$ by resumming the ladder diagrams for the in-medium two-fermion interactions.
Our derivation is based on the many-body formalism of Ref.~\cite{Oller:2001sn},
which we denote as the in-medium many-body quantum field theory, whose main points are briefly reviewed next.

\subsection{Basics of the in-medium many-body formalism of Ref.~\cite{Oller:2001sn}}
\label{sec.201223.1}

 The Ref.~\cite{Oller:2001sn} calculates the generating functional ${\cZ}[J]$ of in-medium Green functions
with external sources $J$'s. 
The Lagrangian of the system consists of the  pure bosonic Lagrangian in vacuum, ${\cal L}_\phi$, and the
bilinear fermion operators that are encoded globally as $\bar{\psi}D\psi$.
The operator $D$ is then split as $D_0-A$, where $D_0$ is the free fermion Lagrangian,
$i\gamma^\mu\partial_\mu-m$ (with $m$ the physical fermion mass) and the operator $A$ incorporates the
boson-fermion interactions.  The bosons could either correspond to light degrees of freedom, e.g. pions in nuclear physics,
as well as to heavier ones. The latter, when taking their masses to infinity, generate 
the contact multi-fermion interactions. This statement is analogous to the essence of the
Hubbard-Stratonovich transformation \cite{Stratonovich:1957dan,Hubbard:1959ub}.

Within a path integral formulation, the main result of Ref.~\cite{Oller:2001sn}, with non-relativistic kinematics for the fermions, can be expressed as
\begin{align}
\label{190521.1}
 e^{i{\cZ}[J]}&=\int[dU]\exp\left (
i\int d^4x {\cal L}_{\phi}+{\rm Tr}\int\frac{d^3p\, n(p)}{(2\pi)^3}\int d^3 x d^3y  e^{-i\vp \vx}{\rm log}{\cal F}(\vx,\vy)e^{i\vp \vy}
\right)~,
\end{align}
where the trace is taken over spin and other internal indices, like
isospin ones for nucleons. 
The integration over the boson fields is represented by the matrix field $U$.
The right-hand side dependence of the equation  on the external sources $J$ is implicit through ${\cal L}_\phi$ and
$A$.  The function $n(p)$ selects the in-medium fermions through a factor $\theta(k_{F\alpha}-|\vp|)$ for each fermion species $\alpha$, where \trr{${k_{F_\alpha}}$ is the corresponding Fermi momentum} and $\theta(x)$ is the Heaviside or step function. We explain next 
other symbols appearing in this equation.

The non-local operator ${\cal F}$ is
\begin{align}
\label{190521.2}
{\cal F}(\vx,\vy)_{\alpha\beta}&=\delta(\vx-\vy)\delta_{\alpha\beta}
-i\int dt\int dt' e^{iH_0t} \big[A[I-D_0^{-1}A]^{-1}\big](\vx,\vy)_{\alpha\beta}e^{-iH_0t'}~,
\end{align}
and it stems from the integration of the fermion fields in the path integral. 
The operator $D_0^{-1}$ is the inverse of $D_0$, and  in momentum space it is given by  
\begin{align}
\label{200904.1}
iD_0^{-1}(p)=\frac{i}{p^0-E(p)+i\ep}~,
\end{align}
corresponding to the vacuum non-relativistic propagator of a fermion.
In Eq.~\eqref{190521.2} \trr{$H_0$ is the free-fermion Hamiltonian associated to $\bar{\psi}D_0\psi$,} and $e^{\pm iH_0t}$ acts on the one-particle
intermediate states made of in-medium fermions with momentum $|\vp|<{k_{F_\alpha}}$. 
As a result of it
these states are multiplied by $e^{\pm i E(p) t}$,  
$E(p)=\vp^2/2m$, which ultimately drives to energy conservation after integrating in $t$ and $t'$ in Eq.~\eqref{190521.2}.

For simplicity, although the generalization is possible, we do not take into account possible polarization phenomena
in the medium and then take the
Fermi momenta independent on the spin of the fermions.  
In the case of isospin symmetry  in nuclear matter it is convenient to introduce the  $2\times 2$ diagonal matrix $n(p)$ in the isospin space to single out the states within the Fermi seas, defined by
\begin{align}
\label{190521.4b}
n(p)&=\left(
\begin{matrix}
\theta({k_{F_1}}-|\vp|) & 0\\
0 & \theta({k_{F_2}}-|\vp|)
\end{matrix}
\right)~.
\end{align}

Next,  $\log {\cal F}$ in Eq.~\eqref{190521.1} is expanded in powers of  
$A[I-D_0^{-1}A]$, and then it results
\begin{align}
\label{190521.7}
&e^{i{\cZ}[J]}=\!\int\! [dU] \exp\Big[i\!\int\! dx\, {\cL}_{\phi}
  -i\!\int\!\! \frac{d\vp}{(2\pi)^3 } 
\!\int\! {\rm Tr}\Big( A[I-D_0^{-1}A]^{-1}\arrowvert_{(x,y)}
n(p)\Big) dx\,dy\, e^{ip(x-y)} \\ 
&-\frac{1}{2}(-i)^2\!\int \frac{d\vp}{(2\pi)^3}\!\int\!  \frac{d\vq}{(2\pi)^3}
\!\int\! {\rm Tr}\Big(
A[I-D_0^{-1}A]^{-1}\arrowvert_{(x,x')}n(q)A[I-  D_0^{-1}A]^{-1}\arrowvert_{(y',y)}n(p)\Big)\nn\\
&\times
 e^{ip(x-y)} e^{-iq(x'-y')} dx\,dx'\,dy\,dy' +...\Big ]~.\nn
\end{align}
\tcr{Compared to Ref.~\cite{Oller:2001sn} we use now Pauli spinors normalized to 1. In this way, the factor $2E(p)$ present in the denominator of Eq.~(8) in Ref.~\cite{Oller:2001sn}, which in the non-relativistic limit would become $2m$, does not appear in Eq.~\eqref{190521.1} because the change of normalization compared to the Dirac spinors used in Ref.~\cite{Oller:2001sn}.  
  There is also another trivial change in the definition of the operator ${\cal F}(\vx,\vy)$ in Eq.~\eqref{190521.2} compared to its analogous one in \cite{Oller:2001sn}, because in the non-relativistic limit instead of the  Dirac Gamma matrix $\gamma^0$ one has the $2\times 2$ identity matrix. Finally, $iD_0^{-1}(p)$ in Eq.~\eqref{200904.1} is the standard non-relativistic propagator of a fermion, being the non-relativistic limit of its analogous one in Ref.~\cite{Oller:2001sn} (except for a global factor $2m$ because of the change of normalization already alluded to). Notice also that the vertex operator $A$ in the non-relativistic case can be built to absorb the relativistic corrections \cite{Bernard:1995dp,Machleidt:2011zz}.}

\tcr{In Eq.~\eqref{190521.7}} each term in the sum containing at least a factor $n(p)$ is called an in-medium generalized vertex (IGV). For interpreting them  
let us introduce the symbol $\Gamma$ as in Ref.~\cite{Oller:2001sn}, corresponding to
the non-local vertex
\begin{align}
\label{190521.8}
\Gamma & \equiv-iA[I-D_0^{-1}A]^{-1}=-iA\sum_{n=0}^\infty (D_0^{-1}A)^n~.
\end{align}
In this way, every IGV in  Eq.~\eqref{190521.7} is composed of $\Gamma$ vertices
which are mutually joined by single Fermi-sea insertions, that can be easily recognized by the factors of $n(p)$, times  a definite numerical coefficient provided by the log series 
\begin{align}
\label{190521.4}
\log(1+\ve)&=\ve-\frac{\ve^2}{2}+\frac{\ve^3}{3}-\frac{\ve^4}{4}+\ldots=\sum_{n=1}^\infty (-1)^{n+1}\frac{\ve^n}{n}~,
\end{align}
and then $n$ $\Gamma$ vertices within an IGV are accompanied by the factor $(-1)^{n+1}/n$. 

The Eq.~\eqref{190521.7} gives rise to Feynman rules such that the associated propagators for the fermion lines are either in-medium insertions of on-shell Fermi-seas,
connecting $\Gamma$ vertices, or fermion vacuum propagators $D_0^{-1}$ joining
vacuum vertices $A$.  
In the following, a pure vacuum fermion propagator is depicted as a solid line and
a Fermi-sea insertion is drawn by a double line with one of them in red (online). 
 A vacuum propagator corresponds to $iD_0^{-1}(p)$, written explicitly in momentum space in Eq.~\eqref{200904.1}, and an in-medium insertion implies the factor $n(p)(2\pi)\delta(p^0-\vp^2/2m)$, since the baryons  are on-shell in the in-medium insertions. In both cases one has to integrate over the intermediate four-momentum $\int d^4p/(2\pi)^4$. In addition, each vertex coming from the fundamental bilinear Lagrangian density $\bar{\psi}D\psi$, and  drawn as a filled circle, implies to multiply by the factor $-iA$. Notice that these vertices emerge by expanding in powers of $D_0^{-1}A$ the non-local $\Gamma$ vertices, Eq.~\eqref{190521.8}, which are drawn as empty circles. In addition, one also has to keep in mind that boson (and source) lines can emerge from the $A$ vertices. Putting together all these pieces standard Feynman diagrams emerge ready for its evaluation. For a more extensive discussion on this many-body formalism we refer to the original Ref.~\cite{Oller:2001sn}, and to Ref.~\cite{Meissner:2001gz} where the first practical applications in perturbation theory were developed for several important observables.

\subsection{Fock and Hartree diagram contributions to ${\cal E}$}
\label{sec.190603.1}

We first discuss the contributions to the in-medium  energy density ${\cal E}$ 
by resumming the Fock ladder diagrams, that involve the exchange part of the fermion-fermion scattering amplitude.
The Fock diagrams  
correspond to IGVs without external lines that stem from Eq.~\eqref{190521.7}. 
In this respect, let us notice that the effective in-medium Lagrangian in Eq.~\eqref{190521.7}
is given by  the exponent of the integrand in this equation.

\begin{figure}
\begin{center}
\includegraphics[width=.6\textwidth]{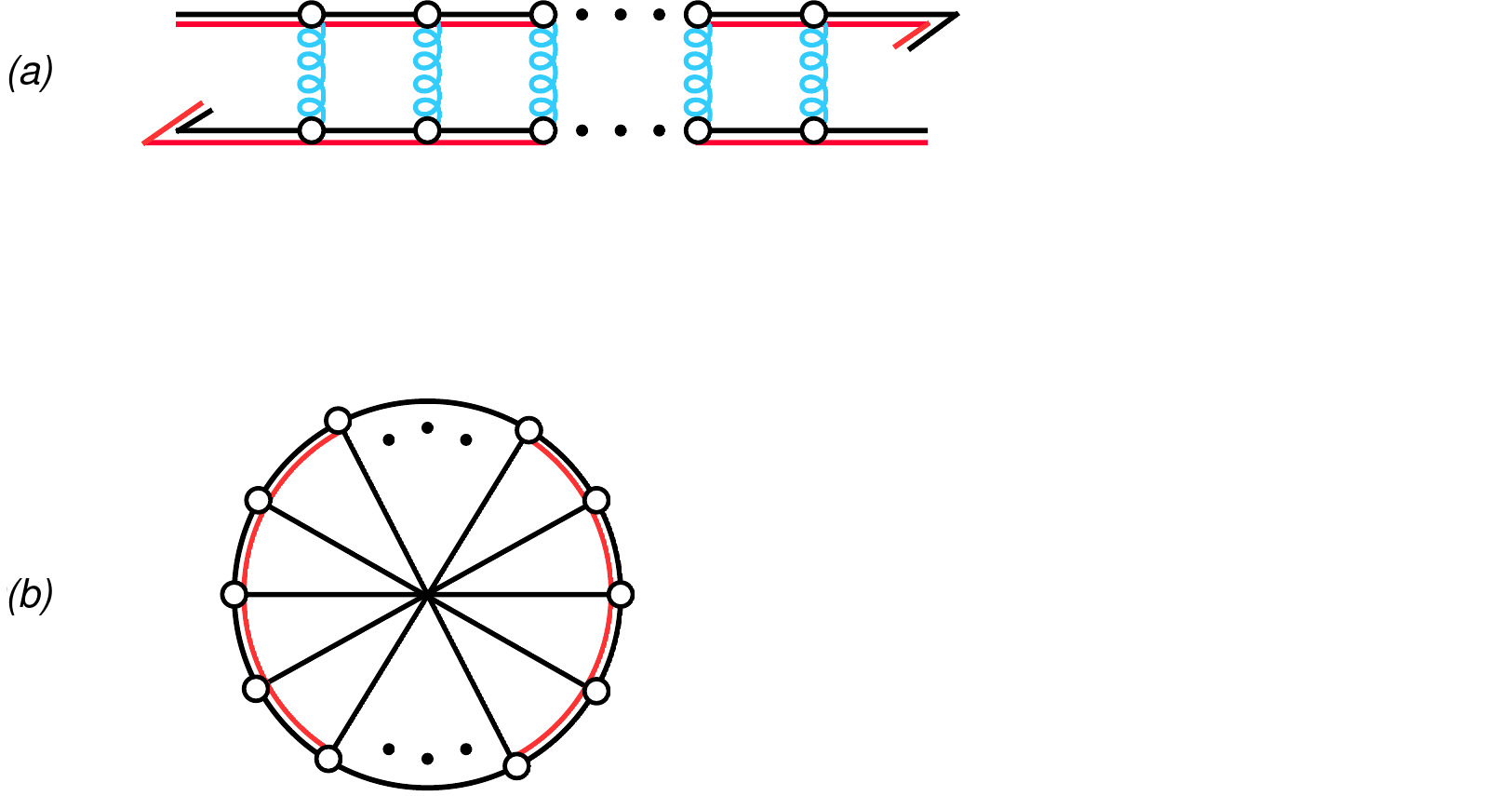}
\end{center}
\caption{{\small The figure (a) represents a Fock diagram for the calculation 
of ${\cal E}$ where all the Fermi-sea insertions 
make up 
intermediate states of the type $\varphi_d$ in which the two fermions belong to their Fermi seas. 
The angle lines at the end and beginning of the top and bottom lines, respectively, indicate that only the exchange part of the fermion-fermion interaction is kept for a Fock diagram. 
The wiggly lines connecting two empty circles ($\Gamma$ vertices)  in (a)  
correspond to the vacuum two-fermion scattering amplitude $t_V$. 
The diagrams (a) and (b) are equivalent to each other.  
In the diagram (b) the connected vertices are those in opposite sides of the radii in the circle.}
\label{fig.190522.1} 
}
\end{figure}

\begin{figure}
\begin{center}
\includegraphics[width=.5\textwidth]{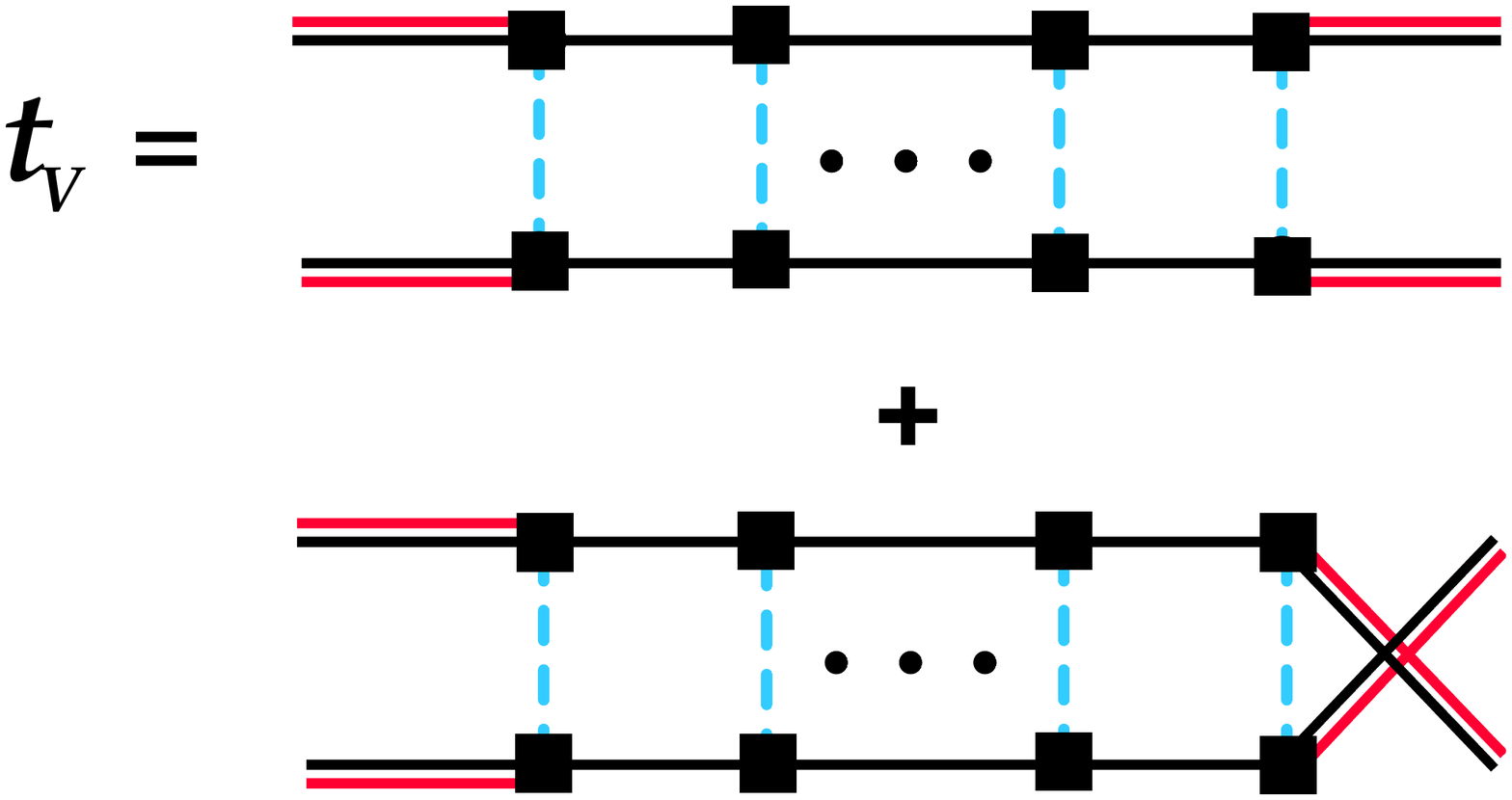}
\end{center}
\caption{{\small Iteration of the vacuum two-fermion intermediate states giving rise to  
the scattering amplitude $t_V$ in vacuum. 
The filled squares on both sides of a vertical dashed line schematically represents a generic fermion-fermion potential
$V$ that results because of the exchange of boson lines between vertices $A$ joined by vacuum fermion propagators $D_0^{-1}$. }
\label{fig.190522.3} 
}
\end{figure} 

A two-fermion intermediate state in which the two fermions belong to  Fermi seas is indicated by $\varphi_d$,
and the total number of them in a Feynman diagram is denoted by $d$. 
A generic Fock diagram in which all the Fermi-sea insertions belong to $\varphi_d$'s
is represented in Fig.~\ref{fig.190522.1}(a), where the two-fermion exchange interaction at the end of the iterative process
is indicated  by the angle lines at the end of the top line and at the beginning of the bottom one.  
In more detail, every wiggly line in Fig.~\ref{fig.190522.1}(a) stemming from a $\Gamma$ vertex (depicted by an empty circle) and ending in another one 
represents the insertion of the vacuum T-matrix $t_V$, which gives rise to the scattering of the two on-shell
fermions belonging to $\varphi_d$. 
The vacuum scattering amplitude $t_V$ results because each $\Gamma$ vertex can be expanded
in a geometric series in powers of $D_0^{-1}A$ involving only vacuum fermion propagators, as explicitly shown in Eq.~\eqref{190521.8}. 
 The connection of the resulting bilinear vertices $A$ by boson lines
 can be interpreted as an interaction potential $V$, that is schematically depicted  in Fig.~\ref{fig.190522.3} 
by a vertical dashed line joining two filled squares. The potential is then iterated by
the two-fermion vacuum intermediate states as shown in the same figure.

 The Fock diagrams can be represented more conveniently in a cartwheel diagram like in Fig.~\ref{fig.190522.1}(b), where 
 the vertices paired by a $t_V$ appear in opposite sides of a radius (plotted as a solid line). 
 This type of circular diagrams is more convenient for visualizing the geometry of the Fock diagrams and their symmetries.

Now, according to the expansion of the in-medium Lagrangian in Eq.~\eqref{190521.7} (namely,  the exponent in its integrand), it is clear that
the diagram of Fig.~\ref{fig.190522.1}, having $2d$ Fermi-sea insertions because there are $d$ $\varphi_d$ intermediate states, is accompanied by a factor $(-1)^{2d+1}/2d=-1/2d$, which results from the expansion of $\log(1+\vep)$  in Eq.~\eqref{190521.4}.
Let us indicate by $L_d$ the loop associated with the insertion of a $\varphi_d$ intermediate state (an explicit
formula is derived below and given in Eq.~\eqref{190615.7}).  
 Then,  the result for Fig.~\ref{fig.190522.1} can be written as
\begin{align}
\label{190521.9}
-\frac{1}{2d}\underbrace{t_VL_d\ldots t_V\bar{L}_d}_{d~\text{factors}~t_V}~,
\end{align}
The last loop associated to $\varphi_d$ is barred because this is a Fock diagram with only
the exchange-particle part of $t_V$ at the end of its iteration in Fig.~\ref{fig.190522.1}(a).

We consider next the 
contributions in which at least one of the two fermions in some of the intermediate states belongs to a Fermi sea and the other corresponds to a vacuum fermion propagator. These mixed intermediate states are denoted by $\varphi_m$ and their total number is $m$. 
A contribution involving $d$ $\varphi_d$'s and $m$ $\varphi_m$'s is represented by $F_{dm}$.
In this notation the contribution of  Eq.~\eqref{190521.9} is $F_{d0}$.

\begin{figure}[H]
\begin{center}
\includegraphics[width=.5\textwidth,angle=0]{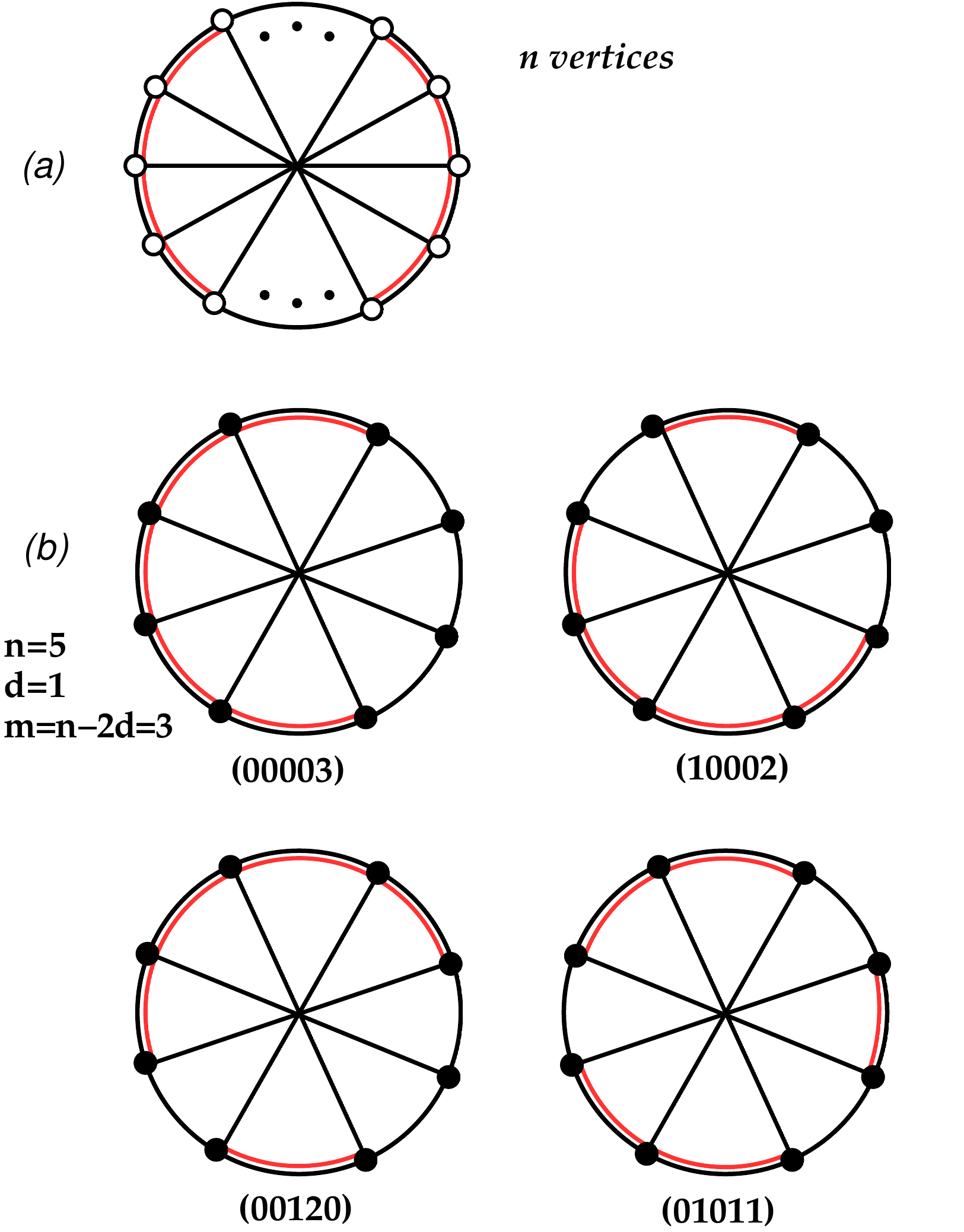}
\end{center}
\caption{{\small In (a) we plot the primordial diagram with only Fermi-sea insertions connecting the 
$\Gamma$ vertices. The generation of vacuum propagators by expanding some of the $\Gamma$ vertices is analyzed in  
  (b) for the case $d=1$ and $n=5$. 
    Below each diagram we give the $5-$tuple in counterclockwise sense with the first $\Gamma$ vertex identified as
    the top one immediately to the left of the vertical diameter. This $5-$tuple indicates how to expand each of the 5
 $\Gamma$ vertices in the primordial diagram for $n=5$, as explained in the text.}
\label{fig.190523.3} 
}
\end{figure}

A general contribution involving $n$ Fermi-sea insertions and $d$ $\varphi_d$
intermediate states, has $m=n-2d$  mixed intermediate states of type $\varphi_m$.   
Our starting point 
is a primordial diagram with $n$ $\Gamma$ vertices
that are connected only through Fermi-sea insertions, which is represented as a ring in Fig.~\ref{fig.190523.3}(a).  
We insert vacuum fermion propagators in the diagram to complete the mixed
$\varphi_m$ intermediate states by expanding an adequate number of $\Gamma$ vertices to the right.\footnote{Notice that the primordial diagram is not necessarily a cartwheel diagram since $n$ could be odd. Once the $\Gamma$ vertices are expanded then it is such.} 
There are typically different possibilities involving $p\leq m$ $\Gamma$ vertices and each of them is  
indicated by the $n-$tuple $(x_{\Gamma_1},x_{\Gamma_2},\ldots,x_{\Gamma_n})$, where $x_{\Gamma_i}$ gives the number
of vacuum vertices that result by expanding the $i_{\rm th}$ $\Gamma$ vertex in the primordial diagram.
As an example, the different possibilities for the case with $n=5$ and $d=1$ are represented in Fig.~\ref{fig.190523.3}(b),
in which extra diagrams that are related by cyclic permutations to the ones shown are not plotted. We recognize diagrams in which:  
i) A $\Gamma$ vertex gives rise to 3 vacuum propagators; ii) a $\Gamma$ vertex does so with 2 vacuum propagators and
another $\Gamma$ vertex with only one; iii) three $\Gamma$ vertices contribute each with one vacuum propagator.

Let us imagine first that all the $x_{\Gamma_i}$ are equal, $i=1,\ldots,n$. The equality $n=2d+nx_{\Gamma_i}$ implies that $n=2d/(1-x_{\Gamma_i})$ and then $x_{\Gamma_i}=0$ since they are either 0 or natural numbers, and the latter option is ruled out because $n$ is a finite natural number. Thus, we end with the contribution $F_{d0}$ already discussed and given by Eq.~\eqref{190521.9}.  

The following  points  fix the combinatorial problem for the calculation of $F_{dm}$, $m\geq 1$:\footnote{Having $m>0$ allows one to distinguish between processes related by rotations of angle $\pi/(d+m)$ in the circle representing a Fock contribution.}

$\bullet$ Given an allowed configuration of vertices $(x_{\Gamma_1},x_{\Gamma_2},\ldots,x_{\Gamma_n})$, we have taken one out of  
the $n$ $\Gamma$ vertices as the first one, and then realize the configuration corresponding to the $n-$tuple.
This could have been done analogously if any other of the $n$ $\Gamma$ vertices were taken as the first one. 
Thus, the representative configuration taken for the
considered $n-$tuple is multiplied by a factor $n$ that cancels the factor $1/n$ from the log expansion in
Eq.~\eqref{190521.7}. Geometrically this is clear from the related cartwheel diagrams,
because  the cyclic permutations between the $\Gamma$ vertices correspond to rotations of angle a multiple of $\pi/(d+m)$.

$\bullet$ By considering the different allowed $n-$tuples one is reproducing all the set of possible
rearrangements of the $m$ mixed $\varphi_m$ intermediate states among the $d$ $\varphi_d$ ones.
All these possibilities also include the mutual exchange in the mixed intermediate states
between  the vacuum propagator and the in-medium insertion.
The loop associated to such symmetrized mixed intermediate state is denoted by $L_m$.\footnote{Its definition also includes the  extra minus sign associated to the accompanying Fermi sea insertion, cf. Eqs.~\eqref{190521.7} and \eqref{190521.4}, 
and its explicit expression is given in Eq.~\eqref{190615.3}.} 
This top bottom symmetry implies that the sum over all these contributions has to be divided by
two.

Again this fact can be most easily seen by employing the cartwheel diagrams and performing a rotation by 180 degrees (so that one exchanges the two medium insertions in the at least one required $\varphi_d$ intermediate state). For explicit examples just consider the action of this rotation on the diagrams in Fig.~\ref{fig.190523.3}(b).

\begin{figure}
\begin{center}
\includegraphics[width=.7\textwidth]{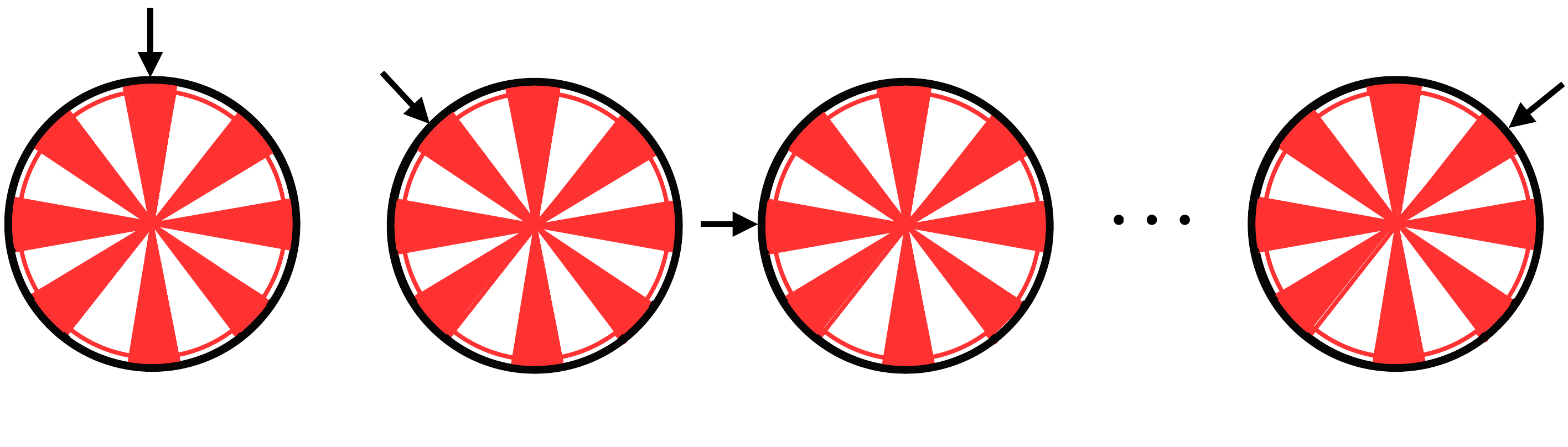}
\end{center}
\caption{{\small Symmetrization process of the $d$ $\varphi_d$ two-fermion intermediate states by taking cyclic permutations among them, corresponding to rotations in the circles by multiples of $\pi/(d+m)$. The filled angular segments correspond to different dispositions of $L_m$ separated by insertions of $t_V$. The arrow is drawn for reference. 
}
\label{fig.201227.1} }
\end{figure}

$\bullet$ We also symmetrize with respect the $d$ $\varphi_d$ two-fermion intermediate states by
taking the $d$ cyclic permutations of their positions in the Feynman diagram.
All these diagrams correspond to the $n$-tuples that are related to each other by cyclic permutations and are already taken into account in the first point. 
We depict this symmetrization process in Fig.~\ref{fig.201227.1}, where the arrow is drawn for reference.
The filled areas correspond to possible different numbers of $L_m$ loops separated by insertions of $t_V$ which, by rotating the circles successively by multiples of the basic angle $\pi/(d+m)$, can be moved cyclically along the circles accomplishing the mentioned symmetrization process.  
This implies that one has to multiply by $1/d$ this symmetrized contributions to avoid double-counting and, then, a factor $-1/2d$ multiply their sum in $F_{dm}$, the same as for $F_{d0}$ given in Eq.~\eqref{190521.9}.

For instance,  if we consider a diagram in which all the $L_m$ are consecutive we would have explicitly
\begin{align}
\label{190523.6}
F_{dm}&=-\frac{1}{2} t_V \underbrace{L_m t_V \ldots t_V L_m}_{m\,L_m} t_V \underbrace{L_d \ldots t_V \bar{L}_d}_{d\, L_d}\\
=-\frac{1}{2d}&\big( t_V L_m t_V \ldots t_V L_m t_V L_d \ldots t_V \bar{L}_d+t_V L_d t_V L_m t_V \ldots t_V L_m t_V \underbrace{L_d \ldots t_V \bar{L}_d}_{d-1\, L_d} + \ldots  \nn\\
+ & t_V \underbrace{L_d t_V \ldots t_V L_d}_{d-1\, L_d} t_V L_m t_V \ldots t_V L_m t_V  \bar{L}_d\big)~.\nn
\end{align}

$\bullet$ The final step is to realize that every $L_m$ is multiplied by a $t_V$ and that one could place any number of
these mixed states in between two $\varphi_d$ ones.  As a result,
between two consecutive $L_d$ loops 
we have the series
\begin{align}
\label{190615.10b}
t_m&=t_V\sum_{m=0}^\infty (L_m t_V)^m=(t_V^{-1}-L_m)^{-1}~.
\end{align}
Therefore, $t_m$  is an in-medium on-shell two-fermion
scattering amplitude that results by iterating $t_V$ between symmetrized intermediate states  of the type $\varphi_m$ (expressed as the loop function $L_m$).

We then conclude that the result for the sum over 
the Fock diagrams, which we call $\cef$, is given by
\begin{align}
\label{190607.4}
\cef&=-\frac{1}{2}{\rm Tr}\left(
\left[\sum_{d=1}^{\infty}\frac{(t_mL_d)^{d-1}}{d}\right] t_m\bar{L}_d \right)~.
\end{align}
with the trace taken in the momentum, isospin and spin spaces.

\begin{figure}[H]
\begin{center}
\includegraphics[width=.6\textwidth,angle=0]{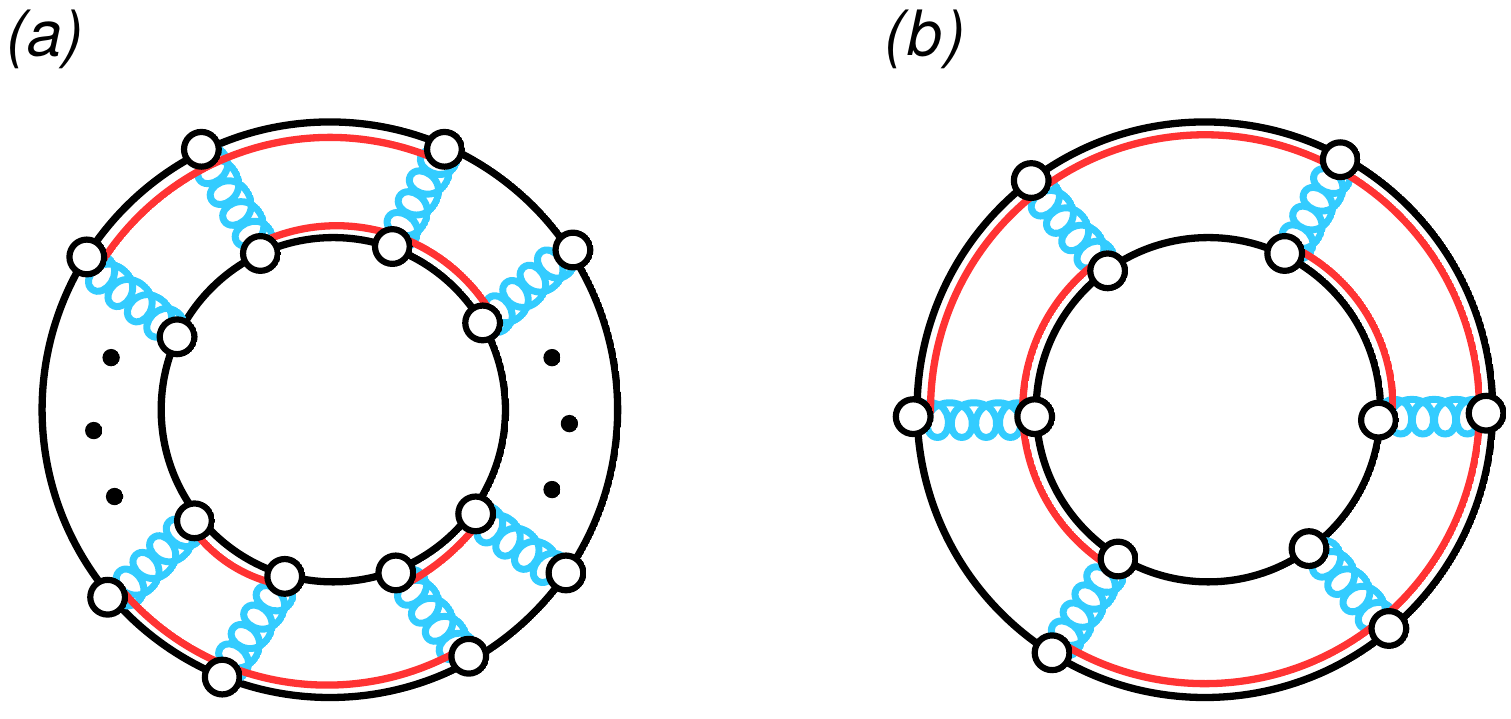}
\end{center}
\caption{{\small (a) Generic Hartree diagram involving $\varphi_d$ and $\varphi_m$ intermediate states. (b) Hartree diagram with $d=m=2$. }
\label{fig.190603.1} 
}
\end{figure}

We now consider the contributions  to $\cE$ from the diagrams of the Hartree type, $\ceh$, that involve the direct part of the fermion-fermion scattering amplitude. The Fig.~\ref{fig.190603.1}(a) represents a generic Hartree diagram with $d$ $\varphi_d$ and $m$ $\varphi_m$ intermediate states, being its contribution called $H_{dm}$. 
These diagrams result by connecting two IGVs of the type given in Eq.~\eqref{190521.7} through the exchange of the vacuum  $T$-matrices $t_V$ (depicted by wiggly lines). 
The fermion lines comprising an intermediate state lie facing each other, and only intermediate states of the type $\varphi_m$ or $\varphi_d$ are plotted.
 In the Fig.~\ref{fig.190603.1}(b) we consider the specific case with $d=2$ and $m=4$.

The expression for the total contribution due to the Hartree diagrams, $\ceh$, can be easily obtained from  Eq.~\eqref{190607.4} for $\cef$, by exploiting the difference in a minus sign between the direct and exchange parts of the  fermion-fermion interaction amplitude, because of the Fermi statistics.  
Thus, 
\begin{align}
\label{190607.5}
\ceh&=\frac{1}{2}{\rm Tr}\left(
\left[\sum_{d=1}^{\infty}\frac{(t_mL_d)^{d-1}}{d}\right] t_m\bar{\bar{L}}_d
\right)~.
\end{align}
The last loop associated to $\varphi_d$ is double-barred because this is a Hartree diagram and only 
the direct part of $t_V$  enters at the end of its iteration in Fig.~\ref{fig.190603.1}(a).

Let us show how the result in Eq.~\eqref{190607.5} arises directly.
First, we notice that any contribution from a
Hartree diagram is multiplied by a factor $1/2$ because of the Dyson series or, in other words, because of the expansion up to
quadratic order of the exponential in Eq.~\eqref{190521.7}. 
Secondly, we denote by $n_1$ and $n_2$ the number of Fermi-sea insertions of the outer and inner rings in a Hartree diagram, respectively, and then  we have the factor $1/2n_1n_2$.
Similarly to the Fock case we can also introduce the idea of a primordial diagram representing schematically the IGVs for the inner and outer rings, so that only Fermi-sea insertions connect $\Gamma$ vertices. 
Vacuum fermion propagators needed to give rise to the $\varphi_m$ intermediate states stem 
from the expansion of these $\Gamma$ vertices to the right. 
As a result, a set of arrangements of Fermi-sea insertions and vacuum propagators generating the needed $d$ $\varphi_d$ and $m$ $\varphi_m$ intermediate states arises. Each of the arrangements is associated to
the pair of tuples $(x_{\Gamma_1}, x_{\Gamma_2},\ldots,x_{\Gamma_{n_1}})$ and 
$(x_{\Gamma_1}, x_{\Gamma_2},\ldots,x_{\Gamma_{n_2}})$, for the outer and inner IGVs, respectively.

Let us consider the Feynman diagrams contributing to $H_{dm}$ in which all the fermion propagators in the inner ring are Fermi-sea insertions, so that $n_2=d+m$. The $n_1$ Fermi-sea insertions in the outer ring give rise all of them to $\varphi_d$ intermediate states, and then $n_1=d$. There are $n_2$ cyclic permutations between the vertices in the inner ring which obviously give the same result, so that these diagrams amount to $n_2$ times one of them. Henceforth, we pick up the expected symmetry factor $n_2/(2dn_2)=1/2d$, that can be recognized already in Eq.~\eqref{190607.5} for $\ceh$. Of course, the same factor arises if we exchange the outer and inner rings between them, and then $n_1=d+m$, $n_2=d$.  
There are other diagrams contributing to $H_{dm}$ that can be generated from the one analyzed 
by exchanging the vacuum propagator and the Fermi-sea insertion comprising every $\varphi_m$ intermediate state  between the different rings. Diagrammatically it is clear that all these diagrams  generate the $L_m$ loops associated to the top-bottom symmetrized mixed intermediate states, and that the same symmetry factor $1/2d$ stays put.  
Indeed, if we have for  $H_{dm}$ that  $n_1=d+m_1$ and $n_2=d+m_2$ ($m_1,\,m_2>0$), clearly       any two $\Gamma$ vertices in the rings of the primordial diagram can be taken as the first ones when considering two possible $n_1$- and $n_2$-tuples. Therefore, one picks up the numerical factor $n_1 n_2/(2 n_1 n_2 d)=1/2d$ after symmetrizing with respect to the $d$ $\varphi_d$ intermediate states, similarly as done for the Fock case in Fig.~\ref{fig.201227.1}.  
It is important to take into account that each $L_m$ is multiplied by a $t_V$ and that any number of 
mixed intermediate states interacting by $t_V$ are allowed in between two $L_d$ loops, so that  the same 
in-medium scattering amplitude $t_m$,  given in Eq.~\eqref{190615.10b}, arises also for the Hartree contribution.
 Then, the Eq.~\eqref{190607.5}  results.

The sum $\cef+\ceh$ corresponds to $-i\cE_L$, with $\cE_L$ the energy density resulting by the sum of the Hartree and Fock
diagrams.\footnote{This extra factor $-i$ arises because the calculation of $\cef$ and $\ceh$ is done 
  directly from the in-medium Lagrangian as read from Eq.~\eqref{190521.1}.}
Thus, $\cE_L$ obtained by resumming the ladder diagrams is given by 
\begin{align}
\label{190615.1}
\cE_L&=\frac{i}{2}{\rm Tr}\left(
\sum_{d=1}^{\infty}\frac{(t_mL_d)^{d}}{d}\right)
\\
\label{190615.1b}
&=-\frac{i}{2}{\rm Tr}
\log\left[I-t_m L_d\right]~.
\end{align}

\subsection{$\cE_L$ is real}
\label{sec.200907.2}

Though $\cE_L$ must be real, this property is not obvious from its expression given in  Eq.~\eqref{190615.1b} because of the complex nature of the operators $t_m$ and $L_d$ involved, and the explicit presence of the imaginary unity. This subsection is dedicated to demonstrate it and, along the way, we also give the explicit formulae for the unitarity loop functions $L_d$ and $L_m$, 
together with the integral equation (IE) satisfied by $t_m$.

\subsubsection{$L_m$, $L_d$ and integral equation for $t_m$}
\label{sec.200907.1}

 In the subsequent we consider that there is only one  Fermi momentum, called $k_F$. This is the case of interest for addressing the properties of normal matter at null temperature at around unitarity. E.g. experiments typically have one fermion component, like alkali atoms of \trr{$^6$Li or $^{40}$K} in optical traps, or neutrons in nature giving a  neutron-matter system.
For symmetric nuclear matter, even though there are neutrons and protons, the Fermi momenta are also the same. 
The general case with different Fermi momenta  will be addressed in Ref.~\cite{incoming}. 

Given two fermions of four-momenta $k_1$ and $k_2$ we define
\begin{align}
\label{190615.2}
a&=\frac{1}{2}(k_1+k_2)~,\\
p&=\frac{1}{2}(k_1-k_2)~,\nn
\end{align}
and the inverse relation is $
k_1=a+p\,$, $
k_2=a-p \,.$  
For on-shell fermions $k_i^0=E(\vk_i)=\vk_i^2/2m$. 

The expression for $L_m(|\vp|,\va)$ is (in the following we use the notation $p=|\vp|$)
\begin{align}
\label{190615.3}
&L_m(p,\va)=-i \int\frac{d^3k_1}{(2\pi)^3}\theta({k_F}-|\vk_1|)
\int\frac{d^4k_2}{(2\pi)^4}\frac{i}{k_2^0-\frac{|\vk_2|^2}{2m}+i\ep}
(2\pi)^4\delta(k_1+k_2-2a)|\vk_1,\vk_2\rangle
\langle \vk_1,\vk_2| \nn \\
&-i\int\frac{d^3k_2}{(2\pi)^3}\theta({k_F}-|\vk_2|)
\int\frac{d^4k_1}{(2\pi)^4}\frac{i}{k_1^0-\frac{|\vk_1|^2}{2m}+i\ep}
(2\pi)^4\delta(k_1+k_2-2a)|\vk_1,\vk_2\rangle\langle \vk_1, \vk_2| \\
&=\int\frac{d^3k}{(2\pi)^3}\left[\theta({k_F}-|\va+\vk|)+\theta({k_F}-|\va-\vk|)\right]
\frac{|\va+\vk,\va-\vk\rangle\langle\va+\vk,\va-\vk|}{2a^0-\frac{|\va+\vk|^2}{2m}
  -\frac{|\va-\vk|^2}{2m}+i\ep} \nn \\
&=-m\int\frac{d^3k}{(2\pi)^3}\left[\theta({k_F}-|\va+\vk|)+\theta({k_F}-|\va-\vk|)\right]
\frac{|\va+\vk,\va-\vk\rangle\langle\va+\vk,\va-\vk|}{\vk^2-(2ma^0-\va^2)-i\ep}~.\nn
\end{align}
Since the total four-momentum is conserved it follows then that $a$ is the same for any
two-fermion intermediate state. 
Furthermore, the initial state setting the trace in Eq.~\eqref{190615.1} is on-shell so that
\begin{align}
\label{190615.4}
2ma^0-\va^2&=m\big(E(\va+\vp)+E(\va-\vp)\big)-\va^2=\vp^2  ~.
\end{align}
In this way, the final integral representation for $L_m(p,\va)$ in Eq.~\eqref{190615.3} can be 
rewritten as
\begin{align}
\label{190615.5}
L_m(p,\va)&=- m\int\frac{d^3k}{(2\pi)^3}\left[\theta({k_F}-|\va+\vk|)+\theta({k_F}-|\va-\vk|)\right]
\frac{|\va+\vk,\va-\vk\rangle\langle\va+\vk,\va-\vk|}{\vk^2-\vp^2-i\ep}~.
\end{align}

We have not explicitly shown the spin indices because they amount to the identity matrix in the spin subspace of the two fermions.
Namely, $L_m$ includes implicitly the spin operator
\begin{align}
\label{190702.1}
\mathbb{I}_S&=\sum_{\sigma_{1,2}=-1/2}^{1/2}|\sigma_1,\sigma_2\rangle \langle \sigma_1,\sigma_2|~.
\end{align}
If needed, like in symmetric nuclear matter, one could treat similarly other discrete indices, e.g. the isospin ones, and there would be also an 
identity operator  $\mathbb{I}_{IS}$ in the isospin vector space, 
\begin{align}
\label{190702.1b}
\mathbb{I}_{IS}&=\sum_{\alpha_{1,2}}|\alpha_1,\alpha_2\rangle \langle \alpha_1,\alpha_2|~.
\end{align}

Due to the fact that the total momentum is always conserved, it is convenient to simplify
the expression for any loop function keeping in mind this fact. Then,
we indicate the two-particle state in terms only of the relative momentum $\vk$ as 
\begin{align}
\label{190624.1}
L_m(p,\va)&=-m\int\frac{d^3k}{(2\pi)^3}\left[\theta({k_F}-|\va+\vk|)+\theta({k_F}-|\va-\vk|)\right]
\frac{| \vk \rangle \langle \vk |}{\vk^2-\vp^2-i\ep}\otimes \mathbb{I}_S\otimes \mathbb{I}_{IS}~.
\end{align}

Let us continue by giving $L_d(p,\va)$, which enters in the expression for
$\cE_L$, Eq.~\eqref{190615.1b}. In this case every fermion in the intermediate state belongs to
Fermi-sea insertions and we have
\begin{align}
\label{190615.7}
L_d(p,\va)&=i\int\frac{d^3k_1}{(2\pi)^3}\frac{d^3k_2}{(2\pi)^3}
\theta({k_F}-|\vk_1|)\theta({k_F}-|\vk_2|)(2\pi)^4\delta(k_1+k_2-2a)|\vk_1,\vk_2\rangle \langle \vk_1,\vk_2|\\
&=i m \int\frac{d^3k}{(2\pi)^2}\theta({k_F}-|\va+\vk|)\theta({k_F}-|\va-\vk|)
\delta(\vk^2-\vp^2)\,|\va+\vk,\va-\vk\rangle\langle\va+\vk,\va-\vk| \nn \\
&=i\frac{m p}{8\pi^2}\int d\hat{\vk}\,\theta({k_F}-|\va+p \hat{\vk}|)
\theta({k_F}+|\va-p\hat{\vk}|)\,|\va+p\hat{\vk},\va-p\hat{\vk}\rangle\langle\va+p\hat{\vk},\va-p\hat{\vk}|~. \nn
\end{align}

Let us indicate the two-particle state in terms only of the relative momentum and rewrite $L_d(p,\va)$ as 
\begin{align}
\label{190624.3}
&L_d(p,\va)=i\frac{m p}{8\pi^2}\int d\hat{\vk}\,\theta({k_F}-|\va+p \hat{\vk}|)
\theta({k_F}+|\va-p\hat{\vk}|)\,
|p\hat{\vk}\rangle\langle p\hat{\vk}| \otimes \mathbb{I}_S \otimes \mathbb{I}_{IS}~,
\end{align}
where the identity operators in the spaces of spin and isospin are explicitly indicated.
Both operators $L_m(p,\va)$ and $L_d(p,\va)$ are symmetric under the exchange of the particles
$1\leftrightarrow 2$.

The two operators entering in Eq.~\eqref{190615.10b} for calculating  the on-shell in-medium scattering operator
$t_m$ are the vacuum $T$-matrix $t_V$ and the unitary loop integral  $L_m$ involving mixed intermediate states.
The in-medium $T$-matrix $t_m$ is a function of the total momentum 
and of the final and initial relative momenta $\vq$ and $\vp$, respectively, with $\vp^2=p^2$ fixed because of energy and
momentum conservation, cf. Eq.~\eqref{190615.4}. 
This scattering operator is denoted either by $t_m(\va)$ or simply as $t_m$. 
We can also rewrite the defining Eq.~\eqref{190615.10b} for $t_m$ as an integral equation (IE) in the form
\begin{align}
\label{190624.2}
t_m(\va)&=t_V+t_V L_m(p,\va) t_m(\va)~.
\end{align}
By including explicitly all the involved arguments this IE becomes
\begin{align}
\label{190624.3b}
t_m(\vq,\vp,\va)_{BA}
&=t_V(\vq,\vp)_{BA}
\\
&-\frac{m}{2}\sum_{C} \int \frac{d^3 k}{(2\pi)^3}t_V(\vq,\vk)_{BC}
\frac{\theta({k_F}-|\va+\vk|)+\theta({k_F}-|\va-\vk|)}{\vk^2-\vp^2-i\ep}
t_m(\vk,\vp,\va)_{CA}~.
\nn
\end{align}
We have denoted here by capital letters  
the spin  $\sigma_i$   and other possible labels $\alpha_i$  (e.g. isospin ones, as taken in the discussion that follows).  
In this way $A\equiv \{\sigma_1\sigma_2\alpha_1\alpha_2 \}$ for the initial state,
$B\equiv \{\sigma'_1\sigma'_2\alpha'_1\alpha'_2\}$ for the final one and then
we have the intermediate-state labels $C\equiv \{\sigma''_1\sigma''_2 \alpha''_1\alpha''_2 \}$.
A factor $1/2$ has been included in the quadratic term in the the previous equation because the two-fermion states are antisymmetric.  
Since $L_m(p,\va)$ is proportional to the identity matrix both in the spin and isospin spaces this implies that $t_m(\va)$ also conserves the total spin $S$ and isospin $I$ of the two fermions.

\subsubsection{Demonstration that $\cE_L$ is real}
\label{sec.200907.2b}

In order to show that $\cE_L$ is real we need first a detour and conveniently relate $t_m$ and the vacuum potential $V$.  The Eq.~\eqref{190624.2} for $t_m(\va)$  can be solved formally as in Eq.~\eqref{190615.10b}, 
\begin{align}
\label{190625.1}
t_m(\va)=\left[I-t_V L_m(p,\va)\right]^{-1}t_V~,
\end{align}
 The inversion of the latter gives 
\begin{align}
\label{190625.2}
t_m(\va)^{-1}&=t_V^{-1}-L_m(p,\va)~.
\end{align}
 It is also the case that the vacuum $T$-matrix $t_V$ satisfies a Lippmann-Schwinger  equation
 in terms of the potential $V$, $t_V=V-VGt_V$. Therefore, 
\begin{align}
\label{190625.3}
t_V^{-1}&=V^{-1}+G(p)~.
\end{align}
Here $G(p)$ is the vacuum unitarity loop function due to the  intermediate states
of two free fermions, given by 
\begin{align}
\label{190625.3b}
 G(p)&=-m\int\frac{d^3k}{(2\pi)^3}\frac{|\vk\rangle\langle \vk|}{\vk^2-\vp^2-i\ep}
  \otimes \mathbb{I}_S\otimes \mathbb{I}_{IS}~.
\end{align}
Then, from Eqs.~\eqref{190625.2} and \eqref{190625.3} we can also write $t_m(\va)$ as
\begin{align}
\label{190625.3c}
  t_m(\va)^{-1}&=V^{-1}+G(p)-L_m(p,\va)~.
\end{align}

Within the notation developed so far along this section we can rewrite the Eq.~\eqref{190615.1b} as
\begin{align}
&{\cE_L}=-i\int \frac{pdp}{m\pi}\int \frac{d^3a}{\pi^3}
{\rm Tr}\left(\log\left[I-t_m(\va) L_d(p,\va)\right]\right)\nn\\
\label{190630.3}
&=-\frac{i}{2}\sum_{C}\int\!\! \frac{pdp}{m\pi}\int \frac{d^3a}{\pi^3}
\int \frac{d^3 q}{(2\pi)^3}
{_A\langle} \vq,C|\log\left[I-t_m(\va) L_d(p,\va)\right]{|\vq,C\rangle_A}~,
\end{align}
where $C$ has the same meaning as set of spin and isospin indices as
in Eq.~\eqref{190624.3b}.
The extra factor of 1/2 in the last equation is introduced due to the antisymmetrized nature of the two-fermion states, indicated by the subscript $A$ in the bra and kets. 
 The factors in the first line of Eq.~\eqref{190630.3} are adjusted so that at first order in $t_mL_d$ the result coincides
 with the plain sum over two Fermi-sea insertions of the expected value of $t_m(\va)$ on the antisymmetric two-fermion states inside their Fermi seas, as
 required by the direct calculation of $F_{10}+H_{10}$ following Sec.~\ref{sec.190603.1}.\footnote{It is only necessary to consider the first order in powers of $t_mL_d$ since 
  the higher orders in the expansion of the $\log$ just corresponds to iterating with coefficients well fixed given the normalization of the states used.}
 Let us notice that the presence of $L_d(p,\va)$ guarantees that the two fermions in the integration over $\vq$ belong to their Fermi seas.

 Now, we are ready to show that $\cE_L$ is real.
 As a result of Eq.~\eqref{190625.3c}  we proceed to rewrite the argument
of the $\log$ in the expression for ${\cal E_L}$ as
\begin{align}
\label{190629.1}
I-t_{m}L_d&=t_m(t_m^{-1}-L_d)\\
&=\left(V^{-1}+G-L_m\right)^{-1}\left(V^{-1}+G-L_m-L_d\right)~.\nn
\end{align}
From the previous expression we introduce the operators $A$ and $B$ defined as 
\begin{align}
  \label{190629.2}
  A&=V^{-1}+G-L_m~,\\
  B&=V^{-1}+G-L_m-L_d~.\nn
\end{align}
We notice the important property that in the subspace
of fermions belonging to the Fermi seas,
in which the trace of Eq.~\eqref{190615.1} is taken, 
the operators $A$ and $B$ satisfy
\begin{align}
\label{190629.3}
A&=B^\dagger~.
\end{align}
To show the validity of this result we make use of the following facts: i) We have the direct observation that $V^\dagger=V$ because the potential is Hermitian. ii) We have for $G-L_m$ that
\begin{align}
\label{190629.4}
G-L_m&=-m\dashint\frac{d^3k}{(2\pi)^3}\frac{|\vk\rangle\langle \vk|}{\vk^2-\vp^2}
+m\dashint \frac{d^3k}{(2\pi)^3}
\frac{\theta({k_F}-|\va+\vk|)+\theta({k_F}-|\va-\vk|)}{\vk^2-\vp^2}
| \vk \rangle \langle \vk |\\
&-i m\pi\int\frac{d^3k}{(2\pi)^3}\delta(\vk^2-\vp^2)
\left\{1-\theta({k_F}-|\va+\vk|)-\theta({k_F}-|\va-\vk|)\right\}
|\vk\rangle\langle \vk|~.\nn
\end{align}
The contributions on the right-hand side (rhs) of this equation that involve the Cauchy principal value (indicated by a dash in the integral symbol)  
are the Hermitian part of $G-L_m$, while the other one in the last line is anti-Hermitian.  
The former are denoted by $R$ and the latter by $iS$, so that both $R$ and $S$ are Hermitian and $G-L_m=R+iS$. 
It is convenient to rearrange the expression for $S$ so as to 
introduce explicitly the particle-particle and hole-hole parts, in which the two fermions have momenta above and below their Fermi seas,
respectively.  
As a result, the anti-Hermitian part of Eq.~\eqref{190629.4} becomes
\begin{align}
\label{190629.5}
iS&=-i\frac{mp}{16\pi^2}\int d\hat{\vk}
\left\{
\left[1-\theta({k_F}-|\va+p\hat{\vk}|)\right]
\left[1-\theta({k_F}-|\va-p\hat{\vk}|)\right]
-\theta({k_F}-|\va+p\hvk|)\theta({k_F}-|\va-p\hvk|)
\right\}\nn \\
&\times |p\hat{\vk}\rangle\langle p\hat{\vk}|~.
\end{align}
In the $B$ function we have the combination $G-L_m-L_d$ of unitary-loop operators.
Its Hermitian part is the same, because $L_d$ is purely anti-Hermitian, cf. Eq.~\eqref{190624.3}.
Thus,
\begin{align}
\label{190629.6}
&G-L_m-L_d=-m\dashint\frac{d^3k}{(2\pi)^3}\frac{|\vk\rangle\langle \vk|}{\vk^2-\vp^2}
+m\dashint \frac{d^3k}{(2\pi)^3}
\frac{\theta({k_F}-|\va+\vk|)+\theta({k_F}-|\va-\vk|)}{\vk^2-\vp^2}
| \vk \rangle \langle \vk |\\
&-i\frac{mp}{16\pi^2}\int d\hat{\vk}
\left\{
\left[1-\theta({k_F}-|\va+p\hat{\vk}|)\right]
\left[1-\theta({k_F}-|\va-p\hat{\vk}|)\right]
+\theta({k_F}-|\va+p\hvk|)\theta({k_F}-|\va-p\hvk|)
\right\}\nn \\
&\times |p\hat{\vk}\rangle\langle p\hat{\vk}|~,\nn
\end{align}
with a change of sign in the hole-hole part of the anti-Hermitian part as compared with
Eq.~\eqref{190629.5}.

When taking the trace to calculate ${\cE_L}$ in Eq.~\eqref{190630.3} the common particle-particle term in
the anti-Hermitian parts of $A$ and $B$ does not give contribution.
This follows because  the particle-particle part in $ S$
requires that 
\begin{align}
\label{190630.1}
(\va+p\hat{\vk})^2+(\va-p\hat{\vk})^2=2(\va^2+\vp^2)> 2{k_F}^2~,
\end{align}
which cannot be fulfilled.
The reason is easily explained by the conservation of the total energy, which
is set by the two fermions inside their Fermi seas that enter in the 
 trace for calculating $\cE_L$. 
If we call their momenta $\vp_1$ and $\vp_2$ then the total energy times $2m$ 
corresponds to
\begin{align}
\label{200907.1}
\vp_1^2+\vp_2^2=(\va+\vp)^2+(\va-\vp)^2=
2(\va^2+\vp^2)\leq 2{k_F}^2~,
\end{align}
in contradiction with Eq.~\eqref{190630.1}.

The next step to show that $\cE_L$ is real is to rewrite Eq.~\eqref{190629.1} as
\begin{align}
\label{191004.2}
&\left(V^{-1}+G-L_m\right)^{-1}\left(V^{-1}+G-L_m-L_d\right)
=(V^{-1}+R+i\,S)^{-1}(V^{-1}+R-i\,S)\\
=&I-2i(V^{-1}+R+i\,S)^{-1}{S}~,\nn
\end{align}
and because of the cyclic property of the trace in Eq.~\eqref{190615.1b}
we can recast the expression for $\cE_L$ as
\begin{align}
\label{191004.3}
{\cE_L}=-\frac{i}{2}{\rm Tr}\log\left[I-2i{S^{\frac{1}{2}}}
(V^{-1}+R+i\,S)^{-1}{S^{\frac{1}{2}}}\right]~,
\end{align}
where we have used that $S$ is positive definite within the Fermi seas of the two fermions,
cf. Eq.~\eqref{190629.5} where only the last term on the rhs of this
equation gives contribution.

An explicit calculation shows that  $I-2iS^{\frac{1}{2}}(V^{-1}+R+i\,S)^{-1}S^{\frac{1}{2}}$ is a unitary operator because  
\begin{align}
\label{191004.4}
&\left[I-2iS^{\frac{1}{2}}(V^{-1}+R+i\,S)^{-1}S^{\frac{1}{2}}\right]\left[I+2iS^{\frac{1}{2}}(V^{-1}+R-i\,S)^{-1}S^{\frac{1}{2}}\right]\\
&=I-2i S^{\frac{1}{2}}\left[(V^{-1}+R+i\,S)^{-1}-(V^{-1}+R-i\,S)^{-1}\right]S^{\frac{1}{2}}
+4S^{\frac{1}{2}}(V^{-1}+R+i\,S)^{-1}S(V^{-1}+R-i\,S)^{-1}S^{\frac{1}{2}}\nn\\
&=I-2i S^{\frac{1}{2}}(V^{-1}+R+i\,S)^{-1}\left[(V^{-1}+R-iS)-(V^{-1}+R+iS)\right](V^{-1}+R-i\,S)^{-1}S^{\frac{1}{2}}\nn\\
&+4S^{\frac{1}{2}}(V^{-1}+R+i\,S)^{-1}S(V^{-1}+R-i\,S)^{-1}S^{\frac{1}{2}}=I~.\nn
\end{align}
One can also show similarly that $\left[I+2iS^{\frac{1}{2}}(V^{-1}+R-i\,S)^{-1}S^{\frac{1}{2}}\right]\left[I-2iS^{\frac{1}{2}}(V^{-1}+R+i\,S)^{-1}S^{\frac{1}{2}}\right]=I$. 
Since a unitary operator can be diagonalized
by a unitary transformation it follows that its eigenvalues are phase factors. 
As a result ${\cal E}_{L}\in \mathbb{R}$
because it is given by $-i/2$ times the trace of the log of
this unitary operator, cf. Eq.~\eqref{190630.3}.\footnote{Incidentally, this is the reason for the always appearing $\arctan$ series, first found in Ref.~\cite{Kaiser:2011cg} when including only the $S$-wave scattering length $a_0$.}

\section{Partial-wave expansion}
\label{sec.190630.1}
\def\theequation{\arabic{section}.\arabic{equation}}
\setcounter{equation}{0}

Since the total momentum $\mathbf{P}=2\va$ is conserved it is appropriate to factorize out in the normalization of the two-particle states
the factor $(2\pi)^3\delta(\mathbf{P}'-\mathbf{P})$. By doing this the normalization of the two-fermion states simplifies to
\begin{align}
\label{190706.1}
\langle \vp'\sigma'_1\sigma'_2|\vp\sigma_1\sigma_2\rangle&=(2\pi)^3\delta(\vp'-\vp)\delta_{\sigma'_1\sigma_1}\delta_{\sigma'_2\sigma_2}~,
\end{align}
involving only the relative momentum.\footnote{In the subsequent we do not refer to isospin.}  It is also interesting to use the angular and modulus Dirac delta functions of $\vp$ separately   
and express the previous equation as
\begin{align}
\label{190706.1b}
\langle \vp'\sigma'_1\sigma'_2|\vp\sigma_1\sigma_2\rangle&=\frac{2\pi^2\delta(p'-p)}{p^2}  4\pi \delta(\hat{\vp}'-\hat{\vp})
\delta_{\sigma'_1\sigma_1}\delta_{\sigma'_2\sigma_2}~.
\end{align}

Another simplification in the notation stems from the fact that $L_d(p,\va)$ is diagonal in
the absolute value of the momentum, as it is clear from Eq.~\eqref{190624.3}.
As a result, the matrix elements of $L_d(p,\va)$ can be expressed as
\begin{align}
\label{190709.3}
\langle \vq'\beta'_1\beta'_2|L_d(p,\va)|\vq\beta_1\beta_2\rangle&=
(2\pi^2)^2\frac{\delta(q'-p)\delta(q-p)}{p^4}\langle \hvq'\beta'_1\beta'_2 |\tilde{L}_d(p,\va)|\hvq\beta_1\beta_2\rangle~, 
\end{align}
so that both $\vq'$ and $\vq$ have their moduli given by $p$ which is conserved.
The matrix element on the right-hand side of the previous equation is  
\begin{align}
\label{190709.4}
\langle \hvq'\beta'_1\beta'_2 |\tilde{L}_d(p,\va)|\hvq\beta_1\beta_2\rangle
&=i\,2 mp \delta(\hvq'-\hvq)\delta_{\beta'_1\beta_1}\delta_{\beta'_2\beta_2}\theta({k_F}-|\va+p\hvq|)\theta({k_F}-|\va-p\hvq|)~.
\end{align}
Nonetheless, in the following we do not distinguish between $L_d(p,\va)$ and $\tilde{L}_d(p,\va)$ and directly use the later extracting out the
factor $2\pi^2\delta(q-p)/p^2$. 
 In this way, we can rewrite Eq.~\eqref{190630.3} as
\begin{align}
\label{200501.2}
{\cE_L}
&=-\frac{i}{2}\sum_{\sigma_{1,2}}\int\!\! \frac{pdp}{m\pi}\int \frac{d^3a}{\pi^3}\int\frac{d\hvp}{4\pi}
\,{_A\langle} \vp,\sigma_1,\sigma_2|\log\left[I-t_m(\va) L_d(p,\va)\right]|\vp,\sigma_1,\sigma_2{\rangle_A}~.
\end{align}

To settle the partial-wave amplitudes (PWAs) in the nuclear medium 
we introduce the states with definite
total angular-momentum quantum numbers and $p$.
For that we follow the Appendix A of Ref.~\cite{Lacour:2009ej}, and an optimized presentation can be
found in the more recent Chapter 2 of the book \cite{oller.book}.
The states with total angular momentum $J$, total spin $S$, orbital angular momentum $\ell$ and
third component $\mu$ of $\mathbf{J}$ are called $|J\mu \ell S p\rangle$ and comprise the partial-wave basis.
Their relation with the plane-wave states is
\begin{align}
\label{200501.3}
|\vp \sigma_1\sigma_2\rangle_A&=\sqrt{4 \pi}\sum_{J\mu\ell m S\sigma_3}
( \sigma_1 \sigma_2 \sis_3| s_1 s_2 S)
(m \sis_3 \mu | \ell S J) {Y_\ell^m}(\hvp)^* \chi( S \ell) |J \mu \ell S \rangle~,\\
\chi(S\ell)&=\frac{1-(-1)^{\ell+S-2s_1}}{\sqrt{2}}~,\nn\\
\label{200501.4}
{_A\langle} \vp'\sigma_1\sigma_2|J\mu\ell S p\rangle&=\chi(S\ell)
\frac{4\pi^\frac{5}{2}\delta(p'-p)}{p^2}(\sigma_1\sigma_2 \sis_3|s_1s_2S)(m\sis_3 \mu|\ell S J)
Y_\ell^m(\hvp)~.
\end{align}
where $s_1=s_2$ are the spins of the two fermions, $\sigma_3=\sigma_1+\sigma_2$ and $m=\mu-\sigma_3$ are the third components of the total spin  and the orbital angular momentum, respectively. 
Because of the presence of the Clebsch-Gordan coefficients $(m_1m_2m_3|j_1j_2j_3)$ one can also sum over $\sigma_3$ and $m$,
as indicated.

When the partial-wave expansion is inserted  in Eq.~\eqref{200501.2} this equation becomes
\begin{align}
\label{200501.5}
      {\cE_L}&=-\frac{i}{2}\sum_{\sigma_{1,2}}\sum_{{\scriptsize\begin{array}{l}m,m' \\ \sis_3,\sis'_3\end{array}}}
\sum_{{\scriptsize\begin{array}{l}\ell,\ell',S\\S',J,J'\end{array}}}
\int\!\! \frac{pdp}{m\pi}\int \frac{d^3a}{\pi^3}\int \frac{d\hvp}{4\pi}
\underbrace{{_A\langle} \vp,\sigma_1,\sigma_2|J'\mu'\ell' S' p\ra}_{\chi(S'\ell')\sqrt{4\pi}
(\sigma_1\sigma_2 \sis'_3|s_1s_2S')(m'\sis_3 \mu'|\ell' S' J')Y_{\ell'}^{m'}(\hvp)}\\
&\times \la J'\mu'\ell' S' p|\log\left[I-t_m(\va) L_d(p,\va)\right]|J\mu\ell S p\ra 
\underbrace{\la J\mu\ell S p |\vp,\sigma_1,\sigma_2{\rangle_A}\,.}_{\chi(S\ell)\sqrt{4\pi}(\sigma_1\sigma_2 \sis_3|s_1s_2S)
  (m\sis_3 \mu|\ell S J)Y_\ell^m(\hvp)^*}\nn
\end{align}
Next, we use the orthogonality properties,
\begin{align}
\label{200501.5b}
\sum_{\sigma_1,\sigma_2}(\sigma_1\sigma_2 \sis'_3|s_1s_2S')(\sigma_1\sigma_2 \sis_3|s_1s_2S)
&=\delta_{SS'}\delta_{\sis'_3 \sis_3}~,\\
\int d\hvp Y_{\ell'}^{m'}(\hvp) Y_\ell^m(\hvp)^*&=\delta_{\ell\ell'}\delta_{mm'} ~,\nn\\
\sum_{m,\sis_3}(m \sis_3 \mu'|\ell S J')(m\sis_3 \mu|\ell S J)&=\delta_{\mu'\mu}\delta_{J'J}~.\nn
\end{align}
Then, Eq.~\eqref{200501.5} simplifies to
\begin{align}
\label{200501.6}
{\cE_L}
&=-\frac{2i}{m\pi^3}\sum_{J\mu \ell S}\chi(S\ell)^2
\int_0^\infty pdp\int_0^\infty a^2da
\la J\mu\ell S p|\log\left[I-t_m(a\hvz) L_d(p,a\hvz)\right]|J\mu\ell S p\ra~.
\end{align}
Where, because of rotational invariance, we have made used of the fact that if $R(\hva)$ is a rotation so that
$R(\hva)\vz=\va$ then
\begin{align}
\label{200501.7}
{\rm Tr}[I-t_m(\va)L_d(p,\va)]={\rm Tr}[R(\hva)(I-t_m(a\hvz)L_d(p,a\hvz)])R(\hva)^\dagger]
={\rm Tr}[I-t_m(a\hvz)L_d(p,a\hvz)]~.
\end{align}
The transformation rules of $t_m(\va)$ and $L_m(p,\va)$ under a rotation on $\va$ are derived in the
Appendix~\ref{app.201228.1}. 

Our final expression for the energy density $\cE$ is given by the sum of the free Fermi-gas result, $\cE_{\rm free}$, plus $\cE_L$, with
\begin{align}
\label{201231.1}
\cE_{\rm free}&=\rho\frac{3k_F^2}{10m}~,\\
\rho&=g\frac{k_F^3}{6\pi^2}~,\nn
\end{align}
where $g$ is the degeneracy factor. In the particular cases analyzed
in this work $g=2$, and in symmetric nuclear matter $g=4$. 
 
\section{Integral equation for $t_m(a\hvz)$ in partial waves}
\label{sec.190808.1}
\def\theequation{\arabic{section}.\arabic{equation}}
\setcounter{equation}{0}

Given the formal solution  for $t_m(\va)$ in Eq.~\eqref{190625.3c}, it is clear that
this in-medium $T$ matrix satisfies also the operator equation
\begin{align}\label{190808.1}
t_m(\va)&=V-V[G(p)-L_m(p,\va)]t_m(\va)~,
\end{align}
with energy  $E=p^2/m$. In components this equation gives rise to the following IE in momentum space, 
\begin{align}
\label{200925.5}
&{_A\la}\vp' \gs'_1 \gs'_2 |t_m(\va)| \vp \gs_1\gs_2 {\ra_A}
={_A\la}\vp' \gs'_1 \gs'_2 |V| \vp \gs_1\gs_2{\ra_A}\\
&+\frac{m}{2}\sum_{\ts_{1,2}}\int\frac{d^3k}{(2\pi)^3}
{_A\la}\vp' \gs'_1 \gs'_2 |V| \vk \ts_1\ts_2 {\ra_A}
\frac{1-\theta({k_F}-|\vk+\va|)-\theta({k_F}-|\vk-\va|)}{k^2-p^2-i\ep} {_A\la}\vk \ts_1 \ts_2  |t_m(\va)| \vp \gs_1\gs_2 {\ra_A}~.\nn 
\end{align}

From the decomposition of the antisymmetrized plane-wave states in the partial-wave basis,
Eqs.~\eqref{200501.3} and Eq.~\eqref{200501.4}, let us rewrite $-V[G(p)-L_m(p,\va)]t_m(\va)$ in Eq.~\eqref{200925.5} in terms of
states in the partial-wave basis:
\begin{align}
\label{200926.2}
&\frac{m}{2}\sum_{\ts_{1,2}}\int\frac{d^3k}{(2\pi)^3}V|\vk\ts_1\ts_2{\ra_A}
\frac{1-\theta({k_F}-|\vk+\va|)-\theta({k_F}-|\vk-\va|)}{k^2-p^2-i\ep}
{_A\la}\vk\ts_1\ts_2|t_m(\va)\\
&=
m\!\sum_{\ts_{1,2}}\!\!\! \sum_{{\scriptsize \begin{array}{l} J \mu \ell m_3\\ S \sis_3 \end{array}} }\!\!\!\!
\sum_{{\scriptsize \begin{array}{l}J'\mu'\ell'm'_3 \\S'\sis'_3\end{array}}}\!\!\!
\int\frac{d^3k}{(2\pi)^2}V|J'\mu'\ell'S' k \ra
\frac{1-\theta({k_F}-|\vk+\va|)-\theta({k_F}-|\vk-\va|)}{k^2-p^2-i\ep}\la J\mu\ell S k |t_m(\va)\nn\\
&\times \chi(S\ell)\chi(S'\ell')(\ts_1\ts_2\sis'_3|s_1 s_2 S')(\ts_1\ts_2\ts_3|s_1 s_2 S)
(m'_3\sis'_3\mu'|\ell' S' J')(m_3\sis_3\mu|\ell S J)
Y_{\ell'}^{m'_3}(\hvk)^*Y_{\ell}^{m_3}(\hvk)~.\nn
\end{align}
The sum over $\ts_1$ and $\ts_2$ can be readily done by taking advantage
of the orthogonality properties of the Clebsch-Gordan coefficients,  Eq.~\eqref{200501.5b}, and it
gives $\delta_{\sis'_3\sis_3}\delta_{S'S}$, so that the total spin $S$ is conserved (as in vacuum).  
 In the following we choose $\displaystyle{\va}$ along the $z$ axis because this is enough to calculate $\cE_L$,
cf. Eq.~\eqref{200501.6}, and it
also induces extra simplifications in the final IE for $t_m(a\hvz)$.
The relationship between $t_m(\va)$ and $t_m(a\hvz)$ is $t_m(\va)=R(\hva)t_m(a\hvz)R(\hva)^\dagger$ as follows from Eq.~\eqref{190702.11}.

Because of this choice  $\va=a\hvz$ it is clear that there is no dependence on the azimuthal angle of $\vk$ in the
integral of Eq.~\eqref{200926.2}, because $|\vk \pm a\hvz|$ only depends on its polar angle.
Thus,
\begin{align}
\label{200926.4a}
\int_0^{2\pi} d\varphi Y_{\ell'}^{m'_3}(\theta,\varphi)^*  Y_{\ell}^{m_3}(\theta,\varphi)\propto
\delta_{m'_3 m_3}~.
\end{align}
As a result $\mu'=\mu$ because $\mu'=s'_3+m'_3=s_3+m_3=\mu$ due to the factor
$\delta_{m'_3m_3}\delta_{s'_3s_3}$.  A simplified version of  Eq.~\eqref{200926.2} then results
\begin{align}
\label{200926.4}
&m\!\sum_{{\scriptsize J\mu\ell m_3 }}
\!\sum_{{\scriptsize J'\ell'  S\sis_3 }}
\int\frac{d^3k}{(2\pi)^2}V|J'\mu\ell'S k \ra
\frac{1-\theta({k_F}-|\vk+a\hvz|)-\theta({k_F}-|\vk-a\hvz|)}{k^2-p^2-i\ep}
\la J\mu\ell S  k |t_m(a\hvz)\\ 
&\times \chi(S\ell)\chi(S\ell') 
(m_3\sis_3\mu|\ell' S J')(m_3\sis_3\mu|\ell S J)Y_{\ell'}^{m_3}(\hvk)^*Y_{\ell}^{m_3}(\hvk)~.\nn
\end{align}

This equation  implies that parity is conserved in the sense that
\begin{align}
\label{200926.5}
(-1)^\ell=(-1)^{\ell'}~.
\end{align}
This result follows by changing the integration  variable $\vk$ to $-\vk$ in
Eq.~\eqref{200926.4} and then taking into account
 the well-known parity property of the spherical harmonics, $Y_\ell^m(-\hvk)=(-1)^\ell Y_\ell^m(\hvk)$.\footnote{For case of symmetric nuclear matter, the conservation of parity also implies that $I'=I$ because Fermi statistics requires that $(-1)^{\ell+S+I}=(-1)^{\ell'+S+I'}=-1$.
Then, $(-1)^I=(-1)^{I'}$ and since $I$, $I'$ are either 0 or 1 it follows that they must be the same.
Therefore, isospin (both $I$ and $i_3$) is conserved in the nuclear-medium scattering process with a common Fermi momentum.}
 In the Table~\ref{tab.200926.1} we express the quantities that are conserved in the
 scattering process of two fermions with spin 
 1/2 in the many-body environment with Fermi momentum ${k_F}$. 
Compared to the vacuum case, the mixing between PWAs is more extreme because $J$ and $J'$ are
different in general.

\begin{table}
\begin{center}
\begin{tabular}{llllll}
  \hline
  \\
  Conserved: & $\mu'=\mu$ & $S'=S$ & $(-1)^{\ell}=(-1)^{\ell'}$ \\     
  \\
  \hline
\end{tabular}
\caption{{\small Set of conserved quantum numbers in a two-body scattering process for a spin 1/2 many-body system with Fermi momentum ${k_F}$.} \label{tab.200926.1}}
\end{center}
\end{table}

We are ready to derive the IE for a PWA in the many-body environment  by sandwiching the Eq.~\eqref{190808.1}
between partial-wave states and using Eq.~\eqref{200926.4}. The result is,
\begin{align}
\label{190809.5}
&\la J'\mu\ell'S p'|t_m(a\hvz)| J\mu\ell S p\ra=
\la J'\mu\ell'Sp'|V| J\mu\ell Sp\ra
+m \!\sum_{{\scriptsize \begin{array}{l}J_1\ell_1 \ell_2 \\ m_3 \sis_3\end{array}}}
\!\! \chi(S\ell_2) \chi(S\ell_1)\\
& \times 
\int\frac{d^3k}{(2\pi)^2}Y_{\ell_2}^{m_3}(\hvk)^*Y_{\ell_1}^{m_3}(\hvk)
 \la J'\mu\ell'Sp'|V|J'\mu \ell_2Sk\ra
(m_3s_3\mu|\ell_2 S J')(m_3s_3\mu|\ell_1 S J_1) \nn\\ 
&\times \frac{1-\theta({k_F}-|\vk+a\hvz|)-\theta({k_F}-|\vk-a\hvz|)}{k^2-p^2-i\ep}
 \la J_1\mu\ell_1 S  k|t_m(a\hvz)| J\mu\ell S p\ra~.\nn
\end{align}
Let us recall that all the orbital angular momenta involved in the previous equation must have same parity, i.e., $(-1)^{\ell'}=(-1)^\ell=(-1)^{\ell_1}=(-1)^{\ell_2}$, and that the total spin $S$ and the third component of angular momentum $\mu$ is conserved,  Table~\ref{tab.200926.1}. In order to solve the on-shell $\la J'\mu\ell'S p|t_m(a\hvz)| J\mu\ell S p\ra$ from Eq.~\eqref{190809.5} we need also to find out the half-off-shell PWA $\la J'\mu\ell'S k|t_m(a\hvz)| J\mu\ell S p\ra$.

Given a value of $\mu$ in Eq.~\eqref{190809.5} only the angular momenta which satisfy that $J,$ $J'$ and $J_1 \geq |\mu|$ can appear in the IE. For computational purposes it is convenient
to write this IE in matrix form as
\begin{align}
\label{190809.6}
{\mathsf t_m}(p',p)&=v(p',p)
+\frac{m}{(2\pi)^2}\int_0^\infty\frac{k^2 dk}{k^2-p^2-i\ep}
v(p',k)\cdot {\cal A}\cdot {\mathsf t_m}(k,p)
\end{align}
with the matrices $v$, ${\mathsf t_m}$ and  ${\cal A}$ given by
\begin{align}
\label{190809.7v}
v(p',k)_{J'\ell',J_2\ell_2}&
=\delta_{J' J_2}\langle J'\mu \ell' S p'|V|J_2\mu\ell_2 S k\rangle~,\\
\label{190809.7t}
{\mathsf t_m}(k,p)_{J_1\ell_1,J\ell}&=
\langle J_1\mu\ell_1 S  k|t_m(a\hvz)|J\mu\ell S  p\rangle~,\\
\label{190809.7}
{\cal A}_{J_2\mu\ell_2, J_1\mu\ell_1}&=
\chi(S\ell_2)\chi(S\ell_1)\sum_{m_3 \sis_3} (m_3\sis_3\mu|\ell_2SJ_2)(m_3\sis_3\mu|\ell_1 S J_1)\\
&\times \int d\hvk Y_{\ell_2}^{m_3}(\hvk)^*Y_{\ell_1}^{m_3}(\hvk)
\left[1-2\theta({k_F}-|\vk-a\hvz|)\right]~.\nn
\end{align}
The free part in $[{\cal A}]$ can be further simplified, so that it also reads
\begin{align}
\label{190809.7b}
{\cal A}_{J_2\mu\ell_2, J_1\mu\ell_1}&=\chi(S\ell_2)\chi(S\ell_1)\left(\delta_{J_2\ell_2,J_1\ell_1}-
  2\sum_{m_3 \sis_3} (m_3\sis_3\mu|\ell_2SJ_2)(m_3\sis_3\mu|\ell_1 S J_1)\right.\\
&\left.\times \int d\hvk Y_{\ell_2}^{m_3}(\hvk)^*Y_{\ell_1}^{m_3}(\hvk)
\theta({k_F}-|\vk-a\hvz|)\right)~.\nn
\end{align}

For the evaluation of $\cE_L$ in Eq.~\eqref{200501.6} one has to sum over $\mu$.
However, there are two relations involving PWAs with different values of $\mu$
that can be used to reduce the burden of needed computations.
We only enumerate them here and their demonstration is given in the Appendix~\ref{app.200925.1}. 
 The first one relates the PWAs with opposite values for $\mu$,
\begin{align}
\label{201228.5}
\langle J_2-\mu_1\ell_2 S_1p'|t_m(a\hvz)|J_1-\mu_1\ell_1 S_1p\rangle
&=(-1)^{J_2+J_1}\langle J_2\mu_1\ell_2 S_1 p'|t_m(a\hvz)|J_1\mu_1\ell_1 S_1  p\rangle~.
\end{align}
As a consequence of this equation for $\mu_1=0$ one has the requirement, 
\begin{align}
\label{201228.6}
(-1)^{J_1+J_2}=+1~,~\mu_1=0~,
\end{align}
otherwise the PWA is zero.

The PWAs of $t_m(a\hvz)$ also satisfy an interesting symmetry property under the exchange of the
initial and final quantum numbers that reads
\begin{align}
\label{201228.7}
\langle J_1\mu_1\ell_1 S_1p|t_m(a\hvz)|J_2\mu_1\ell_2 S_1p'\rangle
&=\langle J_2\mu_1\ell_2 S_1p'|t_m(a\hvz)|J_1\mu_1\ell_1 S_1p\rangle~.
\end{align}
This relation is particularly useful for the on-shell case with $p'=p$,  the one
needed in the evaluation of ${\cal E_L}$. It implies that the in-medium on-shell $T$ matrix is symmetric
under the exchange of the discrete labels. 

The first operator equation for $t_m(\va)$ was written in terms of the vacuum $T$-matrix $t_V$, cf. Eq.~\eqref{190615.10b}. 
In this regard, instead of Eq.~\eqref{190808.1} we could also have equally started with the operator equation of $t_m(\va)$ in terms of $t_V$, Eq.~\eqref{190625.1}, and proceed in a complete analogous way since $V$ and $t_V$ are invariant operators  under rotations. Therefore, in the IE of Eq.~\eqref{190809.5} we have to 
replace $V$ by $t_V$ and remove the free part in the unitarity loop function.
Then, we also have for the calculation of the in-medium PWAs the following IE, 
  \begin{align}
\label{201228.1}
&\la J'\mu\ell'S p'|t_m(a\hvz)| J\mu\ell S p\ra=
\la J'\mu\ell'Sp'|t_V| J\mu\ell Sp\ra
-m \!\!\!\!\sum_{{\scriptsize \begin{array}{l} J_1\ell_1 \ell_2 \\ m_3 \sis_3 \end{array}}}
\!\!\!\! \chi(S\ell_2) \chi(S\ell_1)\!\!
\int\!\!\frac{d^3k}{(2\pi)^2}Y_{\ell_2}^{m_3}(\hvk)^*Y_{\ell_1}^{m_3}(\hvk)\nn\\
& \times 
 \la J'\mu\ell'Sp'|t_V|J'\mu \ell_2Sk\ra
 \frac{\theta({k_F}-|\vk+a\hvz|)+\theta({k_F}-|\vk-a\hvz|)}{k^2-p^2-i\ep}
 (m_3s_3\mu|\ell_2 S J')(m_3s_3\mu|\ell_1 S J_1)\\
 &\times  \la J_1\mu\ell_1 S  k|t_m(a\hvz)| J\mu\ell S p\ra~.\nn
  \end{align}
  This IE can be written in matrix form in a completely analogous way to Eq.~\eqref{190809.6}, with the replacement of $V$ by $t_V$ and the removal of the the Kronecker-delta term on the  rhs of Eq.~\eqref{190809.7b}.
  Therefore, we can write 
  \begin{align}
\label{201228.2}
{\mathsf t_m}(p',p)&={\mathsf t_V}(p',p)
+\frac{m}{(2\pi)^2}\int_0^\infty\frac{k^2 dk}{k^2-p^2-i\ep}
{\mathsf t_V}(p',k)\cdot {\cal B}\cdot {\mathsf t_m}(k,p)
\end{align}
with the matrices ${\mathsf t_V}$ and  ${\cal B}$ given by
\begin{align}
\label{201228.3}
{\mathsf t_V}(p',k)_{J'\ell',J_2\ell_2}&
=\delta_{J' J_2}\langle J'\mu \ell' S p'|t_V|J_2\mu\ell_2 S  k\rangle~,\\
\label{201228.4}
{\cal B}_{J_2\mu\ell_2, J_1\mu\ell_1}&=-2\chi(S\ell_2)\chi(S\ell_1)
\sum_{m_3 s_3} (m_3s_3\mu|\ell_2SJ_2)(m_3s_3\mu|\ell_1 S J_1)\\
&\times \int d\hvk Y_{\ell_2}^{m_3}(\hvk)^*Y_{\ell_1}^{m_3}(\hvk)
\theta({k_F}-|\vk-a\hvz|)~.\nn
\end{align}
  
\section{\tcr{Method} applied to solve the PWAs in the medium for contact interactions}
\label{sec.201228.1}
\def\theequation{\arabic{section}.\arabic{equation}}
\setcounter{equation}{0}

We solve $t_V$ for the case of contact interactions by adapting the method developed in
Sec.~4.1 of Ref.~\cite{Oller:2017alp}, also reviewed in Ref.~\cite{Oller:2019opk}, to the many-body environment. 
Instead of reproducing it here we directly moved to solve the similar IE in Eq.~\eqref{190809.5} for $t_m(a\hvz)$ in PWAs,
which reduces to $t_V$ in the limit in which the Fermi momentum ${k_F}\to 0$. The different coupled partial-waves are called
channels and are fixed by the quantum numbers in the state vector $|J\mu\ell S p\ra$. To avoid confusion in the following the PWAs in the medium and in vacuum are denoted by ${\mathsf t_m}(k,p)$ and ${\mathsf t_V}(k,p)$, respectively, and it should be understood that $\va=a\hvz$. 

\tcr{Let us develop the method} that allows us to calculate ${\mathsf t_m}(k,p)$ given a contact interacting potential.  The potential 
coupling the channels $\alpha$ and $\beta$ is expressed in a polynomial expansion as  
 \begin{align}
\label{180319.1}
v_{\alpha\beta}(k,p)&=k^{\ell_\alpha}p^{\ell_\beta} \sum_{i,j=1}^N v_{\alpha\beta;ij}k^{2(i-1)}p^{2(j-1)}~.
 \end{align} 
In this equation the Greek subscripts denote the channels and run from 1 to $n$ and the factor  
$k^{\ell_\alpha}p^{\ell_\beta}$ keeps track of the right threshold behavior, which factorizes in a zero-range potential. 
We next introduce a matrix notation and write the potential as 
\begin{align}
\label{180319.4}
v_{\alpha\beta}(k,p)&=[k_\alpha]^T\cdot [v]\cdot [p_\beta]~,\\
\label{180319.2}
[v]&=\left(\begin{matrix}
     [v_{11}] & [v_{12}] & \ldots & [v_{1n}]\\
     [v_{21}] & [v_{22}] & \ldots & [v_{2n}]\\
     \ldots  & \ldots  &\ldots & \ldots \\
     [v_{n1}] & [v_{n2}] & \ldots & [v_{nn}]\\
     \end{matrix}
   \right)~,\\
\label{180319.3}
[k_\alpha]^T&=(
\underbrace{0,\ldots,0}_{N(\alpha-1) \text{~places}},k^{\ell_\alpha},k^{\ell_\alpha+2},\ldots,k^{\ell_\alpha+2(N-1)},0,\ldots,0)~.
 \end{align}
Here each $[v_{\alpha\beta}]$ is an $N\times N$ of matrix elements $v_{\alpha\beta;ij}$, $i,j=1,\ldots,N$, and the
$[k_\alpha]$ are $Nn\times 1$ column vectors. 
The solution to the IE in Eq.~\eqref{190809.6} can also be written in a matrix notation as 
\begin{align}
\label{180319.5}
{\mathsf t_{m}}(k,p)&=[k_\alpha]^T\cdot [\hat{t}_m(p)]\cdot [p_\beta]~,
\end{align}
with $[\hat{t}_m(p)]$ a squared $N n\times N n$ matrix, analogous to $[v]$ in Eq.~\eqref{180319.2}.
The fulfillment of  Eq.~\eqref{190809.6} (also for the full-off-shell case) implies that $[\hat{t}_m(p)]$ must satisfy the {\it algebraic} equation
\begin{align}
\label{180319.6}
[\hat{t}_m(p)]&=[v]-[v]\cdot [{\cal G}(p)] \cdot [\hat{t}_m(p)]~.
\end{align}
The $Nn\times Nn$ matrix  $[{\cal G}(p)]$ can be inferred from Eq.~\eqref{190809.6}  to be
\begin{align}
\label{180319.6b}
[{\cal G}(p)_{\alpha\beta}]&=-
\frac{m}{(2\pi)^2}\int_0^\infty \frac{k^2dk}{k^2-p^2-i\ep} {\cal A}_{\alpha\beta} [k_\alpha] [k_\beta]^T~.
\end{align}
The sought solution to Eq.~\eqref{180319.6} can then be expressed as  
\begin{align}
\label{180319.8}
[\hat{t}_m(p)]&=\left(I+[v][{\cal G}(p)]\right)^{-1}[v]~,
\end{align}

The (free part of the) loop functions in Eq.~\eqref{180319.6b} are evaluated by employing a cutoff regularization with a cutoff scale $\Lambda$, which at the end of the renormalization procedure is taken to infinity.
In general, power-like divergences of $\Lambda$ appear through the regularized integrals \cite{vanKolck:1998bw}
\begin{align}
\label{201229.1a}
L_n=-\frac{m}{2\pi^2}\int_0^\infty dk k^{n-1}=\theta_n\Lambda^n~,
\end{align}
where $\theta_n$ is a number that depends on the regularization scheme chosen. E.g. for a sharp cutoff $\theta_n=-m/(2\pi^2 n)$, and for the the case of dimensional regularization (DR) $\theta_n=0$.

The renormalization process consists of reproducing the ERE up to some order for vacuum scattering.
The ERE amounts to a polynomial expansion of the $n\times n$ matrix
\begin{align}
\label{201229.2a}
\frac{4\pi}{m}(p^\ell){\mathsf t_V}(p,p)^{-1}(p^\ell)+i(p^\ell)^2(p)=-(a)^{-1}+\frac{1}{2}(r)p^2+\sum_{i=1}(v^{(2i)})p^{2(i+1)}~,
\end{align}
where $(a)$, $(r)$ and $(v^{(2i)})$ are $n\times n$ matrices corresponding to the scattering lengths, effectives ranges and higher-order shape parameters, respectively. Other diagonal matrices are $(p)$, that is $p$ times the identity matrix, and $(p^\ell)$ given by ${\rm diag}(p^{\ell_1},\ldots,p^{\ell_n})$. Once the couplings $v_{\alpha\beta;ij}$,  Eq.~\eqref{180319.1}, are given as a function of the shape parameters of the ERE  and the cutoff, we finally take the limit $\Lambda\to\infty$. \trr{In this regard, 
  we require that the resulting $T$ matrix of PWAs be finite and independent of the regulator-dependent numbers $\theta_n$, cf. Eq.~\eqref{201229.1a},  for $\Lambda\to\infty$. Reaching  this limit successfully calls for the inclusion of} 
enough numbers of counterterms because of the factor $[k_\alpha][k_\beta]^T$ in the free part of $[{\cal G}(p)_{\alpha\beta}]$, which  gives rise to increasing power-like divergences
with the orbital angular momenta, starting with $\Lambda^{\ell_\alpha+\ell_\beta+1}$.

Interestingly enough, in all the explicit calculations that we have performed making use of 
\tcr{this method} for cutoff regularization ($\theta_n\neq 0$), 
it is found that  the renormalized solution for the off-shell vacuum PWAs 
${\mathsf t_V}(k,q)$ in the limit $\Lambda\to\infty$,  with $k$ and $q$ less than $2{k_F}$, can be written as
\begin{align}
\label{201229.3}
{\mathsf t_V}(k,q)&=\frac{4\pi}{m}(k)^{\ell}\tau(p)(q)^{\ell}~,\\
\tau(p)^{-1}&=-(a)^{-1}+\frac{1}{2}(r)p^2+\sum_{i=1}^M(v^{(2i)}_\ell)p^{2(i+1)}-i (p^\ell)^{2}(p)~,\nn
\end{align}
where $M$ is the order up to which the ERE is reproduced. It is important to remark that $\tau(p)$ is a purely on-shell ($p=mE$) and that 
the off-shellness  is  incorporated in ${\mathsf t_V}$ through its dependence on $(k)^\ell$ and $(q)^\ell$.

A clarifying remark is in order. When solving a Lippmann-Schwinger equation for on-shell scattering one has to provide also the solution for the half-off-shell ${\mathsf t_V}(k,p)$ with $k$ as big as $\Lambda$. However, in the case of the evaluation of the in-medium half-off-shell PWA ${\mathsf t_m}(k,p)$ we need ${\mathsf t_V}(p',k)$ in Eq.~\eqref{201228.2} with $p'$ and $k$ less than $2k_F$ and $p\leq k_F$, with ${k_F}/\Lambda\to 0$.
For such cases the result in Eq.~\eqref{201229.3} has always been found in our calculations for finite $M$ as long as $\theta_n\neq 0$ (i.e. cutoff regularization). 

\trr{Solving}  Eq.~\eqref{201228.2} for ${\mathsf t_m}(k,p)$ 
with ${\mathsf t_V}(k,p)$ expressed as in Eq.~\eqref{201229.3} is straightforward,
and this is indeed another more direct \tcr{way} 
to calculate ${\mathsf t_m}(k,p)$.  
In all the cases that we have explicitly worked out in this research it perfectly agrees with the one obtained by solving directly the IE of Eq.~\eqref{190809.6} with the method \tcr{discussed}. 
This solution adopts \trr{a  form analogous to that already given} for ${\mathsf t_V}(k,q)$ in Eq.~\eqref{201229.3}, and ${\mathsf t_m}(k,p)$ can be written as
\begin{align}
\label{201229.6}
{\mathsf t_m}(k,p)&=\frac{4\pi}{m}(k^\ell) \tau_m(p)(p^\ell)~,\\
\tau_m(p)&=\tau(p)-\tau(p)[{\cal G}_m(p)]\tau_m(p)=\left(\tau(p)^{-1}+[{\cal G}_m(p)]\right)^{-1}~,\nn\\
[{\cal G}_m(p)_{\alpha\beta}]&=-\frac{1}{\pi}\int_0^\infty\frac{k^2 dk}{k^2-p^2-i\ep}
 (k^\ell)\cdot{\cal B}_{\alpha\beta}\cdot (k^\ell)~.\nn
\end{align}

\trr{Due to the non-linear dependence of the couplings (counterterms) $v_{\alpha\beta;ij}$ on the ERE parameters up to the order considered in the matching for a general cutoff regularization, it is well known in the literature that some of the $v_{\alpha\beta;ij}$  could become complex for $\Lambda\to\infty$.
  This is discussed for the specific case of $S$-wave scattering in Refs.~\cite{Phillips:1997xu,Entem:2007jg,Habashi:2020qgw} where the scattering length and a positive effective range are reproduced with two counterterms (in agreement with our own findings). 
  However, the resulting off-shell partial-wave amplitudes from Eq.~\eqref{201229.3} fulfill off-shell unitarity (see Eq.~(2.29) of Ref.~\cite{Oller:2018zts}),
and they gives rise to perfectly meaningful phase shifts for on-shell scattering. 
 In this regard, we should stress that for us the potential has just been an intermediate step of no further use, and the real interesting point for our method 
 is to finally dispose of the non-perturbative Eq.~\eqref{201229.3} for ${\mathsf t_V(k,q)}$, with its explicit functional dependence on $k$,  $q$ and $p$.  
 The reason is because from this equation we can calculate  ${\mathsf t}_m(k,p)$, 
 without any further reference to the potential, as we have just explained. 
 Indeed positive effective ranges are realized in nature, and this is how our non-perturbative method accounts for them.   
 We give explicit examples for the application of this method in the next Sec.~\ref{sec.201229.1} for the case of $S$- and $P$-wave scattering.}

\section{Calculation of $\cE$ in $S$- and $P$-wave interacting systems and related aspects}
\label{sec.201229.1}
\def\theequation{\arabic{section}.\arabic{equation}}
\setcounter{equation}{0}

Given an uncoupled ERE expansion $p^{2\ell+1}\cot\delta=-1/a_\ell+r_\ell p^2/2+\ldots$, we always denote by $a_\ell$, $r_\ell$ the scattering length and effective range despite that for $\ell>0$ they have dimensions of ${\rm length}^{2\ell+1}$ and ${\rm length}^{1-2\ell}$, respectively. The system is said to be at unitarity \cite{tan25,tan26,tan27} when the scattering length $a_\ell\to \infty$ and $p\ll R^{-1}$, where let us recall that $R$ is the typical range of the interactions. For instance, for $S$ wave this requires that \trr{ $p\ll |r_0|^{-1}$}, while for $P$ wave the requirement is \trr{$p\ll |r_1|$}.

An interesting aspect to be studied is the influence of the effective range in the Fermi gases at around unitarity. Indeed, from simple scaling arguments \cite{Maki:2020zsv} one expects that in the ERE for a PWA with orbital angular momentum $\ell$ the first $\ell+1$  terms in the expansion are relevant at low energies and the rest of parameters are irrelevant. To show this explicitly let us scale $p\to\lambda p$ in the ERE 
and factorize $\lambda^{2\ell+1}$ from the phase space term $-ip^{2\ell+1}$. It then results the naive dimensional scaling
\begin{align}
\label{201229.1}
-\frac{1}{\lambda^{2\ell+1}a_\ell}+\frac{1}{2}\frac{r_\ell}{\lambda^{2\ell-1}} p^2
+\sum_{i=1}\frac{v^{(2i)}_\ell}{\lambda^{2\ell-1-2i}}p^{2(i+1)}~.
\end{align}
Attending to the evolution of the different terms for $\lambda\to 0$ \tcr{(the low-momentum limit)} one obtains the expectation claimed.\footnote{\tcr{From a physical point of view this scaling argument is just reflecting the fact that, taking into account 
    the typical range of interactions ($R$), the effective range $r_\ell$ depends on it as $R^{-\ell+1}$, while the shape parameters $v_\ell^{(2i)}$ do so as $R^{-2\ell-1+2i}$ (see Ref.~\cite{Oller:2014uxa} for explicit calculations in nucleon-nucleon scattering in which case $R$ is settled by the inverse of the pion mass). Notice that the exponent in the corresponding power of $R$ is the negative of the exponent of $\lambda$ for the same term in Eq.~\eqref{201229.1}. Therefore, from these scaling arguments, the first $\ell+1$ terms in the ERE in the low-momentum limit, $p\ll R^{-1}$,  can give rise to large contributions, so that it is necessary to consider them.}\label{foot.210907.1}} 
This result can be taken to the scattering of fermions in the medium  because of Eq.~\eqref{201229.6}.
Therefore, it is pertinent to study the dependence of the energy density $\cE$ on the higher-order shape parameters in the ERE, even if the system is at unitarity. In this way, one can ascertain whether there is a perturbative dependence of this magnitude on the effective range and even on other shape parameters.

Regarding this point, we find that $P$-wave results can only be renormalized in cutoff regularization
once both the scattering length and effective range are reproduced, \trr{as already obtained in Ref.~\cite{Bertulani:2002sz}.} 
A paradigmatic example where these results are of interest is neutron matter, where the scattering length has a large magnitude, $a_0=-18.95$~fm, and the effective range $r_0=2.75$~fm is sizeable, both of them compared with the inverse of the pion mass around 1.4~fm.
In ultracold atoms there is also the possibility to tune the scattering length, and also the effective range through the dark-state optical control of Feshbach resonances \cite{Hu:2019wrm,thomas:2012}.

\trr{A non-perturbative fact that we take into account when applying our formalism is to exclude those
  values of the scattering length and effective range which give rise to an unacceptable pole content of $\tau(p)$, Eq.~\eqref{201229.3}, when analytically continued to complex values of $p$. 
  For instance, a pair of resonant poles located in the complex-$p$ plane  with opposite values of their real parts and the same positive imaginary part is a pole disposition to be excluded because it would give rise to normalized eigenstates of the Hamiltonian with complex eigenvalues, which is not allowed for a Hermitian Hamiltonian \cite{gottfried.book}. 
Another configuration that should be avoided is to have two $S$-wave shallow poles along the imaginary axis with positive imaginary part,  such that $|p|$ is clearly smaller than the inverse of the range of the interactions (one of the poles corresponds to a bound state and the other to redundant pole \cite{ma:1946,Ma:1947zz}), as recently  shown in Ref.~\cite{Habashi:2020qgw}.}

In the next subsections we calculate ${\cE}$ by employing Eq.~\eqref{200501.6} in the case of considering  separately $S$ and $P$ waves.  
Each in-medium renormalized PWA calculated obeys Eq.~\eqref{201229.6}, as it has  been  obtained by direct calculation, cf. Eq.~\eqref{180319.8}. 

\subsection{$S$ waves}
\label{sec.201230.1}

Given  a constant potential $v=c_0$ 
the vacuum scattering amplitude then reads ${\mathsf t_V}=\left(1/c_0+\theta_1\Lambda-imp/4\pi\right)^{-1}$. The renormalization is achieved by setting 
$c_0^{-1}=-m/4\pi a_0-\theta_1\Lambda $, so that 
  ${\mathsf t_V}=-4\pi/m \left(1/a_0+ip\right)^{-1}$.  
The expression for $\tau_m(p)$ is equally simple in this case,
\begin{align}
\label{201231.2}
\tau_m(p)&=\left(-\frac{1}{a_0}-ip+{\cal G}_m(p)\right)^{-1}~.
\end{align}
In terms of it, $\cE_L$ in Eq.~\eqref{200501.6} reads 
\begin{align}
\label{201231.3}
\cE_L&=-\frac{4i}{m\pi^3}\int_0^{k_F}a^2 da\int_0^{\sqrt{k_F^2-a^2}}p dp
\log\left(1-\frac{4\pi/m}{-\frac{1}{a_0}-ip+{\cal G}_m(p)}L_d\right)\\
&=-\frac{4i}{m\pi^3}\int_0^{k_F}a^2 da\int_0^{\sqrt{k_F^2-a^2}}p dp
\log\left(\frac{1-a_0(-ip+{\cal G}_m(p)-4\pi L_d/m)}{1-a_0(-ip+{\cal G}_m(p))}\right)~.\nn
\end{align}
Explicit expressions for the functions ${\cal G}_m(p)$ and $L_d(p,a\hvz)$ are given in the Appendix~C of Ref.~\cite{Lacour:2009ej}. The connection with the functions there defined is ${\cal G}_m(p)=4\pi L_{10,m}/m$ and $L_d(p,a\hvz)=-L_{10,d}$. Regarding the one-loop functions  $B_1$, $B_2$, $R$ and $I$ introduced by Kaiser in Ref.~\cite{Kaiser:2011cg} we have the relations: $B_1=a_0{\cal G}_m(p)$, $B_2=-4\pi a_0L_d/m=-i2k_Fa_0I$, $\Re B_1=a_0k_F R/\pi$, such  that $a_0k_F(R+i\pi I)/\pi=a_0(-ip+{\cal G}_m)$ and $a_0k_F(R-i\pi I)/\pi=a_0(-ip+{\cal G}_m-4\pi L_d/m)$. 
Then, Eq.~\eqref{201231.3} can be written as
\begin{align}
\label{201231.4}
\cE_L&=-\frac{4i}{m\pi^3}\int_0^{k_F}a^2 da\int_0^{\sqrt{k_F^2-a^2}}p dp
\log\left(\frac{1-a_0k_F(R/\pi-i I)}{1-a_0k_F(R/\pi+i I)}\right)~,\\
&=\frac{8k_F^5}{m\pi^3}\int_0^{1}s^2ds\int_0^{\sqrt{1-s^2}}\kappa d\kappa  
\arctan\left(\frac{I}{(a_0k_F)^{-1}-R/\pi}\right)~,\nn
\end{align}
where the dimensionless variables $s=a/ k_F$ and $\kappa=p/k_F$ are introduced.
The previous expression is the same as the one from Ref.~\cite{Kaiser:2011cg}, except for the fact that this reference introduces the scattering length with an extra minus sign. \trr{For completeness we reproduce the expressions for the functions $R$ and $I$:}
\trr{\begin{align}
\label{211028.3}
R(s,\kappa)&=2+\frac{1-(s+\kappa)^2}{2s}\log\frac{1+s+\kappa}{|1-s-\kappa|}
+\frac{1-(s-\kappa)^2}{2s}\log\frac{1+s-\kappa}{1-s+\kappa}~,\\
I(s,\kappa)&=\left\{\begin{array}{cc}\kappa & \text{for } 0<\kappa<1-s\\
\frac{1-s^2-\kappa^2}{2s} & \text{for } 1-s<\kappa<\sqrt{1-s^2}~.\end{array}
\right.\nn
\end{align}}

Several interesting expansions of  $\bar{\cE}= {\cE}/\rho$  have been studied in the literature.
The first \cite{tan25,tan26,tan27,Chang:2004zza,Astrakharchik:2004zz,tan30,Tan:2008a,Kaiser:2011cg} consists of an expansion in negative powers of $a_0{k_F}$ around unitarity,
\begin{align}
\label{201229.2}
\bar{\cE}&=\frac{3{k_F}^2}{10m}\left\{\xi-\frac{\zeta}{a_0k_F}-\frac{5\nu}{3(a_0k_F)^2}+\ldots\right\}~.
\end{align}

The first term between brackets on the rhs of the previous equation is the famous Bertsch parameter $\xi$, while the others measure the deviation from the unitary limit.
The expression for $\xi$ that follows from Eq.~\eqref{201231.4} is \cite{Kaiser:2011cg}
\begin{align}
\label{201229.2b}
\xi&=1-\frac{80}{\pi}\int_0^1 ds s^2\int_0^{\sqrt{1-s^2}} d\kappa \kappa \arctan\left(\frac{\pi I}{R}\right) =  0.5066\,.
\end{align}

\begin{figure}[H]
\begin{center}
\includegraphics[width=.6\textwidth]{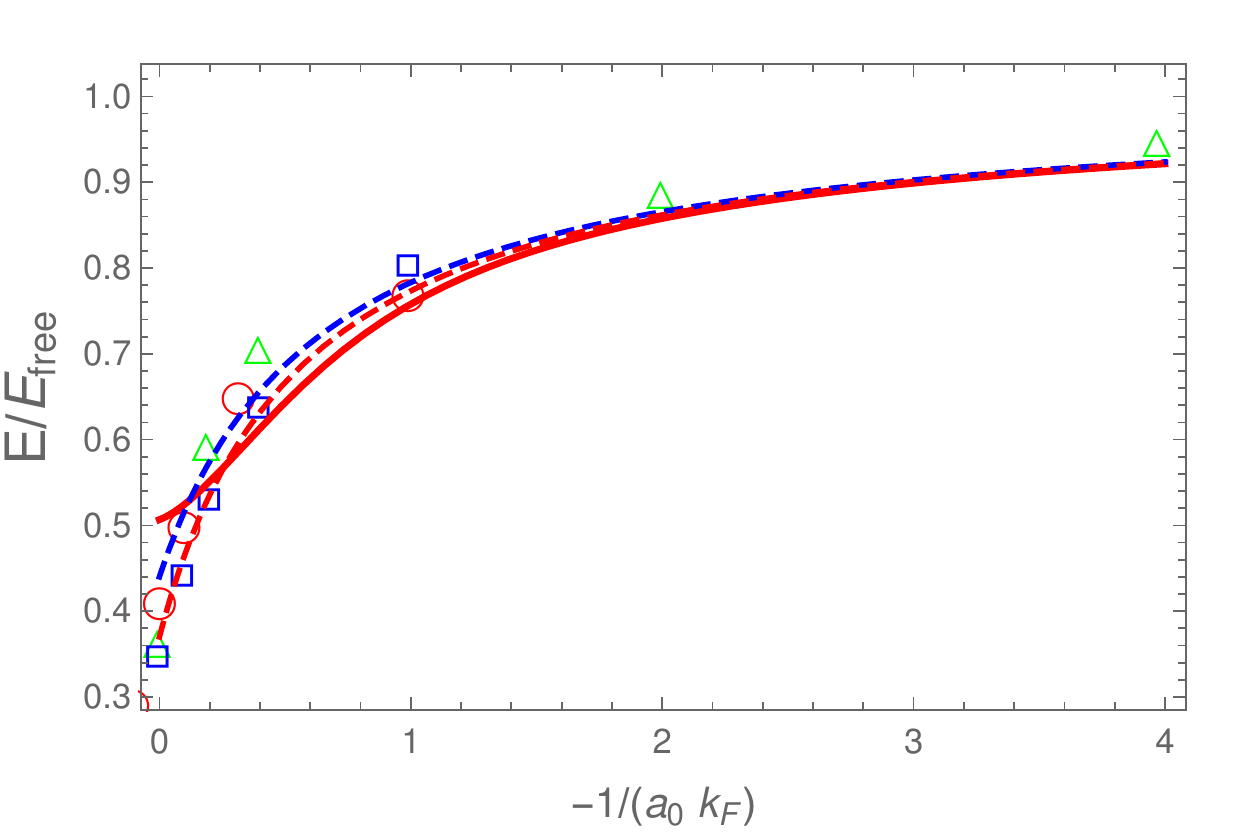}
\end{center}
\caption{{\small Behavior of ${\cE}/{\cE}_\text{free}$ as a function of $-1/ak_F$. The results in our study are plotted by the solid line. The upper and lower  dashed lines are the results of Ref.~\cite{Lacroix:2016dfs} with $\xi_0 = 0.44$ and $\xi_0 = 0.37$, respectively. The circles \cite{Chang:2004zza}, triangles \cite{Astrakharchik:2004zz} and  squares \cite{Gezerlis:2007fs} are Monte-Carlo calculations.}  \label{fig:makfinv} 
}
\end{figure}

The Bertsch parameter has been measured experimentally by Ref.~\cite{Ku2012RevealingTS} with the value
$\xi=0.370(5)(8)$ corresponding to a superfluid phase \cite{Navon2010TheEO}. In the latter reference it is also deduced the value of $\xi$ for normal matter at the unitary limit, $\xi_n\approx 0.45$, by extrapolating the curve for the chemical potential above the critical temperature up to zero temperature.
This value is quite close to our calculation in Eq.~\eqref{201229.2b}.
The parameters $\zeta$ and $\nu$ have been calculated by applying quantum Monte Carlo in Ref.~\cite{Gandolfi:2011} with the values $\zeta=0.901(2)$ and $\nu=0.49(2)$.
Our results obtained for normal matter look very different, 
$\zeta = -0.116$ and $\nu = -0.30$. 
The difference in sign reflects a  different qualitatively behavior of our results as a function of $1/a_0k_F$ compared with others corresponding to a superfluid near the unitary limit.
This is explicitly shown in Fig.~\ref{fig:makfinv} where we plot $\cE/\cE_{\rm free}$ as a function of $-1/(a_0k_F)$ by the solid line and compare it with other calculations.
The dashed lines come from the density-functional theory of Ref.~\cite{Lacroix:2016dfs} (the upper one implements $\xi_0=0.44$ and the lower $\xi_0=0.37$), the points correspond to different Monte-Carlo simulations: circles \cite{Chang:2004zza}, triangles \cite{Astrakharchik:2004zz} and squares \cite{Gezerlis:2007fs}. We also notice that our results and the others presented in Fig.~\ref{fig:makfinv} rapidly converge outside the unitary limit.

\trr{The algebraic expressions derived for $\zeta$ and $\nu$ in Ref.~\cite{Kaiser:2011cg}, and their numerical values thereof, obtained by expanding directly the integrand in Eq.~\eqref{201231.4} are not right. This is due to the emergence of an in-medium pole singularity in the $S$-wave amplitude in Eq.~\eqref{201231.2} sitting at the border of the Fermi seas of the two interacting fermions (namely, for $\kappa=\sqrt{1-s^2}$ and $s\in [0,1]$). This issue is discussed in the Appendix \ref{app.211029.1}.}

An interesting point is to include the effective range in the calculation of $\cE$ and determine how it modifies the unitary limit.  A pertinent question is  whether its effects are perturbative, in line with the scaling arguments in Eq.~\eqref{201229.1} and as obtained in Ref.~\cite{Maki:2020zsv} or, on the contrary, they alter dramatically the unitary limit according to the DR calculations of Ref.~\cite{Kaiser:2012sr} and  Appendix \ref{app.210102.1}. 
Therefore, this digression elaborates on including a scale at unitarity in connection with the range of the interactions. 
In the case of {$^6$Li}  cold-atom experiments $r_0\approx 4.7$~nm and the gas can be cooled at densities of $1/k_F\approx 400$~nm, so that $r_0 k_F\approx 0.01$ and this is a small parameter. However, for neutron matter with $r_n=2.75(11)$~fm \cite{Kaiser:2012sr,Miller:1990iz,Perez:2014waa} 
 and $k_F\sim 1~m_\pi$ then $r_0k_F\sim 3$, and it could impact considerably the limit $a_0\to\infty$, cf. Fig.~\ref{fig:ma0kf}.

We start by considering  the contact $S$-wave potential
\begin{align}
\label{201231.5}
v(k,p)=c_0+\frac{1}{2}c_2(k^2+p^2)~,
\end{align}
and then apply the method in Sec.~\ref{sec.201228.1} to obtain renormalized scattering amplitudes by reproducing given values of $a_0$ and $r_0$. The resulting in-medium ${\mathsf t_m}(k,p)$ in the limit $\Lambda\to\infty$ reads from Eq.~\eqref{201229.6}, 
\begin{align}
\label{201231.6}
{\mathsf t_m}(k,p)&=\frac{4\pi/m}{-\frac{1}{a}+\frac{1}{2}r_0p^2-ip+{\cal G}_m(p)}~.
\end{align}
Let us stress that this result is worked out in cutoff regularization, $\theta_i\neq 0$ for $i=1,\ldots,4$, being the result independent of the particular finite values taken by these parameters. 
However, this is not the same as taking the potential in Eq.~\eqref{201231.5} and solving for $\cE$ in DR as in  Ref.~\cite{Kaiser:2012sr}.
Within our general formalism we can also proceed with DR, which is done in the Appendix~\ref{app.210102.1}, and reproduce Kaiser's results \cite{Kaiser:2012sr}. In the same Appendix we also include $v^{(2)}_0$, one more order considered in the ERE as compared with Ref.~\cite{Kaiser:2012sr}.

We present along the main text  the results with cutoff regularization since we consider it as the correct physical regularization and renormalization procedure for non-perturbative calculations.
The point is that the use of DR is not justified a priori for non-perturbative calculations \cite{Phillips:1997xu}, because of the lose of track of new divergences when iterating the potential that are set to zero in DR, while in cutoff regularization they are  kept explicitly. Several examples of nonsensical results by using DR in the non-perturbative calculations of scattering amplitudes with zero-range potentials are given in Ref.~\cite{Phillips:1997xu}. We also have in mind the implementation of non-perturbative  QFT in  lattice gauge theories \cite{rothe.lattice}, where the spacing $a$ of the grid provides the cutoff and the limit $a\to 0$ has to be performed.

\begin{figure}
\begin{center}
\includegraphics[width=.6\textwidth]{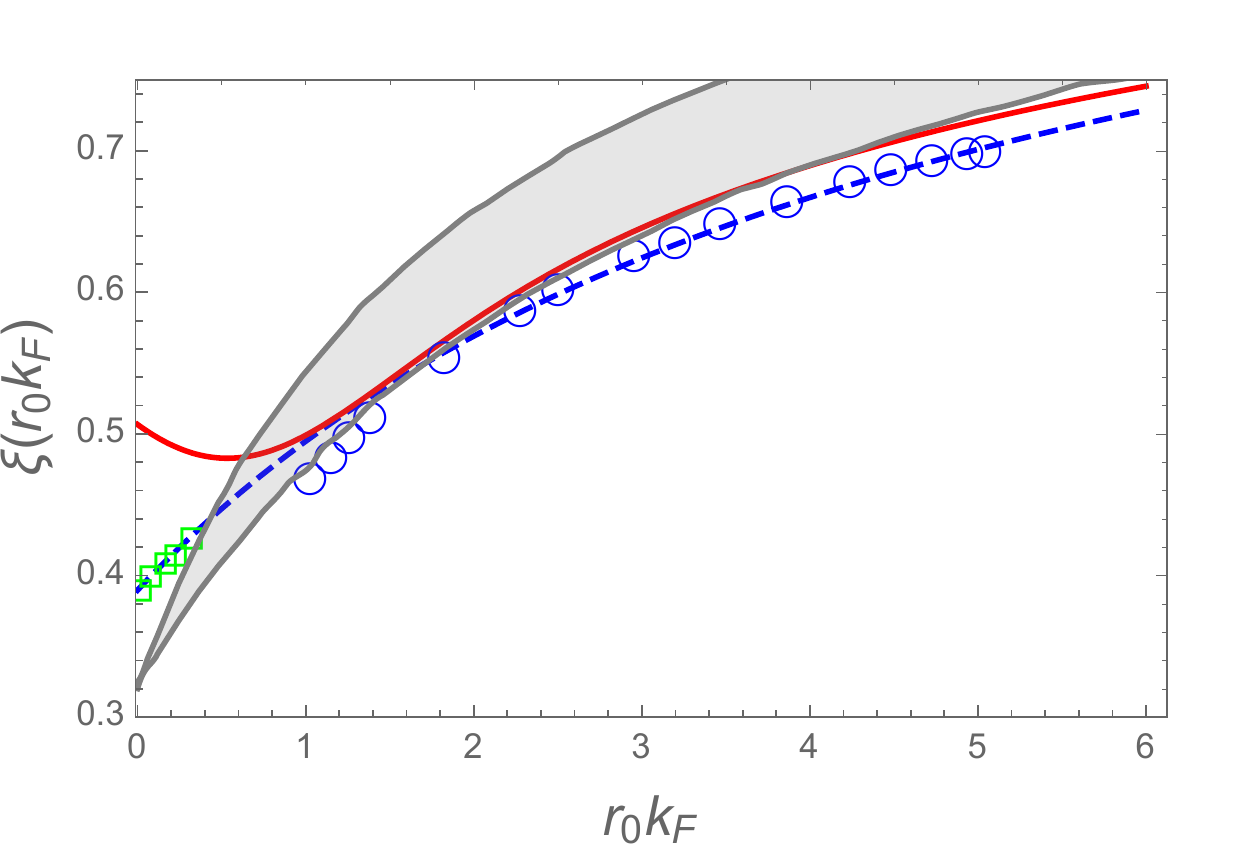}
\end{center}
\caption{{\small Dependence of the Bertsch parameter on $r_0 k_F$, $\xi(r_0k_F)$.  
  Our result is the solid line,  the dashed line corresponds to the density-functional theory applied in Ref.~\cite{Lacroix:2017whm} with $\xi_0 = 0.3897$ and $\eta_e = 0.127$, and we also show the circles from Ref.~\cite{Schwenk:2005ka}, the squares from Refs.~\cite{Forbes:2010gt,Forbes:2012ku}, and the gray area that stems from Ref.~\cite{Schafer:2005kg}. }
  \label{fig:r0kf} 
}
\end{figure}

Now, when Eq.~\eqref{201231.6} is implemented in Eq.~\eqref{200501.6}, the following result is obtained
\begin{align}
\label{201231.7}
\cE_L&=-\frac{4k_F^5i}{m\pi^3}\int_0^{k_F}a^2 da\int_0^{\sqrt{k_F^2-a^2}}p dp
\log\left(\frac{1-a_0r_0k_F^2 \kappa^2/2-a_0k_F(R/\pi-i I)}{1-a_0r_0k_F^2 \kappa^2/2-a_0k_F(R/\pi+i I)}\right)~,\nn\\
&=\frac{8k_F^5}{m\pi^3}\int_0^{1} ds s^2\int_0^{\sqrt{1-s^2}} d\kappa \kappa
\arctan\left(\frac{a_0k_FI}{1-a_0r_0k_F^2 \kappa^2/2-a_0k_FR/\pi}\right)~.
\end{align}

The result in Eq.~\eqref{201231.7} for $\cE$ is clearly perturbative with respect to $r_0$, and for $r_0\to 0$ one recovers $\cE$ as given by the unitarity limit in Eq.~\eqref{201231.4}. This conclusion is 
in agreement with Ref.~\cite{Maki:2020zsv}, that also obtained that $r_0$ is perturbative concerning its role in the bulk viscosity of Fermi gases interacting in $S$-wave.

The explicit dependence on $r_0$ of $\xi$ can be calculated from Eq.~\eqref{201231.7} by taking the limit $a_0\to\infty$. 
Having included a scale related with the range of the interactions then $\xi$ becomes a function of $k_F$ through the dimensionless parameter $r_0k_F$. The result for $\xi(k_F)$ is then,
\begin{align}
\label{210101.1}
\xi(k_F)&=1-\frac{80}{\pi}\int_0^1 ds s^2\int_0^{\sqrt{1-s^2}}d\kappa \kappa  \arctan\left(\frac{\pi I}{\pi r_0 k_F \kappa^2/2+R}\right)~. 
\end{align}
We depict  $\xi(k_F)$ as a function of $k_Fr_0$ in Fig.~\ref{fig:r0kf}, where it is clear the smooth dependence on this parameter.  The Taylor series of $\xi(k_F)$ up to quadratic order in powers of $r_0k_F$ that has been proposed in the literature  \cite{Castin:2012,Forbes:2012ku,Lacroix:2017whm,Kolck2017UnitarityAD} reads, 
\begin{align}
\label{210101.2a}
\xi(k_F)&=\xi_0+\eta_e r_0k_F+\delta_e(r_0k_F)^2+\ldots
\end{align}
Quantum Monte-Carlo simulations give the reference values $\eta_e=0.127$, $\delta_e=-0.055$ \cite{Forbes:2012ku} and $\eta_e=0.12(3)$ \cite{Carlson:2011kv}, which are universal parameters. The numerical values that we obtain from the expansion of $\xi(k_F)$ in Eq.~\eqref{210101.1} are $\eta_e = -8.59\cdot 10^{-2}$, and $\delta_e =6.45\cdot 10^{-2}$.\footnote{These values cannot be obtained by expanding the integrand of Eq.~\eqref{210101.1} in powers of $r_0 k_F$ because the Leibniz rule cannot be applied\trr{, as explained in the Appendix~\ref{app.211029.1}.}\label{leibnitz.2}} These values reflect the $r_0k_F$ dependence of the solid line  near the origin  as seen in Fig.~\ref{fig:r0kf}. First, one has a decrease of $\xi(r_0k_F)$ when the linear term dominates and then an increase when the quadratic terms become more important. Afterwards our results and those from the density-functional theory calculation in Ref.~\cite{Lacroix:2017whm} (dashed line) run rather closely. We also show in the Fig.~\ref{fig:r0kf} by the empty circles the results of Ref.~\cite{Schwenk:2005ka}, by the squares the low-density calculation of Refs.~\cite{Forbes:2010gt,Forbes:2012ku}, and by the gray area the calculation in Ref.~\cite{Schafer:2005kg}, where the width of the band is due to the dependence on the renormalization scale chosen.

 \begin{figure}
\begin{center}
\includegraphics[width=.6\textwidth]{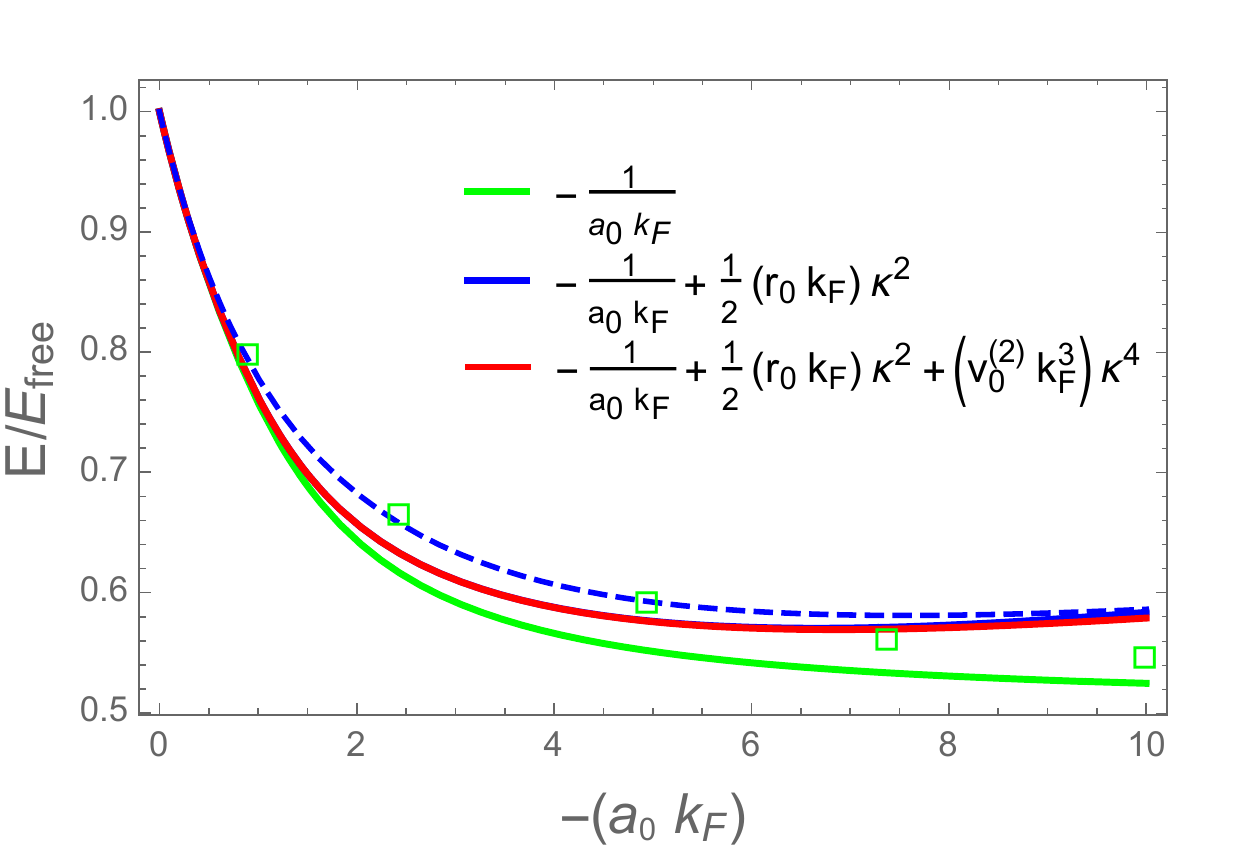}
\end{center}
\caption{{\small ${\cE}/{\cE}_{\text{free}}$ for neutron matter as a function of $- a_0 k_F$. Our results including only $a_0$ are plotted by the lowest lying solid line. The other two solid lines overlap each other and correspond to our results including consecutively  $r_0$ and $v_0^{(2)}$. The dashed line is the density-functional theory result from Ref.~\cite{Lacroix:2016dfs} with $\xi_0 = 0.3897$, $r_0 = 2.75~$fm and $\eta_e = 0.127$. The squares correspond to the quantum Monte Carlo calculation of Ref.~\cite{Gezerlis:2009iw}. }
\label{fig:ma0kf} 
}
\end{figure}

We also consider the  contributions to $\cE$ from $v^{(2)}_0$, the next shape parameter in the ERE,  and determine its impact on  ${\cE}$. 
For evaluating this case we consider the contact potential
\begin{align}
\label{210101.4}
v(k,p)&=c_0+\frac{1}{2}c_2(k^2+p^2)+\frac{1}{2}c_4(k^4+p^4)~.
\end{align}
Here we have taken into account the main result of Ref.~\cite{Beane:2000fi}, which shows that there is only one new operator for every higher order in the low-energy expansion of the potential for contact interactions. This is then completely consistent with the ERE where only one new coefficient is added by increasing the expansion in $p^2$ one  more order. By applying the method of Sec.~\ref{sec.201228.1} we can solve for a renormalized ${\mathsf t_m}(k,p)$ so that its contribution to $\cE_L$ is a straightforward extension of Eq.~\eqref{201231.7} that reads
\begin{align}
\label{220102.1}
\cE_L&=\frac{8k_F^5}{m\pi^3}\int_0^{1} ds s^2\int_0^{\sqrt{1-s^2}} d\kappa \kappa
\arctan\left(\frac{a_0k_FI}{1-a_0r_0k_F^2 \kappa^2/2-a_0v^{(2)}_0k_F^4 \kappa^4-a_0k_FR/\pi}\right)~, 
\end{align}
which is perturbative in $v^{(2)}_0$. The dependence of $\xi(k_F)$ on this parameter is easily obtained by taking the limit $a_0\to\infty$ in the previous equation,
\begin{align}
\label{220102.2}
\xi(k_F)&=1-\frac{80}{\pi}\int_0^1 ds s^2\int_0^{\sqrt{1-s^2}}d\kappa \kappa  \arctan\left(\frac{I}{ r_0 k_F \kappa^2/2+ v^{(2)}_0k_F^3\kappa^4+R/\pi}\right)~.
\end{align}
A Taylor expansion in powers of $r_0k_F$ and $v^{(2)}k_F^3$ of the previous equation gives the first contribution from $v^{(2)}_0$ as
$v^{(2)}_0k_F^3 \gamma_e$. Numerically, we obtain that $\gamma_e = -0.164$.\footnote{This value cannot be obtained by expanding the integrand of Eq.~\eqref{220102.2} in powers of $v_0^{(2)} k_F^3$ because the Leibniz rule cannot be applied\trr{, as explained in the Appendix~\ref{app.211029.1}.}\label{leibnitz.3}}

\begin{figure}
\begin{center}
\includegraphics[width=.6\textwidth]{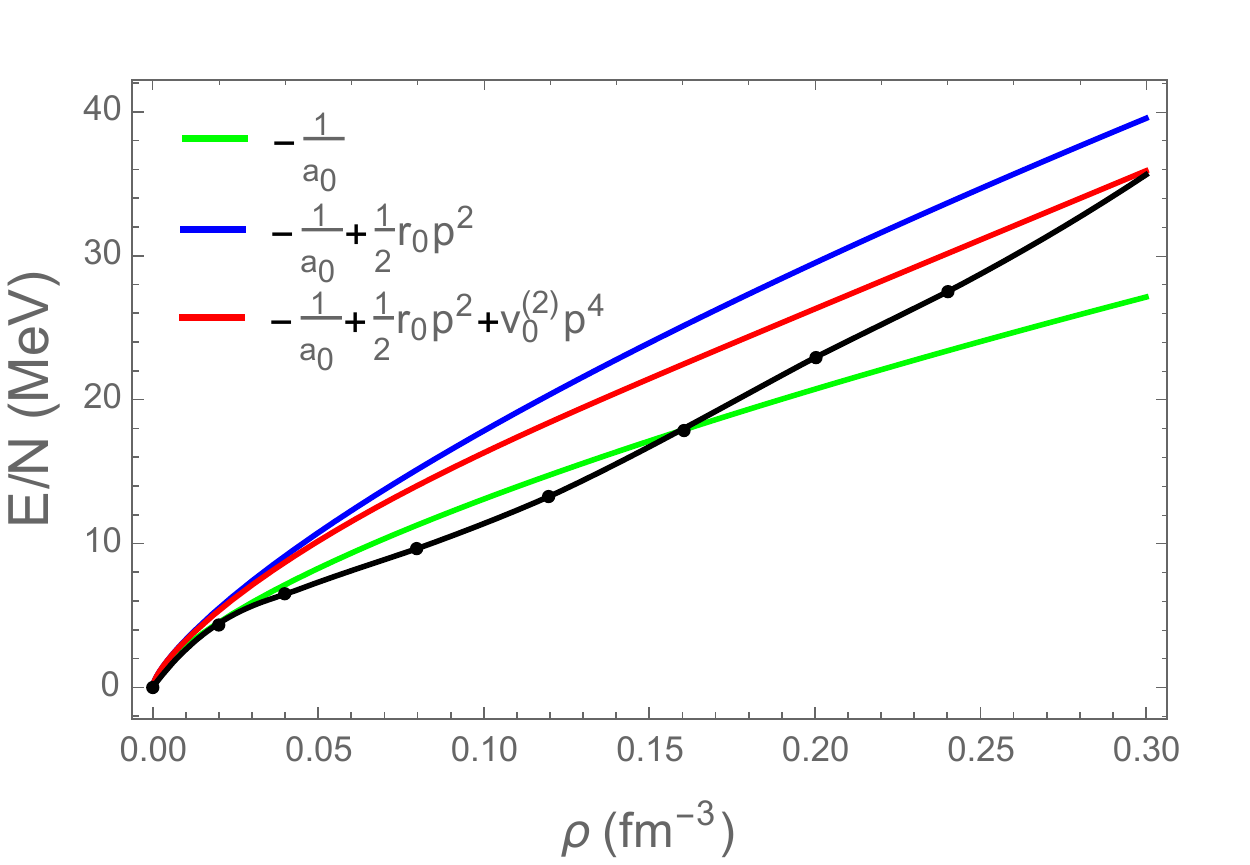}
\end{center}
\caption{{\small Our results for the energy per particle of pure neutron matter calculated with different orders in the ERE: The lowest solid line only includes  $a_0$, the highest line includes $a_0$ and $r_0$, and the solid line in the middle considers the contributions from $a_0$, $r_0$ and $v_0^{(2)}$. The dots correspond to the result of Akmal {\it et al.} \cite{Akmal:1998cf}. } \label{Fig:E_ERE}}
\end{figure}

For the case of neutron matter with only the $S$-wave interactions we plot next $\cE/\cE_{\rm free}$ in Fig.~\ref{fig:ma0kf} as a function of $-ak_F$. We show our results by the  solid lines. The lowest one stems from the contributions of only $a_0$. The other two lines overlap each other within the scale of the figure and include consecutively $r_0$ and $v_0^{(2)}$ (with values $r_0=2.75$~fm and $v_0^{(2)}=-0.50$~fm$^3$ \cite{Perez:2014waa}).
The density-functional theory result from Ref.~\cite{Lacroix:2016dfs} with $\xi_0 = 0.3897$, $r_0 = 2.75$ and $\eta_e = 0.127~$fm is plotted by the dashed line. 
We also give the quantum Monte-Carlo approach of Ref.~\cite{Gezerlis:2009iw} with a finite-range $S$-wave interaction by the squares.

The Fig.~\ref{Fig:E_ERE} shows our results for the energy per particle $\bar{\cE}$ of neutron matter as a function of the density $\rho$. The lowest, upper and middle solid lines include  $a_0$, $r_0$ and $v_0^{(2)}$ consecutively, with these parameters taking the values given already for the $^1S_0$ partial-wave amplitude. The dots are the results from Ref.~\cite{Akmal:1998cf} using variational chain summation methods and sophisticated nucleon-nucleon potentials. From this figure it is clear the sizeable impact of $r_0$ on $\bar{\cE}$  providing extra repulsion, that is reduced to some extent by the inclusion of $v_0^{(2)}$.

\begin{figure}[H]
\begin{center}
\includegraphics[width=.6\textwidth]{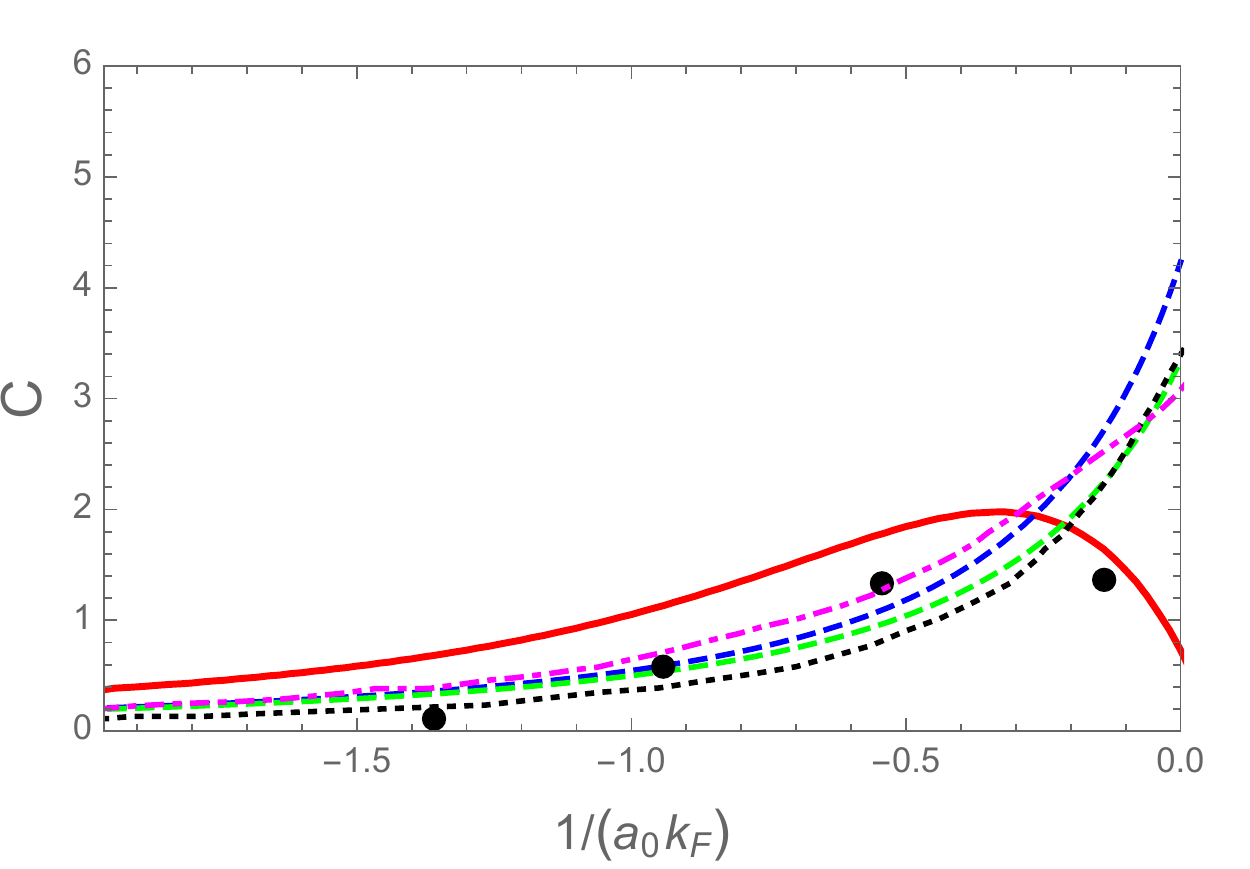}
\end{center}
\caption{{\small Dependence of the Tan density contact parameter on $1/a_0 k_F$. Upper (lower) dashed lines are from Ref.~\cite{Lacroix:2016dfs} with $\xi_0 = 0.37$ ($\xi_0 = 0.44$) and $r_0 = 0$. The dotted and dash-dotted lines are the theoretical results from 
Refs.~\cite{Werner:2009} and \cite{Haussmann:2009}, 
respectively. The dots are the measurements of Ref.~\cite{Kuhnle:2010}. 
The  solid line corresponds to our calculation. 
  }
\label{fig:tanc} 
}
\end{figure}

\begin{figure}[H]
\begin{center}
\includegraphics[width=.6\textwidth]{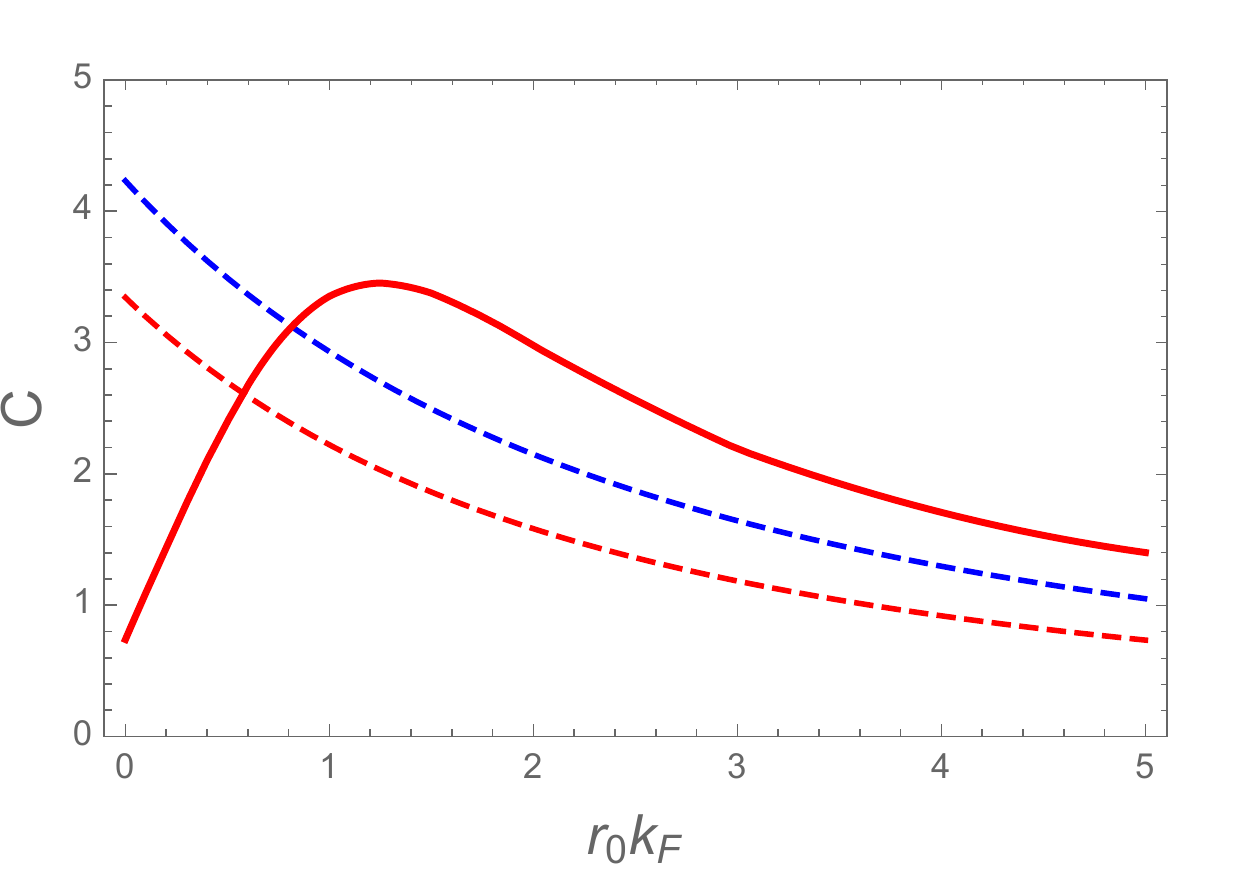}
\end{center}
\caption{{\small Dependence of the density contact parameter ${\mathsf C}$ on $r_0 k_F$ for $a_0 \to  \infty $. Our results are given by the solid line and the upper and lower dashed lines come from Ref.~\cite{Lacroix:2016dfs} with $\xi_0=0.37$ and 0.477, respectively, and $\eta_e=0.127$. 
 }
 \label{fig:tancr0kf}
}
\end{figure}

We also study 
the Tan contact parameter $C$ \cite{Tan:2008a}, which is related to the contact density parameter ${\mathsf C}$  by $C/(Nk_F)=3\pi^2{\mathsf C}/k_F^4$, such that \cite{braaten:2012}
\begin{align}
\label{210105.1}
{\mathsf C}&=4\pi m a_0^2\frac{d{\cal E}}{da_0} .
\end{align}
By taking numerically the derivative with respect to $a_0$ in Eq.~\eqref{220102.1} (the derivative with respect to $a_0$ under the integral symbol is not correct, cf. \trr{Appendix~\ref{app.211029.1}}) we then have the solid line in Fig.~\ref{fig:tanc} that shows ${\mathsf C}$ as a function of $1/a_0k_F$ with $r_0=v_0^{(2)}=0$. 
The  upper and lower dashed lines  are the results from the density-functional theory of Ref.~\cite{Lacroix:2016dfs} with $r_0 = 0$ and $\xi_0 = 0.37$, $0.44$, respectively.
The dotted and dash-dotted lines are the theoretical results from Refs.~\cite{Werner:2009} and
\cite{Haussmann:2009}, 
respectively. The black dots are the experimental measurements of Ref.~\cite{Kuhnle:2010}.
It is interesting to notice that our results are the only theoretical ones that do not increase for $a_0\to \infty$, as it is also the case for the experimental points. Next, we plot in Fig.~\ref{fig:tancr0kf} the dependence of our results (solid line) for the contact density parameter in the unitary limit as a function of $r_0 k_F$. We also include the dashed lines  from Ref.~\cite{Lacroix:2016dfs} with the upper one using the value $\xi_0=0.37$ and the lower $\xi_0=0.44$, both taking $\eta_e= 0.127$.

\trr{We now elaborate on the pole content and its feasibility that results by having included first $r_0$ together with $a_0$, and then additionally $v_0^{(2)}$. The poles in the complex-$p$ plane of $\tau(p)$ when taking the ERE up to and including $r_0$ correspond to the zeroes of $-1/a_0+r_0 p^2/2-ip$ which are located at $p=(i \pm \sqrt{2r_0/a_0-1})/r_0$.
  Regarding the specific cases studied in Figs.~\ref{fig:r0kf}, \ref{fig:ma0kf}, \ref{Fig:E_ERE} and \ref{fig:tancr0kf}  by using Eq.~\eqref{201231.7},
given that we consider the unitary limit or the case $|a_0|\gg r_0$ for neutron matter,
the problematic pole dispositions to be avoided (discussed just before the present  Sec.~\ref{sec.201230.1}),
and associated with a positive effective range \cite{Habashi:2020qgw}, do not affect our results because:}

\trr{i) In the unitary limit $2r_0/a_0\to 0$, and for our results, Figs.~\ref{fig:r0kf} and \ref{fig:tancr0kf}, it does not really matter which is the sign of $a_0\to \infty$ so that we can always avoid having a redundant bound state when $0<2r_0/a_0<1$ by taking $a_0\to-\infty$ for $r_0>0$.}

\trr{ii)  For the case of neutron matter, considered in Figs.~\ref{fig:ma0kf} and \ref{Fig:E_ERE}, one has that $a_0<0$ and $r_0>0$, so that $2r_0/a_0<0$, and neither the presence of two resonant poles
with positive imaginary part nor of a redundant low-energy second pole lying along the positive momentum imaginary axis occur \cite{Habashi:2020qgw}. Instead, one has a bound and a virtual state.}

\trr{Concerning Eq.~\eqref{220102.1} involving also the shape parameter $v_0^{(2)}$ we recall that we use the values for $S$-wave  neutron-neutron scattering $a_0=-18.95$~fm, $r_0=2.75$~fm, and $v^{(2)}_0=-0.5$~fm \cite{Chen:2008zzj,Perez:2014waa} to get the results shown in Figs.~\ref{fig:ma0kf} and \ref{Fig:E_ERE}. 
For determining the pole content of $\tau(p)$ in this case we have solved numerically the zeros of $-\frac{1}{a_0}+\frac{1}{2}r_0 p^2+ v_0^{(2)} p^4-i\,p$ in the complex-$p$ plane and we have found an acceptable disposition of poles located at: $p_1=-i\,0.0706\,m_\pi$, $p_2=i\,0.9627\,m_\pi$, and $p_{3,4}=(\pm 2.50075-i\,0.4460)\,m_\pi$, in units of the pion mass denoted by $m_\pi$. }

\subsection{$P$ waves}
\label{sec.210104.1}

Let us consider a $P$-wave spin-independent zero-range potential given by
\begin{align}
\label{210104.1}
 V(\vk,\vp)&=\vk\cdot\vp\left(d_0+\frac{1}{2}d_2(k^2+p^2)\right)~.
\end{align}
It is necessary at least to include two counterterms in order to achieve renormalization of the PWAs with cutoff regularization, in agreement with our discussion in Eq.~\eqref{201229.1} about the fact that the first $\ell+1$ parameters in the ERE are non-perturbative.  \trr{The need of at least two counterterms for renormalizing  $P$-wave scattering   is also deduced within EFT in Ref.~\cite{Bertulani:2002sz} dedicated to the study of $n\alpha$ scattering at low energies. There,  a dimeron field was introduced \cite{Kaplan:1996nv}  to solve the partial-wave amplitude in vacuum.}

\trr{We follow the method explain in Sec.~\ref{sec.201228.1} and} by considering only  $d_0$ \trr{in Eq.~\eqref{210104.1}} fixed to reproduce the scattering length $a_1$ as a function of $\Lambda$ the resulting PWA vanishes as $\Lambda\to\infty$. 
However, this is not the case in DR where only one counterterm is enough to end with finite non-trivial results 
as worked out in  Ref.~\cite{Kaiser:2012sr}, and in Appendix~\ref{app.210104.1} within our present formalism where we reproduce the results of \cite{Kaiser:2012sr}. We do not dwell
any more on them in the main text since we consider that the suitable process for non-perturbative QFT is to perform 
a cutoff regularization, as it has been discussed above.

The partial-wave projection of the potential in Eq.~\eqref{210104.1} for all the PWAs ${^3P}_2$, ${^3P}_1$ and ${^3P}_0$ is\footnote{This can be obtained by inverting Eq.~\eqref{200501.3}, or from Eq.~(2.31) in Ref.~\cite{Oller:2019opk} implemented without including the isospin indices and with the matrix element $\langle \vk,\sigma'_1\sigma'_2|V|p\hvz,\sigma_1\sigma_2\rangle=\delta_{\sigma'_1\sigma_1}\delta_{\sigma'_2\sigma_2}(d_0+\frac{1}{2}d_2(k^2+p^2)) kp\cos\theta'$.} 
\begin{align}
\label{210104.2}
v(k,p)&=\frac{kp}{3}(d_0+\frac{1}{2}d_2(k^2+p^2))~.
\end{align}
The factor $1/3$ can also be seen to stem from the fact that in a Lippmann-Schwinger equation only the longitudinal component of the unitarity loop function, proportional to $p_ip_j$, is iterated. Thus, of the three possible states of orbital polarization only one is picked up.

\begin{figure}
\begin{center}
\includegraphics[width=.6\textwidth]{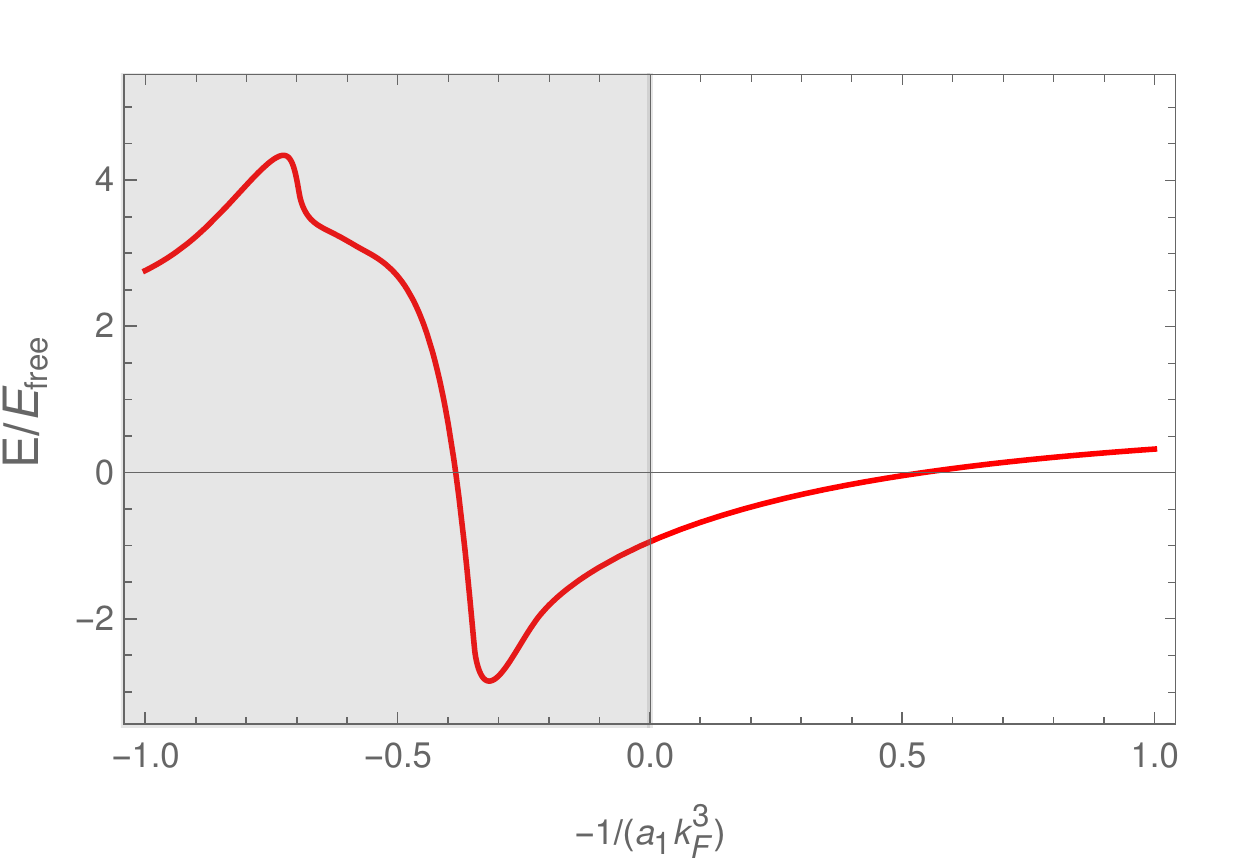}
\end{center}
\caption{{\small Dependence of ${\cE}/{\cE}_{\rm free}$ with $-1/(a_1k_F^3)$ for $r_1 =0$. \trr{The  region for negative values of $-1/a_1 k_F^3$ is shaded because it gives rise to an unacceptable pole content in vacuum scattering. See the text for details.}}
 \label{Fig:P-waves_a1}}
\end{figure} 

With the matrix notation of Eq.~\eqref{180319.2} the partial-wave projected potential reads
\begin{align}
\label{210104.3}
v(k,p)&=[k]^T\cdot[v]\cdot [p]~,\\
[v]&=\left(\begin{array}{ll} d_0 & d_2/2\\
 d_2/2 & 0 \\
\end{array}
\right)~,\nn\\
[k]^T&=(k,k^3)~,\nn
\end{align}
and similarly for $[p]$. Now we apply the method  of Sec.~\ref{sec.201228.1} \tcr{to calculate ${\mathsf t}_m$}.
To study the renormalization of the resulting PWAs we then set the Fermi momentum to zero and 
the matrix $[{\cal G}]$ in Eq.~\eqref{180319.6b} only retains its free part, $[{\cal G}_f]$, which is diagonal in the channel indices $\alpha$ and $\beta$.
This matrix  is given by
\begin{align}
\label{210104.4}
[{\cal G}_f]&=-m\int\frac{d^3k}{(2\pi)^3}\frac{[k][k]^T}{k^2-p^2-i\ep}~,
\end{align}
and it is evaluated with cutoff regularization. Taking the matrices $[{\cal G}]$ and $[v]$ into Eq.~\eqref{180319.8} we then have the following expression for all the 
PWAs in vacuum stemming from the potential in Eq.~\eqref{210104.3},
\begin{align}
\label{210104.5}
{\mathsf t_V}(k,p)&=[k]^T\left([v]^{-1}+[{\cal G}_f]\right)^{-1}[p]=\frac{4\pi}{m}\frac{kp}{-\frac{1}{a_1}+\frac{1}{2}rp^2-ip^3}+{\cal O}(\Lambda^{-1})~.
\end{align}
\trr{For on-shell scattering our resulting $P$-wave partial-wave amplitude is the same as the one obtained with cutoff regularization including a dimeron field in \cite{Bertulani:2002sz}.}

We are then ready to apply Eq.~\eqref{201229.6} for calculating ${\mathsf t_m}(k,p)$, and ${\cE}$ in terms of it. There is an important novelty compared to the vacuum case concerning the fact that the ${^3P}_2$, ${^3P}_1$ and ${^3P}_0$ PWAs mix in the
medium despite having different $J$, and this mixing depends on the value of  $\mu$. 
Because of the property in Eq.~\eqref{201228.5}  we have to calculate explicitly the $P$-waves only for $\mu=0,$ $1$ and $2$.
For the latter value only the ${^3P}_2$ contributes and it is uncoupled.
When $\mu=1$ we have the coupling between the ${^3P}_2$ and ${^3P}_1$ PWAs.
For $\mu=0$ because of the properly in  Eq.~\eqref{201228.6} the ${^3P}_1$ decouples from the ${^3P}_2$ and ${^3P}_0$, and the latter ones are coupled.
Therefore, at most we have the scattering of two coupled channels.

Once this process is accomplished we then proceed with the calculation of ${\cE_L}$ by applying Eq.~\eqref{200501.6}. One has to sum over the allowed values of $\mu$ (in all cases here $\ell=S=1$) and then for each $\mu$ we proceed with the diagonalization of the matrix $I-{\mathsf t_m}(p,p)L_d(p,a\hvz)$, calculate the $\log$ of the eigenvalues, sum over them and perform the double integration.
\tcr{These technicalities are treated and exemplified in full detail in the Appendix~\ref{app.210104.1} to which we refer for further discussions on this respect. There, Kaiser's results \cite{Kaiser:2012sr}, obtained using dimensional regularization and tensorial methods for a $P$-wave interaction keeping only the scattering length, are reproduced within the general formalism derived in our work. }

\begin{figure}[H]
\begin{center}
\includegraphics[width=.6\textwidth]{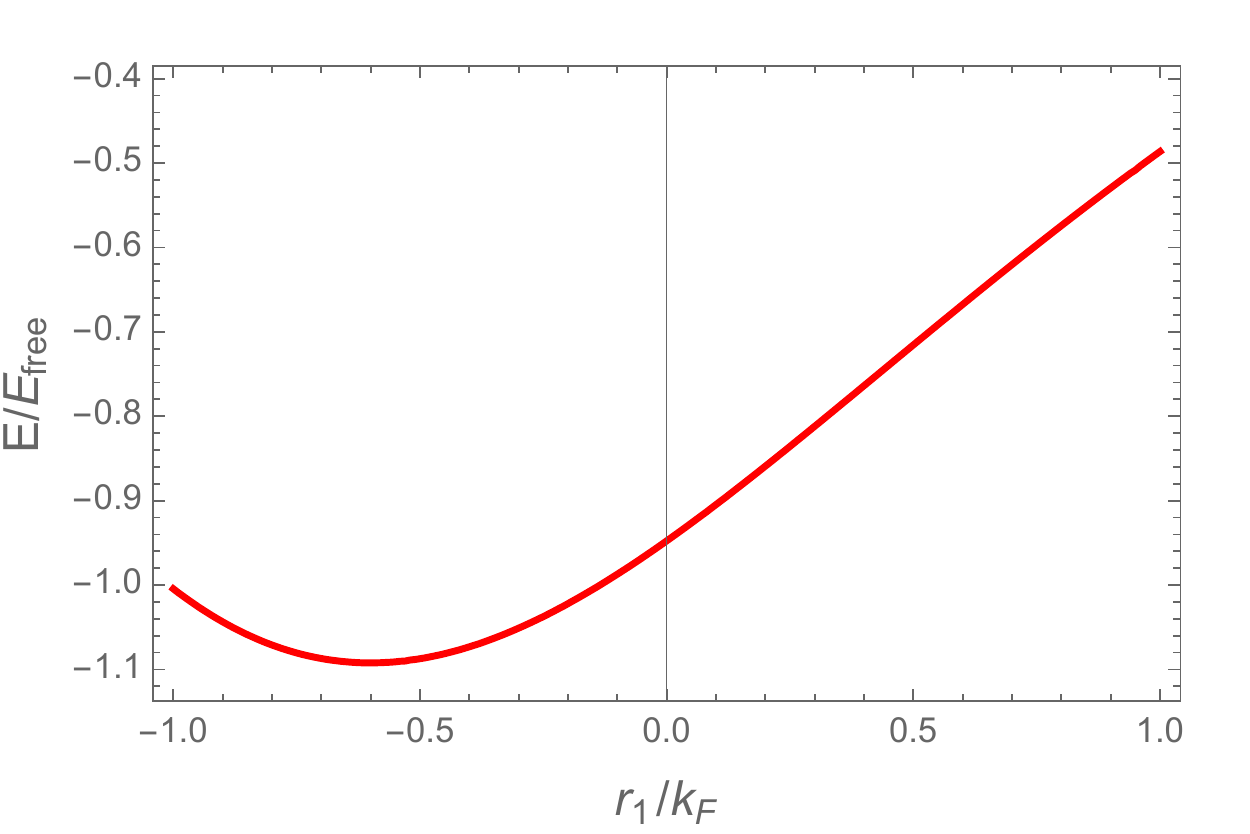}
\end{center}
\caption{{\small Dependence of ${\cE}/{\cE}_{\rm free}$ with $r_1/k_F$ for $a_1 k_F^3 =\infty$. } 
 \label{Fig:P-waves_r1}}
\end{figure}

\trr{In discussing our results for  values of ERE parameters we discard those regions in the parametric space that drives to resonant  poles in ${\mathsf t}_V$ lying in the upper part of the complex-$p$ plane, similarly as already explained for $S$ waves.} 
We first consider our results with $r_1=0$  and plot ${\cal E}/{\cE_{\rm free}}$ as a function of $-1/a_1k_F^3$ in Fig.~\ref{Fig:P-waves_a1}. \trr{The poles are located at the three cubit roots of $(i/a_1)^{1/3}$, with two of them being resonant poles and lying in the upper half complex-$p$ plane for $a_1>0$. Therefore, we should exclude the results for negative $-1/a_1k_F^3$ in Fig.~\ref{Fig:P-waves_a1}, and this is why this region is shaded. We do not completely remove it for comparison with Ref.~\cite{Kaiser:2012sr}, where an analogous figure is plotted including the region with $a_1>0$ too.} 
We consider \trr{next} the limit $a_1\to \infty$ and show the dependence of ${\cal E}/{\cE_{\rm free}}$ on the dimensionless parameter $r_1/k_F$ in Fig.~\ref{Fig:P-waves_r1}. It has a convex form around a minimum for $r_1/k_F\approx -0.6$. 
\trr{Regarding  Fig.~14 calculated for $|a_1|\to  \infty$ all the values of $r_1/k_F$ are kept because the results do not really depend on the sign of $a_1\to\infty$ and by adjusting it appropriately one can avoid the resonant pole positions with positive imaginary part.}\trr{\footnote{\trr{This is clear from the Appendix~\ref{app.211027.1}, cf. Eqs.~\eqref{211027.2} and \eqref{211028.1}, because if $|a_1|\to \infty$ it is enough to  choose that the sign of $a_1$ is opposite to that of $r_1$,  such that the product $r_1\alpha_1\to-\infty$. In such situation no resonant poles are generated.}}}

\begin{figure}
\begin{center}
\includegraphics[width=.5\textwidth]{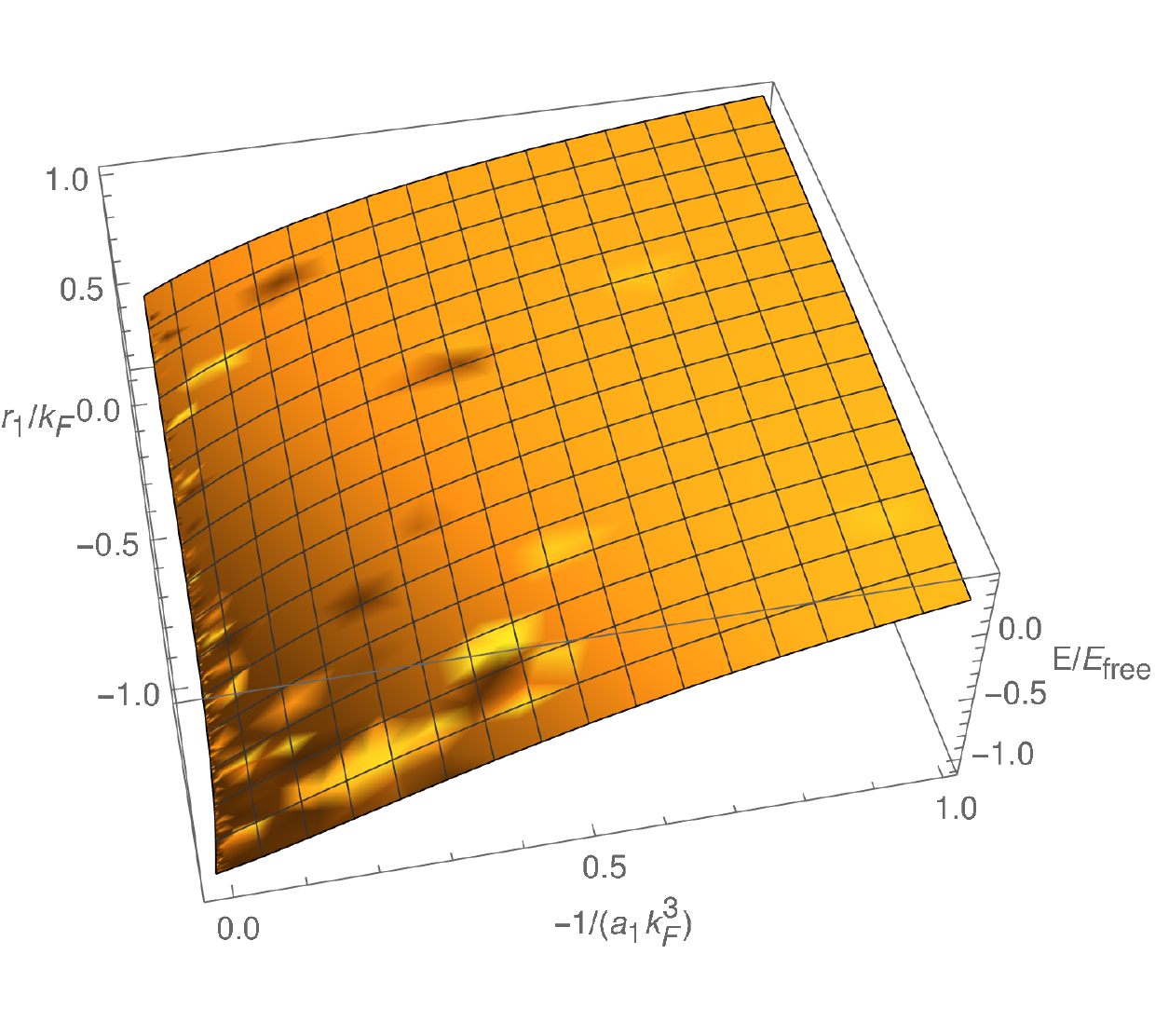}
\end{center}
\caption{{\small Three dimensional plot of ${\cE}/{\cE}_{\rm free}$ as a function of $-1/(a_1k_F^3)$ and $r_1/k_F$. 
 }
 \label{Fig:P-waves_3D}}
\end{figure} 

\begin{figure}
\begin{center}
\includegraphics[width=.5\textwidth]{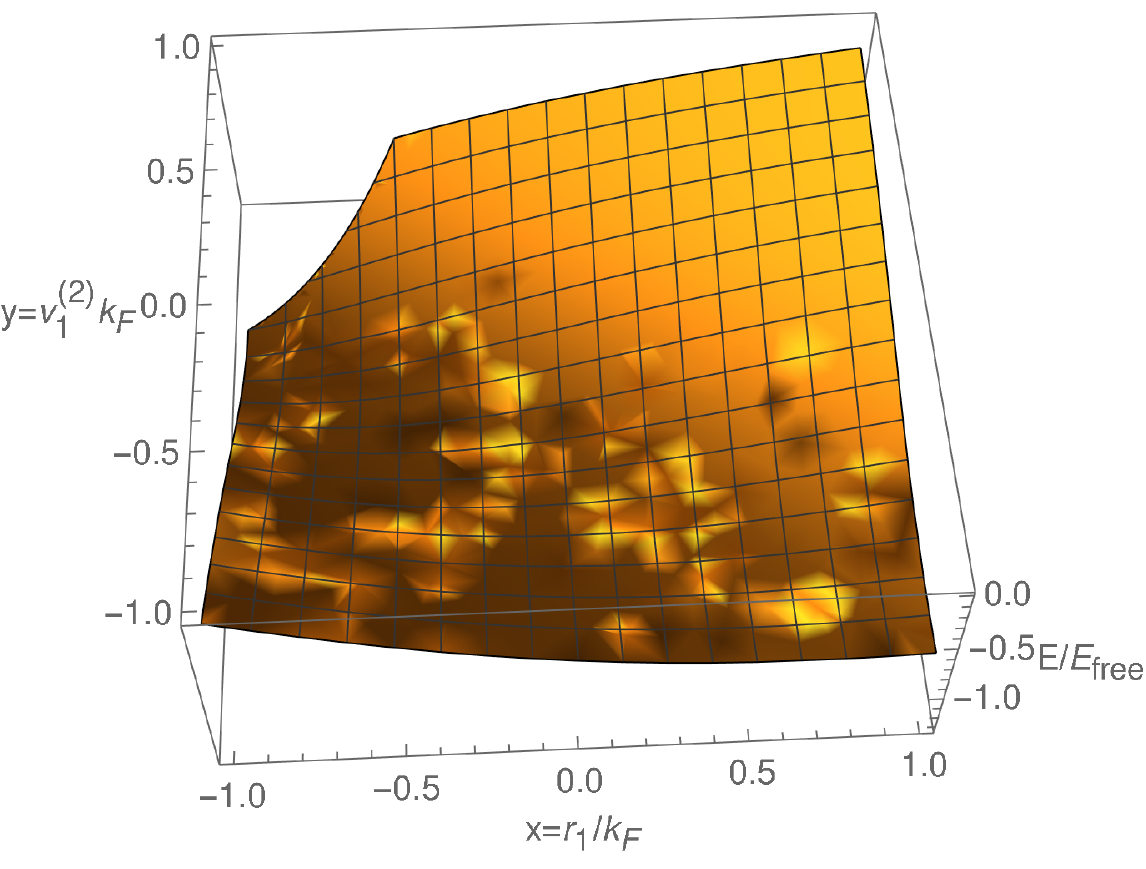}
\end{center}
\caption{{\small ${\cE}/{\cE}_{\rm free}$ as a function of $r_1/k_F$ and $v_1^{(2)}k_F$. }
 \label{Fig:P-waves_3D_r1_v1}}
\end{figure}

Next, we draw a three-dimensional plot  
of ${\cE}/{\cE}_{\rm free}$ as a function of the dimensionless parameters $-1/a_1 k_F^3$ and $r_1/k_F$ in Fig.~\ref{Fig:P-waves_3D}. \trr{In the parametric space of $(a_1,r_1)$ we find a region that is excluded because $P$-wave resonant poles with positive imaginary part occur in $\tau(p)$. As deduced in the Appendix~\ref{app.211027.1}, this region corresponds to $-1/a_1k_F^3<0$ and $r_1>-(54/|a_1|k_F^3)^{1/3}$.}
 We also show another three-dimensional plot in the limit $a_1\to\infty$ again by plotting ${\cE}/{\cE_{\rm free}}$ as a function of $r_1/k_F$ and $v_1^{(2)}k_F$ in Fig.~\ref{Fig:P-waves_3D_r1_v1}. The long-distance limit 
 $k_F R\ll 1$ (with $R$ the range of the interactions) corresponds to $|v_1^{(2)}|k_F\ll 1$ and $|r_1|/k_F\gg 1$  in which the shape of the surface is rather structureless. \trr{In Fig.~\ref{Fig:P-waves_3D_r1_v1} the excluded region is the part of the top left quadrant such that $v_1^{(2)}>0$ and $r_1<-1/2v_1^{(2)}$, and we refer again to the Appendix~\ref{app.211027.1} for its derivation.} \tcr{Let us finish this section by clarifying that we have not introduced an expansion for $\bar{\cE}$ like that in Eq.~\eqref{201229.2} for the $S$ wave interactions because  its direct application to $P$ waves is not possible as the relevant dimensionless parameters are $1/a_1r_1^3$ and $k_F/r_1$, and both have to be taken into account in a power expansion around $a_1\to \infty$ and $R\to 0$.   
}

\section{Conclusions}

We have presented the derivation of the resummation of the ladder diagrams for the evaluation of the energy density $\cE$ of a spin 1/2 fermion many-body system in terms of arbitrary vacuum two-body interactions. All in-medium two-body intermediate states are accounted for. In standard many-body notation this means that we have resummed this iteration by taking into account both particle-particle and hole-hole intermediate states. In our derivations it has been essential to use the formulation of many-body quantum field theory from Ref.~\cite{Oller:2001sn} \tcr{because of the resulting framework in which the many-body theory is organized. This fact provides us with a rearrangement of the diagrams involved 
in the calculation of $\bar{\cE}$ that allows the solution of the non-trivial combinatoric problem  associated.}  
The resulting expression provides a real value for  $\cE$ because of in-medium unitarity, as we prove. At the practical level, since it is usually the case that the two-body finite-range  interactions are studied in partial waves,  it is interesting to express $\cE$ in a partial-wave amplitude expansion.
This requires a special treatment in  the many-body environment because of extra mixing among the partial-wave amplitudes 
due to the breaking of rotational invariance within the relative degrees of freedom. The reason is because the scattering amplitudes also depend on the total momentum.
The necessary formalism for the partial-wave expansions has been worked out in detail.

The case of contact interactions has been fully solved, providing explicit renormalized results within a cutoff regularization in a wide class of schemes. First concrete examples have been considered involving $S$- and $P$-wave interactions and including up to the first three terms in the effective-range expansion, namely, $a_\ell$, $r_\ell$ and $v_\ell^{(2)}$ with $\ell=0$ or 1. Special attention has been payed to the study of the parametric region around the unitary limit (in normal matter). It is shown that the effective range in $S$ waves, $r_0$,  plays there a perturbative role as expected. We also study the $S$-wave contributions to ${\cE}$ for neutron matter and the Tan density contact parameter ${\mathsf C}$. The case of a spin-independent $P$-wave interaction is also analyzed and \tcr{we show that the effective range $r_1$ is needed together the scattering length $a_1$ to obtained renormalized results.} We have given several plots for ${\cE}/{\cE_{\rm free}}$ where we show its dependence with $-1/a_1 k_F^3$ for $r_1=0$, and then move to the limit $a_1\to \infty$ and give it as a function of the dimensionless parameter $r_1/k_F$. We also plot ${\cE}/{\cE_{\rm free}}$ as a function of $-1/a_1 k_F^3$ and $r_1/k_F$.
The next shape parameter $v_1^{(2)}$ is also considered and  we plot  ${\cE}/{\cE_{\rm free}}$ within the plane 
$r_1/k_F$ and $v_1^{(2)}k_F$ in the limit $a_1\to\infty$. 
In all cases interesting non-trivial shapes for ${\cE}/{\cE_{\rm free}}$ are found. 

\tcr{We have largely generalized  the previous Kaiser's pivotal analyses in Refs.~\cite{Kaiser:2011cg,Kaiser:2012sr}, regarding the resummation of the ladder series. Apart from what was mentioned previously, it is worthwhile to list here for concreteness the attainment of several other important achievements of this work, as explained along the paper:  
 \begin{enumerate}
 \item[i)] We have derived a general formula for the resummation of the ladder series expressed in terms of  arbitrary fermion-fermion interactions in vacuum.
 \item[ii)] We have developed a partial-wave expansion for calculating  the  general formula referred in i).
 \item[iii)] We can consider any number of partial waves contributing, either separately or simultaneously.
 \item[iv)] For every partial wave we can include any number of terms in the effective-range expansion.
  \item[v)] We can apply our method with  cutoff regularization in a generic scheme obtaining always scheme independent renormalized results. 
  \item[vi)] We can also work out the case with dimensional regularization (the one used  by Kaiser exclusively in Refs.~\cite{Kaiser:2011cg,Kaiser:2012sr}). 
 However, this method is generally not correct for non-perturbative calculations  as discussed in the manuscript.
\item[vii)] We have shown that the ansatz used by Kaiser \cite{Kaiser:2012sr} to resum the ladder series when including  up to the effective range in $S$ waves is correct.
\end{enumerate} }

As a concluding remark, we consider that further applications of this powerful approach should  be pursued in many-body systems,
like  nuclear matter or ultracold trapped atoms and ions.

\section*{Acknowledgements}

We would like to thank   Felipe~J. Llanes-Estrada for introducing the term cartwheel to refer to the Fock diagrams and  feedback on the manuscript. We also thank  Manuel Valiente  for interesting discussions. This work has been partially funded by MICIIN AEI (Spain) under Contracts No. PID2019-106080GB-C22/AEI/10.13039/501100011033,  PID2019-106080GB-C21/AEI/10.13039/501100011033,  and by EU Horizon 2020 research and innovation program, STRONG-2020 project, under grant agreement No 824093.

\newpage

\appendix

\section{Technical material on rotational symmetry}
\label{app.201228.1}
\setcounter{equation}{0}
\def\theequation{\Alph{section}.\arabic{equation}}

Let us first study the transformation of $L_m(p,\va)$ and $L_d(p,\va)$ under a rotation $R$ 
on the vector $\va$. One has that
\begin{align}
\label{190702.2}
L_m(p,R \va)&=R L_m(p,\va) R^\dagger~,\\
\label{190702.2b}
L_d(p,R \va)&=R L_m(p,\va) R^\dagger~.  
\end{align}

We first notice the trivial fact that $\mathbb{I}_S$ is invariant under any rotation since
\begin{align}
\label{190702.3}
R\IS R^\dagger&=\IS~,
\end{align}
so that we do not show $\IS$ explicitly in the following. 
We proceed with the derivation in detail for $L_m(p,R\va)$, and for $L_d(p,R\va)$ we quote the final
result because the procedure is completely analogous.
\begin{align}
\label{190702.4}
L_m(p,R\va)&=-m\int\frac{d^3k}{(2\pi)^3}\left[\theta({k_{F_1}}-|R\va+\vk|)+\theta({k_{F_2}}-|R\va-\vk|)\right]
\frac{|\vk\rangle\langle \vk|}{k^2-p^2-i\ep}\\  
&=-m\int\frac{d^3k}{(2\pi)^3}\left[\theta({k_{F_1}}-|\va+R^{-1}\vk|)+\theta({k_{F_2}}-|\va-R^{-1}\vk|)\right]
\frac{|\vk\rangle\langle \vk|}{k^2-p^2-i\ep}~.\nn
\end{align}
We perform next the change of integration variable $R^{-1}\vk\to \vk$ and take into account that $|R\vk\rangle=R|\vk\rangle$.
Thus, Eq.~\eqref{190702.2} results. Analogous steps also lead to Eq.~\eqref{190702.2b}.

We use now Eq.~\eqref{190702.2} to study the transformation properties of $t_m(\va)$ under a rotation of $\va$.
We also take into account the rotational invariance of the  vacuum $T$-matrix $t_V$, 
\begin{align}
\label{190702.7}
R t_V R^\dagger&=t_V~.
\end{align}
Multiplying both sides of Eq.~\eqref{190624.2} by $R$ and $R^\dagger$ to the left and 
right, respectively, we have that 
\begin{align}
\label{190702.8}
Rt_m(\va)R^\dagger&=t_V+R t_V L_m(p,\va)t_m(\va)R^\dagger\\
&=t_V+t_V  L_m(p,R\va)R t_m(\va)R^\dagger~,\nn
\end{align}
so that $Rt_m(\va)R^\dagger$ and $t_m(R\va)$ satisfy the same IE and then they must coincide, 
\begin{align}
\label{190702.11}
t_m(R \va)&=R t_m(\va) R^\dagger~.
\end{align}

\section{Some symmetry properties of the  PWAs in the many-body environment}
\label{app.200925.1}
\setcounter{equation}{0}
\def\theequation{\Alph{section}.\arabic{equation}}

We proceed with the demonstration of the relations in Eqs.~\eqref{201228.5}, \eqref{201228.6} and \eqref{201228.7}.
To demonstrate the equality  
\begin{align}
\label{190809.8}
\langle J_2{-\mu_1}\ell_2 S_1p'|t_m(a\hvz)|J_1{-\mu_1}\ell_1 S_1p\rangle
&=(-1)^{J_2+J_1}\langle J_2\mu_1\ell_2 S_1  p'|t_m(a\hvz)|J_1\mu_1\ell_1 S_1 p\rangle~,
\end{align}
let us write down the IE of Eq.~\eqref{190809.5} for the PWAs with $-\mu_1$,
\begin{align}
\label{190809.9}
&\langle J_2-\mu_1\ell_2 S_1p'|t_m(a\hvz)|J_1-\mu_1\ell_1 S_1p\rangle=
\langle J_2-\mu_1\ell_2S_1p'|V|J_1-\mu_1\ell_1 S_1 p\rangle\\
&+\sum_{J_4\ell_3 \ell_4 m_3\sigma_3}\chi(S_1\ell_3)\chi(S_1\ell_4) \frac{m}{(2\pi)^2}\int\frac{k^2dk}{k^2-p^2-i\ep}
\langle J_2-\mu_1 \ell_2 S_1 p'|V|J_2-\mu_1\ell_3 S_1k\rangle\nn\\
&\times
\langle J_4-\mu_1\ell_4 S_1 k|t_m(a\hvz)|J_1-\mu_1\ell_1 S_1p\rangle
(m_3\sigma_3-\mu_1|\ell_3S_1J_2)(m_3\sigma_3-\mu_1|\ell_4 S_1 J_4)\nn\\
&\times 
\int d\hvk Y_{\ell_3}^{m_3}(\hvk)^*Y_{\ell_4}^{m_3}(\vk)
\left[1-\theta({k_F}-|\vk+a\hvz|)-\theta({k_F}-|\vk-a\hvz|)\right]~.\nn
\end{align}
The next step is to take into account that the matrix elements of $V$ are independent of $\mu_1$ and 
use the symmetry properties of the Clebsch-Gordan coefficients 
\begin{align}
\label{190809.10}
(-m_1-m_2-m_3|j_1j_2j_3)&=(-1)^{j_1+j_2-j_3}  (m_1m_2m_3|j_1j_2j_3)~,
\end{align}
so that
\begin{align}
\label{190809.11}
(m_3\sigma_3-\mu_1|\ell_3S_1J_2)(m_3\sigma_3-\mu_1|\ell_4S_1J_4)&=(-1)^{J_2+J_4}
(-m_3-\sigma_3 \mu_1|\ell_3S_1J_2)(-m_3-\sigma_3 \mu_1|\ell_4S_1J_4)~,
\end{align}
where we have used that $(-1)^{\ell_3+\ell_4}=+1$ because of Eq.~\eqref{200926.5}. 
The sum over the dummy indices $m_3$ and $\sigma_3$ is symmetric around 0, so that by exchanging their signs simultaneously the only  change is in the
the angular integration. But this does not introduce any change in the final result
because of the well-known property of the spherical harmonics
\begin{align}
\label{190809.12c}
Y_\ell^m(\vk)^*&=(-1)^mY_\ell^{-m}(\vk)~.
\end{align}
Therefore, 
\begin{align}
&\int d\hvk Y_{\ell_3}^{-m_3}(\hvk)^*Y_{\ell_4}^{-m_3}(\vk)
\left[1-\theta({k_F}-|\vk+a\hvz|)-\theta({k_F}-|\vk-a\hvz|)\right]
=\int d\hvk Y_{\ell_3}^{m_3}(\hvk)Y_{\ell_4}^{m_3}(\vk)^*\\
&\times
\left[1-\theta({k_F}-|\vk+a\hvz|)-\theta({k_F}-|\vk-a\hvz|)\right]~. 
\nn
\end{align}
However, the integrand in the previous equation is real, and by taking its complex conjugate, we recover again
the original expression in Eq.~\eqref{190809.9}.
We can then rewrite it as 
\begin{align}
\label{190809.12}
&\langle J_2-\mu_1\ell_2 S_1p'|t_m(a\hvz)|J_1-\mu_1\ell_1 S_1p\rangle=
\langle J_2\mu_1\ell_2S_1p'|V|J_1\mu_1\ell_1 S_1 p\rangle\\
&+\sum_{m_3\sigma_3 J_4\ell_3 \ell_4} \chi(S_1\ell_3)\chi(S_1\ell_4)\frac{m}{(2\pi)^2}\int\frac{k^2dk}{k^2-p^2-i\ep}
\langle J_2\mu_1 \ell_2 S_1 p'|V|J_2\mu_1\ell_3 S_1k\rangle\nn\\
&\times \langle J_4-\mu_1\ell_4 S_1 k|t_m(a\hvz)|J_1-\mu_1\ell_1 S_1p\rangle
(-1)^{J_2+J_4}(m_3\sigma_3\mu_1|\ell_3S_1J_2)(m_3\sigma_3\mu_1|\ell_4 S_1 J_4)\nn\\
&\times \int d\hvk Y_{\ell_3}^{m_3}(\hvk)^*Y_{\ell_4}^{m_3}(\vk)
\left[1-\theta({k_F}-|\vk+a\hvz|)-\theta({k_F}-|\vk-a\hvz|)\right]~,\nn
\end{align}
where we have taken into account again that the matrix elements of $V$ between PWAs are $\mu$ independent
because it is an scalar operator.
Let us multiply both sides of Eq.~\eqref{190809.12} by $(-1)^{J_2+J_1}$ and then we conclude that
$(-1)^{J_2+J_1}\langle J_2-\mu_1\ell_2 S_1p'|t_m(a\hvz)|J_1-\mu_1\ell_1 S_1p\rangle$ satisfies the same
IE as $\langle J_2\mu_1\ell_2 S_1p'|t_m(a\hvz)|J_1\mu_1\ell_1 S_1p\rangle$, which implies Eq.~\eqref{190809.8}.
Notice that when multiplying $\langle J_2\mu_1\ell_2S_1p'|V|J_1\mu_1\ell_1 S_1 p\rangle$   
by $(-1)^{J_1+J_2}$ we get the same result 
because the matrix element is proportional to $\delta_{J_1J_2}$.
Furthermore, for the matrix element of $t_m(a\hvz)$ inside the integrand
since the $J_i$ are integers it results that  $(-1)^{J_1+J_2}(-1)^{J_2+J_4}=(-1)^{J_1+J_4}$.

For the demonstration of the symmetric relation in Eq.~\eqref{201228.7}, 
\begin{align}
\label{190809.13}
\langle J_1\mu_1\ell_1 S_1p|t_m(a\hvz)|J_2\mu_1\ell_2 S_1p'\rangle
&=\langle J_2\mu_1\ell_2 S_1p'|t_m(a\hvz)|J_1\mu_1\ell_1 S_1p\rangle~, 
\end{align}
we start with the IE for $\langle J_1\mu_1\ell_1 S_1 p|t_m(a\hvz)|J_2\mu_1\ell_2 S_1p'\rangle$,
cf. Eq.~\eqref{190809.5},
\begin{align}
\label{190809.14}
&\langle J_1\mu_1 \ell_1S_1p|t_m(a\hvz)|J_2\mu_1\ell_2S_1p'\rangle=
\langle J_1\mu_1 \ell_1S_1 p|V|J_2\mu_1\ell_2S_1 p'\rangle\\
&+\sum_{ J_4\ell_3 \ell_4 m_3\sigma_3}\chi(S_1\ell_3)\chi(S_1\ell_4)\frac{m}{(2\pi)^2}
\int_0^\infty\frac{k^2 dk}{k^2-p^2-i\ep}\langle J_1\mu_1\ell_1S_1p|V|J_1\mu_1\ell_3S_1k\rangle \nn\\
&\times 
\langle J_4\mu_1\ell_4S_1k|t_m(a\hvz)|J_2\mu_1\ell_2S_1p'\rangle
(m_3\sigma_3\mu_1|\ell_3S_1J_1)(m_3\sigma_3\mu_1|\ell_4 S_1J_4)\nn\\
&\times
\int d\hvk Y_{\ell_3}^{m_3}(\hvk)^*Y_{\ell_4}^{m_3}(\hvk)
[1-\theta({k_F}-|\vk+a\hvz|)-\theta({k_F}-|\vk-a\hvz|)]\nn~.
\end{align}
Because of time-reversal invariance the matrix elements of $V$ in PWAs are symmetric. 
In addition, since the product of the two spherical harmonics is real we take its complex
conjugate. 
Then, we can rewrite Eq.~\eqref{190809.14} as
\begin{align}
\label{190810.1}
&\langle J_1\mu_1 \ell_1S_1p|t_m(a\hvz)|J_2\mu_1\ell_2S_1p'\rangle=
\langle J_2\mu_1\ell_2S_1p'|V|J_1\mu_1 \ell_1S_1p\rangle \\
&+\sum_{ J_4\ell_3 \ell_4 m_3\sigma_3}\chi(S\ell_3)\chi(S\ell_4)\frac{m}{(2\pi)^2}\int_0^\infty\frac{k^2 dk}{k^2-p^2-i\ep}
\langle J_4\mu_1\ell_4S_1k|t_m(a\hvz)|J_2\mu_1\ell_2S_1p'\rangle\nn\\
&\times \langle J_1\mu_1\ell_3S_1k|V|J_1\mu_1\ell_1S_1p\rangle (m_3\sigma_3\mu_1|\ell_3S_1J_1)(m_3\sigma_3\mu_1|\ell_4 S_1J_4)\nn\\
&\times
\int d\hvk Y_{\ell_4}^{m_3}(\hvk)^*Y_{\ell_3}^{m_3}(\hvk) 
[1-\theta({k_F}-|\vk+a\hvz|)-\theta({k_F}-|\vk-a\hvz|)]\nn~,
\end{align}
which is the same IE as the one satisfied by
$\langle J_2\mu_1\ell_2S_1p'|t_m(a\hvz)|J_1\mu_1 \ell_1S_1p\rangle$ as we wanted to show.
In order to arrive to this conclusion we have used
the fact that the IE for $t_m(\va)$ of
Eq.~\eqref{190808.1} can also be written as
\begin{align}
\label{190810.2}
t_m(\va)&=V-t_m(\va)[G-L_m(p,\va)]V~.
\end{align}

\section{Calculations employing dimensional regularization}
\label{app.210102.1}
\setcounter{equation}{0}
\def\theequation{\Alph{section}.\arabic{equation}}

Here we discuss the calculation with DR of the $S$-wave and  $P$-wave potentials up to ${\cal O}(p^4)$ and ${\cal O}(p^2)$, respectively.

\subsection{$S$ waves}
\label{app.210102.2}

We take the potential in Eq.~\eqref{210101.4},
\begin{align}
\label{210102.4}
v(k,p)&=c_0+\frac{1}{2}c_2(k^2+p^2)+\frac{1}{2}c_4(k^4+p^4)~,
\end{align}
and proceed with the calculation of ${\mathsf t_m}(k,p)$ by applying the method of Sec.~\ref{sec.201228.1}. The coefficients in the expansion of the potential in powers of $k^2$ and $p^2$ are denoted in this case $v_{ij}$, which is enough since it is uncoupled. Therefore, the only non-zero coefficients are
\begin{align}
\label{210102.5}
v_{11}=c_0~,~v_{12}=v_{21}=\frac{1}{2}c_2~,~v_{13}=v_{31}=\frac{1}{2}c_4~.
\end{align}
Next one needs to implement the matrix $[\cal G]$, Eq.~\eqref{180319.6b}, given in this case by
\begin{align}
\label{210102.6}
[{\cal G}]&=-
\frac{m}{(2\pi)^3}\int_0^\infty \frac{d^3k}{k^2-p^2-i\ep}
\left(
\begin{array}{lll}
1 & k^2 & k^4 \\
k^2 & k^4 & k^6 \\
k^4 & k^6 & k^8
\end{array}
\right)\left(1-2\theta(k_F-|\vk-a\hvz|)\right)~.
\end{align}
In DR the calculation of the free part of $[{\cal G}]$, called $[{\cal G}_f]$,  is straightforward because $\int_0^\infty d^Dk k^n=0$ for $n\geq 0$. The result is
\begin{align}
\label{210102.6b}
[{\cal G}_f]&=-i\frac{m p}{4\pi}
\left(
\begin{array}{lll}
1 & p^2 & p^4 \\
p^2 & p^4 & p^6 \\
p^4 & p^6 & p^8
\end{array}
\right)~.
\end{align}
Applying Eq.~\eqref{180319.8} with $[{\cal G}_f]$ one obtains for ${\mathsf t_V}(k,p)$ the following expression
\begin{align}
{\mathsf t_V}(k,p)&=[k]^T\left([v]^{-1}+[{\cal G}_f]^{-1}\right)^{-1}[p]\\
  &=v(k,p)\left(1-i\frac{mp}{4\pi}v(p,p)\right)^{-1}~.\nn
\end{align}
Indeed, the last formula is a general one for any uncoupled vacuum PWA calculated with DR.
Taking the explicit expression for $v(k,p)$ we have
\begin{align}
\label{210102.10}
{\mathsf t_V}(k,p)&=(c_0+\frac{1}{2}c_2(k^2+p^2)+\frac{1}{2}c_4(k^4+p^4))\left(1-i\frac{mp}{4\pi}(c_0+c_2p^2+c_4p^4)\right)^{-1}~.
\end{align}
The matching with the ERE expansion, Eq.~\eqref{201229.2a}, of the on-shell ${\mathsf t_V}(p,p)$ up to ${\cal O}(p^4)$ is straightforward and then 
\begin{align}
\label{210102.11}
\frac{m}{4\pi}c_0&=-a_0~,\\
\frac{m}{4\pi}c_2&=-\frac{a_0^2 r_0}{2}~,\nn\\
\frac{m}{4\pi}c_4&=-\frac{a_0^2}{4}(a_0r_0^2+4v_0^{(2)})~.\nn
\end{align}
With these expressions for the couplings
\begin{align}
\label{210102.12}
{\mathsf t_V}(p,p)&=-\frac{4\pi}{m}\left(\frac{1}{a_0+\frac{1}{2}a_0^2 r_0 p^2+a_0^2(v_0^{(2)}+\frac{1}{4}a_0r_0^2)p^4}+ip\right)^{-1}~.
\end{align}

The $k_F$ dependence of the matrix elements 
of $[\cal G]$, Eq.~\eqref{210102.6},  can be written in terms of the basic integrals
\begin{align}
\label{210102.8}
J_n&=\frac{m}{4\pi^3}\int d^3k\,k^{2n}\theta(k_F-|\vk-a\hvz|)=\frac{m}{4\pi^3}\int d^3k\,(\vk+a\hvz)^{2n}\theta(k_F-k)~,\\
I_n&=\frac{m}{4\pi^3}\int \frac{d^3k}{k^2-p^2-i\ep}k^{2n}\theta(k_F-|\vk-a\hvz|)
=\sum_{l=0}^{n-1}p^{2(n-1-l)}J_l +p^{2n}I_0~,\nn
\end{align}
such that  
\begin{align}
\label{210102.9}
{[\cal G]}_{jl}&=-i\frac{mp^{2(j+l)-3}}{4\pi}+I_{j+l-2}~.
\end{align}
For the basic $I_0$ integral we have the relation
\begin{align}
\label{210102.7}
      {[\cal G]}_{11}&=-i\frac{mp}{4\pi}+I_0
  =\frac{m k_F}{4\pi}(\frac{R}{\pi}+i\,I)~,
\end{align}
where we have related $I_0$ with the functions $R$ and $I$ of Ref.~\cite{Kaiser:2011cg}, also used in Sec.~\ref{sec.201229.1}.

To calculate ${\mathsf t_m}(p,p)$ we just have to express the couplings as in Eq.~\eqref{210102.11} in terms of the ERE parameters and  apply Eq.~\eqref{180319.8} with $[\cal G]$ as in Eq.~\eqref{210102.9}. Then,
\begin{align}
\label{210103.1}
&{\mathsf t_m}(p,p)=[p]^T[\hat{t}_m(p)][p]=\frac{4\pi}{m}\left(W_0^{-1}+\frac{4\pi}{m}I_0-ip\right)^{-1}~,\\
&W_0^{-1}=\frac{A}{B}~,\\
&A=-28(-30\pi+a_0^2k_F^3(10r_0+a_0k_F^2r_0^2(3+5s^2+5\kappa^2)+4k_F^2v_0^{(2)}(3+5s^2+5\kappa^2) ) )^2~,\nn\\
&B=5a_0\pi(
1260\pi(4+2a_0k_F^2 r_0 \kappa^2+a_0k_F^4 (
a_0r_0^2+4v_0^{(2)} )\kappa^4 )+
a_0^3k_F^5 (60 a_0 k_F^2r_0^3 (3+7s^2(2+s^2)-7\kappa^4 )\nn\\
&+240k_F^2 r_0 v_0^{(2)}(3+7s^2(2+s^2)-7\kappa^4)
+4r_0^2(21(3+5s^2)+14 a_0k_F^4(5+45s^2+63s^4+15s^6)v_0^{(2)}\nn\\
&+15(-7+2a_0k_F^4(3+7s^2(2+s^2))v_0^{(2)})\kappa^2
-42 a_0k_F^4(3+5s^2)v_0^{(2)}\kappa^4-210 a_0 k_F^4 v_0^{(2)} \kappa^6)
+a_0^2 k_F^4 r_0^4 (5 (7 + 9 \kappa^2)\nn\\
&+21(5 s^6 - 5 s^2 (-3 + \kappa^2) (1 + \kappa^2) - \kappa^4(3+5 \kappa^2)
+ s^4 (21 + 5 \kappa^2))) + 16 k_F^4 {v_0^{(2)2}} (5 (7 + 9 \kappa^2)\nn\\
&+21 (5 s^6 - 5 s^2 (-3 + \kappa^2) (1 + \kappa^2) - \kappa^4 (3 + 5 \kappa^2)
+ s^4 (21 + 5 \kappa^2)))))~.\nn
\end{align}
Here we have used the dimensionless variables $s=a/k_F$ and $\kappa=p/k_F$, already introduced in Sec.~\ref{sec.201230.1}. 

Applying Eq.~\eqref{200501.6}, and recalling that $4\pi L_d/m=2iI k_F$, as discussed in Sec.~\ref{sec.201230.1}, we then have that
\begin{align}
\label{210103.2}
\bar{\cE}_L&=-\frac{4i}{m\pi^3}
\int_0^{k_F} a^2 da \int_0^{\sqrt{k_F^2-a^2}} p dp
\log\left(1-\frac{4\pi/m}{W_0^{-1}+\frac{4\pi}{m}I_0-ip}L_d \right)
\\
\label{210103.2b}
&=-\frac{8k_F^5}{m\pi^3}\int_0^{1}s^2 ds\int_0^{\sqrt{1-s^2}}\kappa d\kappa
\arctan\left(\frac{\pi I}{\frac{\pi}{k_F}W_0^{-1}+R}\right)~.
\end{align}

If in Eq.~\eqref{210102.11} we take $v_0^{(2)}=-a_0^2r_0/4$ then $c_4=0$ and the potential reduces to $v(k,p)=c_0+c_2(k^2+p^2)/2$. The expression for $W_0^{-1}$ simplifies to 
\begin{align}
\label{210103.3}
W_0^{-1}&=-\frac{20(-3\pi+a_0^2 r_0k_F^3)^2}{3a_0\pi(a_0^3r_0^2k_F^5(3+5s^2-5\kappa^2)+30\pi(2+a_0r_0k_F^2\kappa^2))}~.
\end{align}
The case up to including $r_0$ was studied by Kaiser in \cite{Kaiser:2012sr} to ascertain the effects of the effective range on $\cE$ and employing DR as here. The expression for $W_0^{-1}$ times $\pi/k_F$, cf. Eq.~\eqref{210103.2b}, is the same as his function $\Omega_0^{-1}$ and we reproduce his results, confirming the correctness of his combinatorial conjecture on the proper resummation of the potential to calculate $\bar{\cE}$ and getting the $\arctan$ as in Eq.~\eqref{210103.2b}. 

We would like to stress that the proper results are those obtained with cutoff regularization of any sort ($\theta_n\neq 0)$, as derived in Sec.~\ref{sec.201230.1}, due to the non-perturbative nature of the calculations. Of course, for perturbative ones, DR is perfectly suited, see also Ref.~\cite{Phillips:1997xu} for related discussions. In particular, our results in Sec.~\ref{sec.201230.1} clearly show that the inclusion of $r_0$ is a perturbative effect, 
such that the corrections due to the effective range vanishes as $r_0\to 0$, and one recovers the result for $a_0\to\infty$ with $r_0=0$. The opposite conclusion reached in Ref.~\cite{Kaiser:2012sr}, so that the two limits $a_0\to\infty$ and $r_0\to 0$ do not commute, 
comes entirely from the use of DR in the calculation. This is why $\lim_{a_0\to\infty}\left.\xi(k_F)\right|_{r_0\neq 0}=0.876$ 
compared with the value $\lim_{a_0\to\infty}\left.\xi(k_F)\right|_{r_0=0}=0.507$, as follows from Eqs.~\eqref{210103.2} and \eqref{210103.3}.  

\subsection{$P$ waves}
\label{app.210104.1}

We consider the $P$-wave spin-independent potential given in Eq.~\eqref{210104.1} with $d_2=0$, 
\begin{align}
\label{210112.1}
V(\vk,\vp)&=\vk\cdot\vp d_0~.
\end{align}
Its partial wave projection for $^3P_0$, $^3P_1$ and $^3P_2$ can be read from Eq.~\eqref{210104.2},
\begin{align}
\label{210112.2}
v(k,p)&=\frac{d_0}{3}kp~.
\end{align}
Instead of cutoff regularization as in Sec.~\ref{sec.210104.1} we proceed with DR and reproduce  the results already obtained in Ref.~\cite{Kaiser:2012sr}. However, while this reference develops a specific method for the resummation of the ladder diagrams for the $P$-wave potential of Eq.~\eqref{210112.1}, we obtain it as a particular case of our general method, cf. Eq.~\eqref{200501.6} and Sec.~\ref{sec.201228.1}. 

For vacuum scattering in our present case the free part of the unitarity loop function is just $-imp^3/4\pi$ and $[v]=d_0/3$.
Then, we apply Eq.~\eqref{180319.8} with $[{\cal G}(p)]\to -imp^3/4\pi$ in order to calculate ${\mathsf t_V}(k,p)$ with the result
\begin{align}
\label{210112.3}
{\mathsf t_V}(k,p)&=\frac{kp}{\frac{3}{d_0}-i\frac{mp^3}{4\pi}}~.
\end{align}
The reproduction of the ERE with $a_1$ the scattering volume implies  
\begin{align}
\label{210112.4}
d_0&=-\frac{12\pi}{m}a_1~.
\end{align}
To simplify the notation we introduce the constant $\alpha_1\equiv (-a_1)^{1/3}$.

In order to work out ${\mathsf t_m}(p,p)$ in the medium by applying Eq.~\eqref{180319.8} we need the full unitarity loop function $[{\cal G}_{J_2\mu 1,J_1\mu 1}(p)]$, Eq.~\eqref{180319.6b}, expressed in terms of ${\cal A}_{J_2\mu 1,J_1\mu 1}$, Eq.~\eqref{190809.7b},
\begin{align}
\label{210112.5}
{\cal A}_{J_2\mu 1,J_1\mu 1}&=2\left(\delta_{J_2J_1}
-2\sum_{m_3\sigma_3}(m_3\sigma_3\mu|11J_2)(m_3\sigma_3\mu|11J_1)\int d\hvk |Y_{1}^{m}(\hvk)|^2\theta(k_F-|\vk-a\hvz|)
\right)~.
\end{align}
Here we have taken into account that $S=\ell=1$ and  $\chi(11)^2=2$. 

Let us also recall that because of the symmetry relations in Eq.~\eqref{201228.5} we only need to obtain explicitly the PWAs for $\mu=2$, 1 and 0, and  for the latter in addition $(-1)^{J_2}=(-1)^{J_1}$, cf.~Eq.~\eqref{201228.6}. The different coupled values of the total angular momentum as a function of $\mu$ are: $J_2=J_1=2$ for $\mu=2$; $J_2,J_1=1$ or 2 for $\mu=1$; $J_2,J_1=0$ or 2 or $J_2=J_1=1$ for $\mu=0$. Thus, either we have one or two-coupled channel scattering.

We derive in detail the coupled channel case of $\mu=1$ and for the other cases, since they can be worked out in complete analogy, just give their contributions to ${\cE_L}$. The two possible values of $m_3$ and $\sigma_3$ for $\mu=1$ are 0 or 1. With respect to $\hvz$ we have for  $m_3=0$ the longitudinal angular integration 
\begin{align}
\label{210112.6}
\int d\hvk |Y_1^0(\hvk)|^2\theta(k_F-|\hvk-a\hvz|)~,
\end{align}
and for $m_3=\pm 1$ the transversal one
\begin{align}
\label{210112.6}
\int d\hvk |Y_1^1(\hvk)|^2\theta(k_F-|\hvk-a\hvz|)~.
\end{align}
 The two basic in-medium integrals that one needs here in connection with the two previous equations are
\begin{align}
\label{210112.8}
C_\parallel&=\frac{2m}{\pi}\int_0^\infty dk\frac{k^4}{k^2-p^2}\int_{-1}^{1}d\!\cos\theta\,\cos^2\!\theta\,\,\theta\!\left(k_F-\sqrt{k^2+a^2-2ak\cos\theta}\right)~,\\
C_\perp&=\frac{2m}{\pi}\int_0^\infty dk\frac{k^4}{k^2-p^2}\int_{-1}^{1}d\!\cos\theta\,\sin^2\!\theta\,\,\theta\!\left(k_F-\sqrt{k^2+a^2-2ak\cos\theta}\right)~.\nn
\end{align}
In terms of the real functions $R_{\perp}(s,\kappa)$, $I_{\perp}(s,\kappa)$, $R_{\parallel}(s,\kappa)$ and $I_{\parallel}(s,\kappa)$  introduced by Kaiser in Ref.~\cite{Kaiser:2012sr}, cf.~Eqs.~(27)--(30) in this reference, we have
\begin{align}
\label{210112.9}
C_\parallel&=\frac{mk_F^3}{3\pi}\left\{R_\parallel(s,\kappa)+i\pi (I_\parallel(s,\kappa)+\kappa^3)\right\}~,\\
C_\perp&=\frac{2mk_F^3}{3\pi}\left\{R_\perp(s,\kappa)+i\pi (I_\perp(s,\kappa)+\kappa^3)\right\}~.\nn
\end{align}
When inserted in $[{\cal G}_{J_211,J_111}]$ this function becomes 
\begin{align}
\label{210113.1}
[{\cal G}_{211,211}]&=[{\cal G}_{111,111}]=\frac{mk_F^3}{8\pi^2}\left\{R_{\perp}+R_{\parallel}+i\pi(I_{\perp}+I_{\parallel})\right\}~,\\
[{\cal G}_{211,111}]&=[{\cal G}_{111,211}]=\frac{mk_F^3}{8\pi^2}\left\{R_{\perp}-R_{\parallel}+i\pi(I_{\perp}-I_{\parallel})\right\}~.\nn
\end{align}
We are interested in the eigenvalues $\lambda_\parallel$ and $\lambda_\perp$ of $([v]^{-1}+[{\cal G}])^{-1}$ which read
\begin{align}
\label{210113.2}
\lambda_\parallel&=\frac{4\pi k_F^{-3}/m}{(\alpha_1k_F)^{-3}+\pi^{-1}R_\parallel+iI_\parallel}~,\\
\lambda_\perp&=\frac{4\pi k_F^{-3}/m}{(\alpha_1k_F)^{-3}+\pi^{-1}R_\perp+iI_\perp}~.\nn
\end{align}
Notice that the argument of the $\log$ in the calculation of $\cE_L$, Eq.~\eqref{200501.6}, is $([v]^{-1}+[{\cal G}])^{-1}([v]^{-1}+[{\cal G}]-p^2L_d)=([v]^{-1}+[{\cal G}])^{-1}([v]^{-1}+[{\cal G}])^*$,  because of Eq.~\eqref{190629.3}, so that its eigenvalues are $e^{-2i\phi_a}$, where $\phi_a$ is the principal argument of the eigenvalues in Eq.~\eqref{210113.2} and $a=\parallel$ or $\perp$. Then, the contribution to  $\cE_L$ per particle from $\mu=\pm 1$, $\bar{\cE}_{\cL}^{\mu=\pm 1}$, is
\begin{align}
\label{210113.3}
\bar{\cE}_\cL^{\mu=\pm 1}&=-\frac{48k_F^2}{m\pi}
\int_0^1ds s^2\int_0^{\sqrt{1-s^2}}d\kappa \kappa \left(
\arctan\frac{I_\parallel}{(\alpha_1 k_F)^{-3}+\pi^{-1}R_\parallel}+
\arctan\frac{I_\perp}{(\alpha_1 k_F)^{-3}+\pi^{-1}R_\perp}
\right)~.
\end{align}
For the other values of $\mu$ we have with similar notation,
\begin{align}
\label{210113.4}
\bar{\cE}_{\cL}^{\mu=\pm 2}&=
-\frac{48k_F^2}{m\pi}\int_0^1ds s^2\int_0^{\sqrt{1-s^2}}d\kappa \kappa 
\arctan\frac{I_\perp}{(\alpha_1 k_F)^{-3}+\pi^{-1}R_\perp}~,\\
\bar{\cE}_{\cL}^{\mu=0}&=\frac{1}{2}\left(\bar{\cE}_{\cL}^{\mu=\pm 2}+\bar{\cE}_{\cL}^{\mu=\pm 1}\right)~.\nn
\end{align}
Summing over all the values of $\mu$ the total result is
\begin{align}
\label{210113.5}
\bar{\cE}_{\cL}&=-\frac{72k_F^2}{m\pi}\int_0^1ds s^2\int_0^{\sqrt{1-s^2}}d\kappa \kappa \left(
\arctan\frac{I_\parallel}{(\alpha_1 k_F)^{-3}+\pi^{-1}R_\parallel}+
2\arctan\frac{I_\perp}{(\alpha_1 k_F)^{-3}+\pi^{-1}R_\perp}
\right)~,
\end{align}
which is the one obtained in Ref.~\cite{Kaiser:2012sr}. Let us recall that our conclusions on the impact of
the $P$ waves in $\cE_\cL$ are those derived in Sec.~\ref{sec.210104.1} making use of cutoff regularization (with the cutoff sent to infinity) in an unspecified scheme.

\section{\trr{Poles in the in-medium  $S$ wave}}
\label{app.211029.1}
\setcounter{equation}{0}
\def\theequation{\Alph{section}.\arabic{equation}}

\trr{The algebraic expressions for $\zeta$ and $\nu$ (and their numerical values after performing the integrations)
  provided by Kaiser in Ref.~\cite{Kaiser:2011cg} by expanding the integrand in Eq.~\eqref{201231.4} are not right, as we have checked numerically (e.g. when used in Eq.~\eqref{201229.2} they fail to reproduce $\bar{\cE}$ around the unitary limit).  Mathematically this is due to fact that the conditions for the application of the Leibniz rule  (Theorem 10.39 of Ref.~\cite{apostol.211016}) for differentiation with respect to $(a_0 k_F)^{-1}$ under the signs of integration in Eq.~\eqref{201231.4} are not met.  Therefore, one has to use the integral representation of $\bar{\cE}$ in Eq.~\eqref{201231.4} and {\it after} the integration  evaluate the derivatives at the unitary limit.  The singularity stems from the fact that one finds a pole of $\tau_m(p)$, Eq.~\eqref{201231.2}, at the border of the integration region, where $\kappa=\sqrt{1-s^2}$ and $s\in[0,1]$,
  when $R(s,\sqrt{1-s^2})=\pi(a_0 k_F)^{-1}$ (notice also that $I(s,\sqrt{1-s^2})=0$).  The solution can be obtained by solving the transcendental equation }
\trr{\begin{align}
    \label{211029.1}
    \kappa=\tanh\left(\frac{1}{\kappa}\left[1-\frac{\pi}{2a_0 k_F}\right]\right)~,~\kappa\in[0,1]~.
\end{align}}
\trr{In the unitary limit the solution is  $\kappa=\sqrt{1-s^2}=0.833557$ and then $s=0.552434$. The fact that $s\neq 0$ implies that the fermionic pair has a non-vanishing total momentum equal to $2s k_F$.}

\trr{More generally Eq.~\eqref{211029.1} has solution as long as $(a_0k_F)^{-1}<2/\pi$. Notice that for $(a_0k_F)^{-1}>2/\pi$ the $\tanh$ in the right-hand side of Eq.~\eqref{211029.1} becomes negative, which cannot be because $\kappa\in[0,1]$. We plot in Fig.~\ref{fig.211029.1} the resulting values for $\kappa$ and $s$ as a function of $(a_0 k_F)^{-1}$. We notice an interesting continuous transition as $(a_0 k_F)^{-1}$ grows from $-\infty$ up to $2/\pi$, from a Cooper-pair like situation \cite{Heiselberg:2001} with total momentum $2sk_F= 0$ and relative momentum $\kappa k_F=k_F$,  passing through the unitary limit in which $2sk_F\approx 1.11 k_F$, $\kappa k_F\approx  0.88 k_F$, up to $(a_0 k_F)^{-1}=2/\pi$ where $2sk_F=2k_F$ and $\kappa k_F=0$.   Since this pole singularity in ${\mathsf t}_m$ happens exactly at the boundary of the two Fermi seas the total energy of the pair is always equal to the Fermi energy, $(s^2+\kappa^2)k_F^2/m=E_F$. 
It is also worth noticing that for $0^+<a_0 k_F<\pi/2$ there is
  no pole that could spoil a perturbative low-density calculation of ${\cal E}$ \cite{Hammer:2000xg}. }

\trr{\begin{figure}
\begin{center}
\includegraphics[width=.6\textwidth]{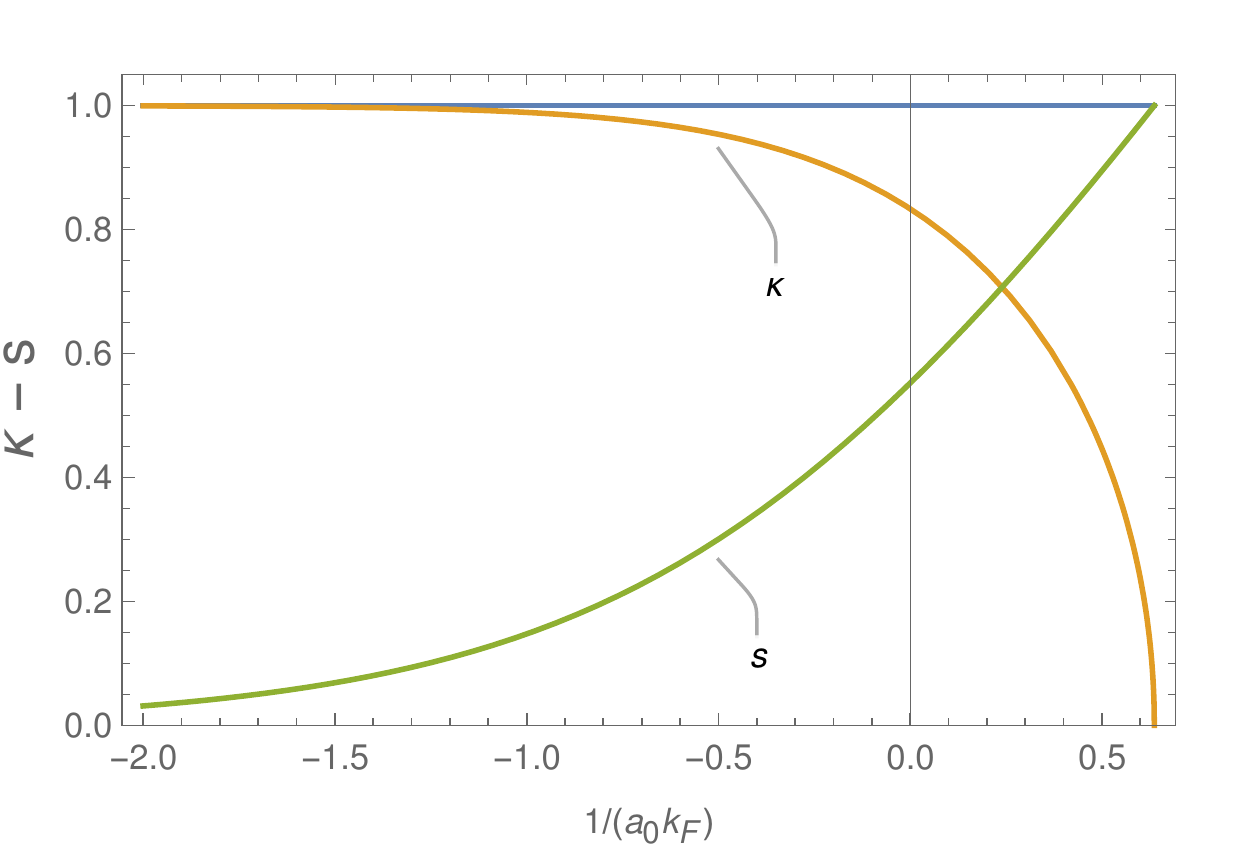}
\end{center}
\caption{{\small \trr{Pole of the in-medium $S$ wave including the scattering length, Eq.~\eqref{201231.2}, at the border of two Fermi surfaces, $\kappa=\sqrt{1-s^2}$, $s\in [0,1]$, as as function of $(a_0 k_F)^{-1}<2/\pi$. The total momentum is $2s k_F$ and the relative momentum is $\kappa k_F$.} 
 \label{fig.211029.1}}}
\end{figure} }

\section{\trr{Poles in the vacuum $P$ wave}}
\label{app.211027.1}
\setcounter{equation}{0}
\def\theequation{\Alph{section}.\arabic{equation}}

\trr{Let us consider the ERE in $P$ wave up to including the effective range $r_1$, Eq.~\eqref{210104.5}, and look for its poles
in the complex-$p$ plane by determining the zeroes of $-1/a_1+r_2 p^2/2-ip^3$. We include the auxiliary variables
\begin{align}
  \label{211027.1}
\alpha_1&={\rm sign}(a_1)|a_1|^{1/3}=\frac{a_1}{|a_1|^\frac{2}{3}}~,\\
t&=\frac{r_1 \alpha_1}{3\cdot 2^\frac{2}{3}}~,\nn\\
u(t)&=\frac{t}{(1+t^3+\sqrt{1+2 t^3})^\frac{1}{3}}~.\nn
\end{align}
The square and cubic roots in the function $u(t)$ are defined such that $z^{1/2}=|z|^{1/2}\exp(i\,{\rm arg}z/2)$ and $z^{1/3}=|z|^{1/3}\exp(i\,{\rm arg}z/3)$, respectively, with ${\rm arg}z\in (-\pi,\pi]$ and ${\arg (-1)}=\pi$. Denoting the real and imaginary parts of $u$ as $u_r$ and $u_i$, respectively,
we can write the pole positions $p_i$, $i=1,\,2,\,3$,  as
\begin{align}
  \label{211027.2}
  p_1&=-\frac{i r_1}{6}(1+u+\frac{1}{u})=\frac{r_1 u_i}{6(u_r^2+u_i^2)}(u_r^2+u_i^2-1)
-i\frac{r_1}{6(u_r^2+u_i^2)}\left((1+u_r)(u_r^2+u_i^2)+u_r\right)~,\\
p_2&=\frac{r_1}{6}\left(e^{-i\frac{\pi}{6}}-e^{i\frac{\pi}{6}}+\frac{e^{i\frac{\pi}{6}}}{u}-u e^{-i\frac{\pi}{6}}\right)\nn\\
&=-\frac{r_1 (u_i+\sqrt{3}u_r)(u_r^2+u_i^2-1)}{12(u_r^2+u_i^2)}
+i\frac{r_1}{12(u_r^2+u_i^2)}\left(u_r+(-2+u_r)(u_r^2+u_i^2)-\sqrt{3}u_i(1+u_r^2+u_i^2)\right)~,\nn\\
p_3&=\frac{r_1}{6}\left(-e^{i\frac{\pi}{6}}+e^{-i\frac{\pi}{6}}-\frac{e^{-i\frac{\pi}{6}}}{u}+u e^{i\frac{\pi}{6}}\right)\nn\\
&=-\frac{r_1 (u_i-\sqrt{3}u_r)(u_r^2+u_i^2-1)}{12(u_r^2+u_i^2)}
+i\frac{r_1}{12(u_r^2+u_i^2)}\left(u_r+(-2+u_r)(u_r^2+u_i^2)+\sqrt{3}u_i(1+u_r^2+u_i^2)\right)~.\nn
\end{align}
It is clear that $u$ is real for $t>-1/2^{1/3}$ and then $p_2$ and $p_3$ are two resonant poles with the same imaginary part and opposite
real parts.
For $t<-1/2^{1/3}$ it results that $|u(t)|=1$ since }
\trr{\begin{align}
  \label{211028.1}
  |u(t)^3|&=\sqrt{\frac{t^6}{(1+t^3)^2-1-2t^3}}=1~, 
\end{align}}
\trr{Thus, it follows from Eq.~\eqref{211027.2} that all the $p_i$ are  purely imaginary since then $u_r^2+u_i^2=1$.}

\trr{We are interested in the forbidden regions of the parameter space $(a_1,r_1)$ that drives to resonant poles with positive imaginary part that could happen for $t>-1/2^{1/3}$. 
For real $u$ the imaginary parts of $p_{2,3}$ can also be written from Eq.~\eqref{211027.2} as
\begin{align}
  \label{211027.3}
  \Im p_{2,3}&=\frac{r_1}{12 u}(1-u)^2~.
\end{align}
Then, for $u>0$, which happens for $r_1 \alpha$ or $r_1 a_1>0$, the forbidden regions are those with $r_1>0$.
In the case of $u<0$ with $-1/2^{1/3}<t<0$, i.e. for $0>r_1\alpha_1>-(54)^{1/3}$, it is not allowed a negative $r_1$.
In terms of the dimensionless variables $-1/a_1 k_F^3$ and $r_1/k_F$ used in Fig.~\ref{Fig:P-waves_3D} this implies that  the regions that are forbidden are the top left quadrant and  $0>r_1/k_F>-(54/|a_1|k_F^3)^{1/3}$ in the  bottom left quadrant. 
We give the signs of $\Im p_{1,2}$ for $t>-1/2^{1/3}$ in  Table~\ref{tab.211027.1} as a function of the signs of $r_1$ and $a_1$.
}

\begin{table}
\trr{  \begin{center}
    \begin{tabular}{lll}
      \hline
      ${\rm sign}(r_1 a_1)$ & $({\rm sign}(r_1),{\rm sign}(a_1))$ & ${\rm sign}(\Im p_{2,3})$ \\
      $+$ & $(+,+)$ & $+$ \\
      & $(-,-)$  & $-$ \\
      $-$& $(+,-)$ & $-$ \\
      & $(-,+)$ & $+$\\
      \hline
    \end{tabular}
    \caption{{\small Sign of $\Im p_{2,3}$ for $t>-1/2^{1/3}$ giving rise to a real $u(t)$ and resonant poles $p_2$ and $p_3$. The signs of $a_1$ and $\alpha_1$ are the same.}\label{tab.211027.1}}
  \end{center}}
\end{table}

\trr{We also considered for generating the Fig.~\ref{Fig:P-waves_3D_r1_v1} the case with $a_1=\infty$ and finite $r_1$ and $v_2^{(1)}$. In this case, the pole positions are given by the zeroes of $r_1p^2/2+v_1^{(2)}p^4-ip^3$. Apart from the double zero at $p=0$ there are two other zeros when $r_1/2+v_1^{(2)}p^2-ip=0$\,, at the positions
\begin{align}
\label{211027.4}
  p_{1,2}=\frac{i}{2v_1^{(2)}}\pm \frac{1}{2v_1^{(2)}}\sqrt{-1-2r_1v_1^{(2)}}~.
\end{align}
The region to be excluded because of the appearance of resonant poles with
positive imaginary parts in their associated momenta stems from the conditions
i) $v_1^{(2)}>0$ and ii) $r_1 <-1/2v_1^{(2)}$. }

\bibliographystyle{ieeetr}
\bibliography{unitary.bib}

\begin{thebibliography}{10}

\bibitem{Oller:2001sn}
J.~A. Oller, ``{Chiral Lagrangians at finite density},'' {\em Phys. Rev. C},
  vol.~65, p.~025204, 2002.

\bibitem{ketterle.200831.1}
W.~Ketterle and M.~Zwierlein, ``{Making, probing and understanding ultracold
  {Fermi} gases},'' {\em Rivista del Nuovo Cimento}, vol.~31, pp.~247--422,
  2002.

\bibitem{Zwierlein2005VorticesAS}
M.~Zwierlein, J.~Abo-Shaeer, A.~Schirotzek, C.~Schunck, and W.~Ketterle,
  ``{Vortices and superfluidity in a strongly interacting {Fermi} gas},'' {\em
  Nature}, vol.~435, pp.~1047--1051, 2005.

\bibitem{Schunck2007SuperfluidEO}
C.~Schunck, M.~Zwierlein, A.~Schirotzek, and W.~Ketterle, ``{Superfluid
  expansion of a rotating {Fermi} gas},'' {\em Phys. Rev. Lett.}, vol.~98 5,
  p.~050404, 2007.

\bibitem{Anderson1995ObservationOB}
M.~Anderson, J.~Ensher, M.~Matthews, C.~Wieman, and E.~Cornell, ``{Observation
  of {Bose-Einstein} Condensation in a Dilute Atomic Vapor},'' {\em Science},
  vol.~269, pp.~198 -- 201, 1995.

\bibitem{davis.200901.1}
K.~Davis, M.~Mewes, M.~Andrews, N.~van Druten, D.~Durfee, D.~Kurn, and
  W.~Ketterle, ``{{Bose-Einstein} Condensation in a Gas of Sodium Atoms},''
  {\em Phys. Rev. Lett.}, vol.~75, 1995.

\bibitem{zwerger.200901.1}
E.~W.~Zwerger, {\em {The BCS-BEC Crossover and the Unitary {Fermi} Gas}}.
\newblock Lecture Notes in Physics 836, Springer-Verlag Berlin Heidelberg,
  2012.

\bibitem{Giorgini2008TheoryOU}
S.~Giorgini, L.~Pitaevskiĭ, and S.~Stringari, ``{Theory of ultracold atomic
  {Fermi} gases},'' {\em Rev. Mod. Phys.}, vol.~80, pp.~1215--1274, 2008.

\bibitem{Randeria2013BCSBECCA}
M.~Randeria and E.~Taylor, ``{Crossover from {Bardeen-Cooper-Schrieffer} to
  {Bose-Einstein} Condensation and the Unitary {Fermi} Gas},'' {\em Annu. Rev.
  Condens. Matter. Phys}, vol.~5, pp.~209--32, 2014.

\bibitem{Bethe.200901.1}
H.~Bethe, ``{Theory of the Effective Range in Nuclear Scattering},'' {\em Phys.
  Rev.}, vol.~76, p.~38, 1949.

\bibitem{Navon2010TheEO}
N.~Navon, S.~Nascimb{\`e}ne, F.~Chevy, and C.~Salomon, ``{The Equation of State
  of a Low-Temperature {Fermi} Gas with Tunable Interactions},'' {\em Science},
  vol.~328, pp.~729 -- 732, 2010.

\bibitem{Ku2012RevealingTS}
M.~Ku, A.~T. Sommer, L.~Cheuk, and M.~Zwierlein, ``{Revealing the Superfluid
  Lambda Transition in the Universal Thermodynamics of a Unitary {Fermi}
  Gas},'' {\em Science}, vol.~335, pp.~563 -- 567, 2012.

\bibitem{Zuern2013PreciseCO}
G.~Zuern, T.~Lompe, A.~Wenz, S.~Jochim, P.~Julienne, and J.~Hutson, ``{Precise
  characterization of {$^6$Li} Feshbach resonances using trap-sideband-resolved
  RF spectroscopy of weakly bound molecules.},'' {\em Phys. Rev. Lett.},
  vol.~110 13, p.~135301, 2013.

\bibitem{Kolck2017UnitarityAD}
U.~van Kolck, ``{Unitarity and Discrete Scale Invariance},'' {\em Few-Body
  Systems}, vol.~58, pp.~1--12, 2017.

\bibitem{tan27}
T.-L. Ho, ``{Universal Thermodynamics of Degenerate Quantum Gases in the
  Unitarity Limit},'' {\em Phys. Rev. Lett.}, vol.~92, p.~090402, 2004.

\bibitem{Forbes:2010gt}
M.~M. Forbes, S.~Gandolfi, and A.~Gezerlis, ``{Resonantly Interacting Fermions
  In a Box},'' {\em Phys. Rev. Lett.}, vol.~106, p.~235303, 2011.

\bibitem{Forbes:2012ku}
M.~M. Forbes, S.~Gandolfi, and A.~Gezerlis, ``{Effective-Range Dependence of
  Resonantly Interacting Fermions},'' {\em Phys. Rev. A}, vol.~86, p.~053603,
  2012.

\bibitem{Carlson:2011kv}
J.~Carlson, S.~Gandolfi, K.~E. Schmidt, and S.~Zhang, ``{Auxiliary Field
  Quantum {Monte Carlo} for Strongly Paired Fermions},'' {\em Phys. Rev. A},
  vol.~84, p.~061602, 2011.

\bibitem{Nishida:2006br}
Y.~Nishida and D.~T. Son, ``{An Epsilon expansion for {Fermi} gas at infinite
  scattering length},'' {\em Phys. Rev. Lett.}, vol.~97, p.~050403, 2006.

\bibitem{Nishida:2006eu}
Y.~Nishida and D.~T. Son, ``{{Fermi} gas near unitarity around four and two
  spatial dimensions},'' {\em Phys. Rev. A}, vol.~75, p.~063617, 2007.

\bibitem{Nishida:2008mh}
Y.~Nishida, ``{Ground-state energy of the unitary {Fermi} gas from the epsilon
  expansion},'' {\em Phys. Rev. A}, vol.~79, p.~013627, 2009.

\bibitem{Papenbrock:2005bd}
T.~Papenbrock, ``{Density-functional theory for fermions in the unitary
  regime},'' {\em Phys. Rev. A}, vol.~72, p.~041603, 2005.

\bibitem{Lacroix:2016dfs}
D.~Lacroix, ``{Density-functional theory for resonantly interacting fermions
  with effective range and neutron matter},'' {\em Phys. Rev. A}, vol.~94,
  no.~4, p.~043614, 2016.

\bibitem{Boulet:2019wfd}
A.~Boulet and D.~Lacroix, ``{Approximate self-energy for {Fermi} systems with
  large s-wave scattering length: a step towards density functional theory},''
  {\em J. Phys. G}, vol.~46, no.~10, p.~105104, 2019.

\bibitem{Grasso:2018pen}
M.~Grasso, ``{Effective density functionals beyond mean field},'' {\em Prog.
  Part. Nucl. Phys.}, vol.~106, pp.~256--311, 2019.

\bibitem{fetter}
A.~Fetter and J.~Walecka, {\em {Quantum Theory of Many-Particle Systems}}.
\newblock McGraw-Hill, New York, 1971.

\bibitem{Huang:1957im}
K.~Huang and C.~Yang, ``{Quantum-mechanical many-body problem with hard-sphere
  interaction},'' {\em Phys. Rev.}, vol.~105, pp.~767--775, 1957.

\bibitem{Lee:1957zza}
T.~Lee and C.~Yang, ``{Many-Body Problem in Quantum Mechanics and Quantum
  Statistical Mechanics},'' {\em Phys. Rev.}, vol.~105, pp.~1119--1120, 1957.

\bibitem{efimov:1965}
V.~N. Efimov and M.~Y. Amusya, ``{Ground state of a rarefied {Fermi} gas of
  rigid spheres},'' {\em Sov. Phys. JETP}, vol.~20, p.~388, 1965.

\bibitem{efimov:1968}
M.~Y. Amusya and V.~N. Efimov, ``{Pair collisions in a low-density fermi
  gas},'' {\em Ann. Phys. (NY)}, vol.~47, pp.~377--403, 1968.

\bibitem{Baker:1971vm}
G.~A. Baker, ``{Singularity structure of the perturbation series for the
  ground-state energy of a many-fermion system},'' {\em Rev. Mod. Phys.},
  vol.~43, pp.~479--531, 1971.

\bibitem{bishop:1973}
B.~F. Bishop, ``{Ground-state energy of a dilute fermi gas},'' {\em Ann. Phys.
  (NY)}, vol.~77, pp.~106--138, 1973.

\bibitem{Hammer:2000xg}
H.~Hammer and R.~Furnstahl, ``{Effective field theory for dilute {Fermi}
  systems},'' {\em Nucl. Phys. A}, vol.~678, pp.~277--294, 2000.

\bibitem{Chen:2008zzj}
Q.~Chen {\em et~al.}, ``{Measurement of the neutron-neutron scattering length
  using the pi-d capture reaction},'' {\em Phys. Rev. C}, vol.~77, p.~054002,
  2008.

\bibitem{Lacour:2009ej}
A.~Lacour, J.~A. Oller, and U.-G. Mei{\ss}ner, ``{Non-perturbative methods for
  a chiral effective field theory of finite density nuclear systems},'' {\em
  Annals Phys.}, vol.~326, pp.~241--306, 2011.

\bibitem{Dobado:2011gd}
A.~Dobado, F.~J. Llanes-Estrada, and J.~A. Oller, ``{The existence of a
  two-solar mass neutron star constrains the gravitational constant $G_N$ at
  strong field},'' {\em Phys. Rev. C}, vol.~85, p.~012801, 2012.

\bibitem{bethebrueckner}
H.~Bethe, ``{Introduction to the Brueckner theory},'' {\em Physics}, vol.~22,
  pp.~987 -- 993, 1956.

\bibitem{Brueckner:1954zz}
K.~Brueckner, C.~Levinson, and H.~Mahmoud, ``{Two-Body Forces and Nuclear
  Saturation. I. Central Forces},'' {\em Phys. Rev.}, vol.~95, pp.~217--228,
  1954.

\bibitem{Brueckner:1955nst}
K.~Brueckner, ``{Nuclear Saturation and Two-Body Forces. II. Tensor Forces},''
  {\em Phys. Rev.}, vol.~96, pp.~508--516, 1954.

\bibitem{Brueckner:1955zze}
K.~Brueckner, ``{Two-Body Forces and Nuclear Saturation. III. Details of the
  Structure of the Nucleus},'' {\em Phys. Rev.}, vol.~97, pp.~1353--1366, 1955.

\bibitem{Thouless:1960anp}
D.~Thouless, ``{Perturbation Theory in Statistical Mechanics and the Theory of
  Superconductivity},'' {\em Ann. Phys.}, vol.~10, pp.~553--588, 1960.

\bibitem{Steele:2000qt}
J.~V. Steele, ``{Effective field theory power counting at finite density},''
  nucl-th/0010066, 2000.

\bibitem{Kaiser:2011cg}
N.~Kaiser, ``{Resummation of fermionic in-medium ladder diagrams to all
  orders},'' {\em Nucl. Phys. A}, vol.~860, pp.~41--55, 2011.

\bibitem{Kaiser:2012sr}
N.~Kaiser, ``{Resummation of in-medium ladder diagrams: s-wave effective range
  and p-wave interaction},'' {\em Eur. Phys. J. A}, vol.~48, p.~148, 2012.

\bibitem{pines}
D.~Pines and P.~Nozieres, {\em {Theory of Quantum Liquids: Normal {Fermi}
  Liquids}}.
\newblock W.A. Benjaim, INC., New York, 1966.

\bibitem{migdal}
A.~B. Migdal, {\em {Nuclear Theory: The Quasiparticle Method}}.
\newblock W.A. Benjaim, INC., New York, 1968.

\bibitem{Carlson:2003wm}
J.~Carlson, J.~Morales, J., V.~Pandharipande, and D.~Ravenhall, ``{Quantum
  {Monte Carlo} calculations of neutron matter},'' {\em Phys. Rev. C}, vol.~68,
  p.~025802, 2003.

\bibitem{Gezerlis:2009iw}
A.~Gezerlis and J.~Carlson, ``{Low-density neutron matter},'' {\em Phys. Rev.
  C}, vol.~81, p.~025803, 2010.

\bibitem{Schwenk:2005ka}
A.~Schwenk and C.~Pethick, ``{Resonant {Fermi} gases with a large effective
  range},'' {\em Phys. Rev. Lett.}, vol.~95, p.~160401, 2005.

\bibitem{Konig:2016utl}
S.~König, H.~W. Grießhammer, H.~Hammer, and U.~van Kolck, ``{Nuclear Physics
  Around the Unitarity Limit},'' {\em Phys. Rev. Lett.}, vol.~118, no.~20,
  p.~202501, 2017.

\bibitem{vanKolck:1998bw}
U.~van Kolck, ``{Effective field theory of short range forces},'' {\em Nucl.
  Phys. A}, vol.~645, pp.~273--302, 1999.

\bibitem{Schafer:2005kg}
T.~Schäfer, C.-W. Kao, and S.~R. Cotanch, ``{Many body methods and effective
  field theory},'' {\em Nucl. Phys. A}, vol.~762, pp.~82--101, 2005.

\bibitem{Kaplan:1998we}
D.~B. Kaplan, M.~J. Savage, and M.~B. Wise, ``{Two nucleon systems from
  effective field theory},'' {\em Nucl. Phys. B}, vol.~534, pp.~329--355, 1998.

\bibitem{Stratonovich:1957dan}
R.~Stratonovich, ``{On a Method of Calculating Quantum Distribution
  Functions},'' {\em Soviet Physics Doklady}, vol.~2, p.~416, 1957.

\bibitem{Hubbard:1959ub}
J.~Hubbard, ``{Calculation of partition functions},'' {\em Phys. Rev. Lett.},
  vol.~3, pp.~77--80, 1959.

\bibitem{Bernard:1995dp}
V.~Bernard, N.~Kaiser, and U.-G. Meissner, ``{Chiral dynamics in nucleons and
  nuclei},'' {\em Int. J. Mod. Phys. E}, vol.~4, pp.~193--346, 1995.

\bibitem{Machleidt:2011zz}
R.~Machleidt and D.~R. Entem, ``{Chiral effective field theory and nuclear
  forces},'' {\em Phys. Rept.}, vol.~503, pp.~1--75, 2011.

\bibitem{Meissner:2001gz}
U.-G. Mei{\ss}ner, J.~A. Oller, and A.~Wirzba, ``{In-medium chiral perturbation
  theory beyond the mean field approximation},'' {\em Annals Phys.}, vol.~297,
  pp.~27--66, 2002.

\bibitem{incoming}
J.~M. Alarc\'on and J.~A. Oller, ``{Properties of nuclear matter},'' {\em in
  preparation}, 2021.

\bibitem{oller.book}
J.~A. Oller, {\em {A Brief Introduction to Dispersion Relations}}.
\newblock SpringerBriefs in Physics, Springer, 2019.

\bibitem{Oller:2017alp}
J.~A. Oller, ``{New results from a number operator interpretation of the
  compositeness of bound and resonant states},'' {\em Annals Phys.}, vol.~396,
  pp.~429--458, 2018.

\bibitem{Oller:2019opk}
J.~A. Oller, ``{{Coupled-channel approach in hadron--hadron scattering}},''
  {\em Prog. Part. Nucl. Phys.}, vol.~110, p.~103728, 2020.

\bibitem{Phillips:1997xu}
D.~R. Phillips, S.~R. Beane, and T.~D. Cohen, ``{Nonperturbative regularization
  and renormalization: Simple examples from nonrelativistic quantum
  mechanics},'' {\em Annals Phys.}, vol.~263, pp.~255--275, 1998.

\bibitem{Entem:2007jg}
D.~R. Entem, E.~Ruiz~Arriola, M.~Pavon~Valderrama, and R.~Machleidt,
  ``{Renormalization of chiral two-pion exchange NN interactions. momentum
  versus coordinate space},'' {\em Phys. Rev. C}, vol.~77, p.~044006, 2008.

\bibitem{Habashi:2020qgw}
J.~B. Habashi, S.~Sen, S.~Fleming, and U.~van Kolck, ``{Effective Field Theory
  for Two-Body Systems with Shallow S-Wave Resonances},'' {\em Annals Phys.},
  vol.~422, p.~168283, 2020.

\bibitem{Oller:2018zts}
J.~A. Oller and D.~Entem, ``{{The exact discontinuity of a partial wave along
  the left-hand cut and the exact $N/D$ method in non-relativistic
  scattering}},'' {\em Annals Phys.}, vol.~411, p.~167965, 2019.

\bibitem{tan25}
G.~A. {Baker Jr.}, ``{Neutron matter model},'' {\em Phys. Rev. C}, vol.~60,
  p.~054311, 1999.

\bibitem{tan26}
K.~O’Hara and {\it et al}, ``{Observation of a strongly interacting
  degenerate {Fermi} gas of atoms},'' {\em Science}, vol.~298, p.~2179, 2002.

\bibitem{Maki:2020zsv}
J.~Maki and S.~Zhang, ``{The Role of the Effective Range in Resonantly
  Interacting {Fermi} Gases: How Breaking Scale Symmetry Affects the Bulk
  Viscosity},'' {\em Phys. Rev. Lett.}, vol.~125, p.~240402, 2020.

\bibitem{Oller:2014uxa}
J.~A. Oller, ``{Nucleon-Nucleon scattering from dispersion relations:
  next-to-next-to-leading order study},'' {\em Phys. Rev. C}, vol.~93,
  p.~024002, 2016.

\bibitem{Bertulani:2002sz}
C.~A. Bertulani, H.~W. Hammer, and U.~Van~Kolck, ``{Effective field theory for
  halo nuclei},'' {\em Nucl. Phys. A}, vol.~712, pp.~37--58, 2002.

\bibitem{Hu:2019wrm}
J.~Hu, F.~Wu, L.~He, X.-J. Liu, and H.~Hu, ``{Theory of strongly paired
  fermions with arbitrary short-range interactions},'' {\em Phys. Rev. A},
  vol.~101, no.~1, p.~013615, 2020.

\bibitem{thomas:2012}
W.~Haibin and J.~Thomas, ``{Optical Control of Feshbach Resonances in {Fermi}
  Gases Using Molecular Dark States},'' {\em Phys. Rev. Lett.}, vol.~108,
  p.~010401, 2012.

\bibitem{gottfried.book}
K.~Gottfried, {\em {Quantum Mechanics. Vol. I. Fundamentals}}.
\newblock Advanced Book Classics. Addison-Wesley Publishing Company, Reading,
  Massachusetts, 1989.

\bibitem{ma:1946}
S.~T. Ma, ``{Redundant Zeros in the Discrete Energy Spectra in Heisenberg's
  Theory of Characteristic Matrix},'' {\em Phys. Rev.}, vol.~69, p.~668, 1946.

\bibitem{Ma:1947zz}
S.~T. Ma, ``{On a General Condition of Heisenberg for the S Matrix},'' {\em
  Phys. Rev.}, vol.~71, pp.~195--200, 1947.

\bibitem{Chang:2004zza}
S.~Chang, V.~Pandharipande, J.~Carlson, and K.~Schmidt, ``{Quantum {Monte
  Carlo} Studies of Superfluid {Fermi} Gases},'' {\em Phys. Rev. A}, vol.~70,
  p.~043602, 2004.

\bibitem{Astrakharchik:2004zz}
G.~Astrakharchik, J.~Boronat, J.~Casulleras, and S.~Giorgini, ``{Equation of
  State of a {Fermi} Gas in the BEC-BCS Crossover: A Quantum {Monte Carlo}
  Study},'' {\em Phys. Rev. Lett.}, vol.~93, p.~200404, 2004.

\bibitem{tan30}
A.~Bulgac and G.~F. Bertsch, ``{Collective Oscillations of a Trapped {Fermi}
  Gas near the Unitary Limit},'' {\em Phys. Rev. Lett.}, vol.~94, p.~070401,
  2004.

\bibitem{Tan:2008a}
S.~Tan, ``{Energetics of a strongly correlated {Fermi} gas},'' {\em Ann. Phys.
  (N.Y.)}, vol.~323, pp.~2925--2970, 2008.

\bibitem{Gandolfi:2011}
S.~Gandolfi, K.~E. Schmidt, and J.~Carlson, ``{BEC-BCS crossover and universal
  relations in unitary {Fermi} gases},'' {\em Phys. Rev. A}, vol.~83,
  p.~041601(R), 2011.

\bibitem{Gezerlis:2007fs}
A.~Gezerlis and J.~Carlson, ``{Strongly paired fermions: Cold atoms and neutron
  matter},'' {\em Phys. Rev. C}, vol.~77, p.~032801, 2008.

\bibitem{Miller:1990iz}
G.~Miller, B.~Nefkens, and I.~Slaus, ``{Charge symmetry, quarks and mesons},''
  {\em Phys. Rept.}, vol.~194, pp.~1--116, 1990.

\bibitem{Perez:2014waa}
R.~Navarro~P\'erez, J.~E. Amaro, and E.~Ruiz~Arriola, ``{The low-energy
  structure of the nucleon\textendash{}nucleon interaction: statistical versus
  systematic uncertainties},'' {\em J. Phys. G}, vol.~43, no.~11, p.~114001,
  2016.

\bibitem{rothe.lattice}
H.~Rothe, {\em {Lattice Gauge Theories: An Introduction}}.
\newblock World Scientific, Singapure, 1992.

\bibitem{Lacroix:2017whm}
D.~Lacroix, A.~Boulet, M.~Grasso, and C.-J. Yang, ``{From bare interactions,
  low-energy constants, and unitary gas to nuclear density functionals without
  free parameters: Application to neutron matter},'' {\em Phys. Rev. C},
  vol.~95, no.~5, p.~054306, 2017.

\bibitem{Castin:2012}
F.~Werner and Y.~Castin, ``{General relations for quantum gases in two and
  three dimensions: Two-component fermions},'' {\em Phys. Rev. A}, vol.~86,
  p.~013626, 2012.

\bibitem{Beane:2000fi}
S.~R. Beane and M.~J. Savage, ``{Rearranging pionless effective field
  theory},'' {\em Nucl. Phys. A}, vol.~694, pp.~511--524, 2001.

\bibitem{Akmal:1998cf}
A.~Akmal, V.~Pandharipande, and D.~Ravenhall, ``{The Equation of state of
  nucleon matter and neutron star structure},'' {\em Phys. Rev. C}, vol.~58,
  pp.~1804--1828, 1998.

\bibitem{Werner:2009}
F.~Werner, L.~Tarruell, and Y.~Castin, ``{Number of closed-channel molecules in
  the BEC-BCS crossover},'' {\em Eur. Phys. J. B}, vol.~68, p.~401, 2009.

\bibitem{Haussmann:2009}
R.~Haussmann, M.~Punk, and W.~Zwerger, ``{Spectral Functions and rf Response of
  Ultracold Fermionic Atoms},'' {\em Phys. Rev. A}, vol.~80, p.~063612, 2009.

\bibitem{Kuhnle:2010}
E.~{\it et al.}. Kuhnle, ``{Universal behavior of pair correlations in a
  strongly interacting {Fermi} gas},'' {\em Phys. Rev. Lett.}, vol.~105,
  p.~070402, 2010.

\bibitem{braaten:2012}
E.~Braaten, {\em {{\rm In} The BCS-BEC Crossover and the Unitary {Fermi} Gas,
  {\rm edited by W. Zwerger, Lectures Notes in Physics}}}, vol.~86, {\rm
  Chap.~6}.
\newblock Springer-Verlag, Berlin, 2012.

\bibitem{Kaplan:1996nv}
D.~B. Kaplan, ``{More effective field theory for nonrelativistic scattering},''
  {\em Nucl. Phys. B}, vol.~494, pp.~471--484, 1997.

\bibitem{apostol.211016}
T.~M. Apostol, {\em {Mathematical analysis, 2nd edition}}.
\newblock Pearson, 1974.

\bibitem{Heiselberg:2001}
H.~Heiselberg, ``{{Fermi} systems with long scattering lengths},'' {\em Phys.
  Rev. A}, vol.~63, p.~043606, 2001.

\end{thebibliography}

\end{document}